\documentclass[longauth]{aa}  
\usepackage{graphicx}
\usepackage{txfonts}
\usepackage{lscape}
\usepackage{natbib}
\usepackage[bookmarks=true]{hyperref}% To add links in your PDF file, use the package "hyperref"
\makeatletter
\renewcommand*\aa@pageof{, page \thepage{} of \pageref*{LastPage}}
\makeatother

\begin{document}

   \title{REsolved ALMA and SMA Observations of Nearby Stars}

   \subtitle{REASONS: A population of 74 resolved planetesimal belts at millimetre wavelengths}

   \author{L. Matr\`a
          \inst{1}
          \and
          S. Marino\inst{2}
          \and 
          D. J. Wilner\inst{3}
          \and
          G. M. Kennedy\inst{4}
          \and
          M. Booth\inst{5,6}
          \and
          A. V. Krivov\inst{6}
          \and
          J. P. Williams\inst{7}
          \and
          A. M. Hughes\inst{8}
          \and
          C. del Burgo\inst{9,10}
          \and
          J. Carpenter\inst{11}
          \and
          C. L. Davies\inst{2}
          \and
          S. Ertel\inst{12,13}
          \and
          Q. Kral\inst{14}
          \and
          J.-F. Lestrade\inst{15}
          \and
          J. P. Marshall\inst{16}
          \and
          J. Milli\inst{17}
          \and
          K. I. \"Oberg\inst{3}
          \and
          N. Pawellek\inst{18,19}
          \and
          A. G. Sepulveda\inst{20}
          \and
          M. C. Wyatt\inst{21}
          \and
          B. C. Matthews\inst{22,23}
          \and
          M. MacGregor\inst{24}
          }

   \institute{School of Physics, Trinity College Dublin, the University of Dublin, College Green, Dublin 2, Ireland\\
              \email{lmatra@tcd.ie}
         \and
             Department of Physics and Astronomy, University of Exeter, Stocker Road, Exeter EX4 4QL, UK
         \and
             Center for Astrophysics $|$ Harvard \& Smithsonian, 60 Garden Street, Cambridge, MA 02138, USA
         \and
             Department of Physics, University of Warwick, Gibbet Hill Road, Coventry, CV4 7AL, UK
         \and
             UK Astronomy Technology Centre, Royal Observatory Edinburgh, Blackford Hill, Edinburgh EH9 3HJ, UK
         \and
             Astrophysikalisches Institut und Universit$\ddot{\rm a}$tssternwarte, Friedrich$-$Schiller$-$Universit$\ddot{\rm a}$t \ Jena, \ Schillerg$\ddot{\rm a}$\ss chen \ 2$-$3, \ D$-$07745 Jena, Germany
          \and
             Institute for Astronomy, University of Hawaii, Honolulu, HI 96822, USA
          \and
             Department of Astronomy, Van Vleck Observatory, Wesleyan University, 96 Foss Hill Dr., Middletown, CT 06459, USA
          \and
             Instituto de Astrof\'\i sica de Canarias, V\'\i a L\'actea S/N, La Laguna, E-38200, Tenerife, Spain
          \and
             Departamento de Astrof\'\i sica, Universidad de la Laguna, La Laguna, E-38200, Tenerife, Spain
          \and
             Joint ALMA Observatory, Avenida Alonso de C\'ordova 3107, Vitacura, Santiago, Chile
          \and
             Department of Astronomy and Steward Observatory, University of Arizona, 933 N. Cherry Avenue, Tucson, AZ 85721-0065, USA
          \and
             Large Binocular Telescope Observatory, University of Arizona, 933 N. Cherry Avenue, Tucson, AZ 85721-0065, USA
          \and
             LESIA, Observatoire de Paris, Universit\'e PSL, CNRS, Universit\'e Paris Cit\'e, Sorbonne Universit\'e, 5 place Jules Janssen, 92195 Meudon, France 
          \and
             LERMA, Observatoire de Paris, PSL Research University, CNRS, Sorbonne Universit\'e, UPMC, 75014 Paris, France
          \and
             Academia Sinica Institute of Astronomy and Astrophysics, 11F of AS/NTU Astronomy-Mathematics Building, No.1, Section 4, Roosevelt Road, Taipei 106216, Taiwan
          \and
             Univ. Grenoble Alpes, CNRS, IPAG, 38000, Grenoble, France
          \and
             Institut f\"ur Astrophysik, Universit\"at Wien, T\"urkenschanzstra\ss{}e 17, 1180 Vienna, Austria
          \and
             Konkoly Observatory, HUN-REN Research Centre for Astronomy and Earth Sciences, Konkoly-Thege Mikl\'os \'ut 15-17, 1121 Budapest, Hungary
          \and
             The University of Texas School of Law. 727 E. Dean Keeton Street Austin, Texas 78705, USA
          \and
             Institute of Astronomy, University of Cambridge, Madingley Road, Cambridge CB3 0HA, UK
          \and
             Herzberg Astronomy \& Astrophysics, National Research Council of Canada, 5071 West Saanich Road, Victoria, BC, V9E 2E9, Canada
          \and
             Department of Physics \& Astronomy, University of Victoria, 3800 Finnerty Rd, Victoria, BC V8P 5C2, Canada
          \and
             Department of Physics and Astronomy, Johns Hopkins University, Baltimore, MD 21218, USA
             }

   \date{Received 8 July 2024; Accepted 9 October 2024}

% \abstract{}{}{}{}{} 
% 5 {} token are mandatory
 
  \abstract
  % context heading (optional)
  % {} leave it empty if necessary  
   {Planetesimal belts are ubiquitous around nearby stars, and their spatial properties hold crucial information for planetesimal and planet formation models.}
  % aims heading (mandatory)
   {We present resolved dust observations of 74 planetary systems as part of the REsolved ALMA and SMA Observations of Nearby Stars (REASONS) survey and archival reanalysis.}
  % methods heading (mandatory)
   {We uniformly modelled interferometric visibilities for the entire sample to obtain the basic spatial properties of  each belt, and combined these with constraints from multi-wavelength photometry.}
  % results heading (mandatory)
   {We report key findings from a first exploration of this legacy dataset: (1) Belt dust masses are depleted over time in a radially dependent way, with dust being depleted faster in smaller belts, as predicted by collisional evolution. (2) Most belts are broad discs rather than narrow rings, with much broader  fractional widths than rings in protoplanetary discs. We link broad belts to either unresolved substructure or broad planetesimal discs produced if protoplanetary rings migrate. (3) The vertical aspect ratios ($h=H/R$) of 24 belts indicate orbital inclinations of $\sim$1-20$^{\circ}$, implying relative particle velocities of $\sim$0.1-4 km/s, and no clear evolution of heights with system age. This could be explained by early stirring within the belt by large bodies (with sizes of at least $\sim$140 km to the size of the Moon), by inheritance of inclinations from the protoplanetary disc stage, or by a diversity in evolutionary pathways and gravitational stirring mechanisms. We release the REASONS legacy multidimensional sample of millimetre-resolved belts to the community as a valuable tool for follow-up multi-wavelength observations and population modelling studies.}
  % conclusions heading (optional), leave it empty if necessary 
   {}


   \maketitle

%
%-------------------------------------------------------------------
\section{Introduction}
\label{sec:intro}
%\LEt{*** The English in this work was already very good, and it did not require full language editing. I have corrected the abstract, introduction, conclusion, and table and the first ten figure captions in full, made non-exhaustive edits elsewhere, and added instructions for the paper that we ask you to apply throughout. I have edited to the UK spelling convention, as this is the convention typically used already here.***}
The planet formation process efficiently produces planetesimal belts, or debris discs, which are extrasolar analogues of the Kuiper and asteroid belts of the Solar System. Their ubiquity is typically inferred from surveys of infrared excess above the stellar photospheric Rayleigh-Jeans tail \citep[]{Aumann1985} around stars in the solar neighbourhood (typically within $\sim$150 pc of Earth). These surveys indicate an occurrence rate for cold Kuiper belt analogues of at least $\sim17-33$\% \citep{Su2006,Eiroa2013, Thureau2014,Sibthorpe2018}, and of potentially as high as $\sim75$\% as observed in the younger, less collisionally evolved belts \citep{Pawellek2021}.

The short lifetime of the observable dust, which is rapidly removed by the combined effect of collisions and radiation pressure from the central  star, implies that a replenishment mechanism is necessary \citep[][and references therein]{Backman1993}. Dust in planetesimal belts is thus of second generation, being produced by collisions of larger bodies within a collisional cascade \citep{Wyatt2002, DominikDecin2003} and eventually removed, typically by radiation pressure \citep[e.g.][]{Thebault2003, Krivov2006, Wyatt2007b}. Overall, mass is expected to be lost through the collisional cascade, with infrared excesses eventually decaying with planetary system age ---although the steepness of this mass decay and its initial time evolution are dependent on the details of the belt evolution model \citep[e.g.][]{Wyatt2002,Krivov2008, Lohne2008,Kenyon2008,Kenyon2010, KobayashiLohne2014,Najita2022}. Surveys generally show dust mass loss (dimming of IR excess) over time \citep[e.g.][]{Carpenter2009, Holland2017}, which, at present, can be explained by a simple, steady state collisional evolution model, where detectable belts start bright, keep their brightness until the largest planetesimals in the cascade have collided, and subsequently decay in brightness following a mass depletion of roughly  $t^{-1}$ with time $t$ \citep[e.g.][]{Wyatt2007a,Najita2022}; though some models predict a shallower $\sim t^{-0.4}$ mass evolution \citep[e.g.][]{Lohne2008, Kral2013}.

Multi-wavelength photometry from mid-infrared (MIR) to millimetre (mm) wavelengths constrains the dust temperature in the majority of belts to approximately a few tens of Kelvin to 120 K \citep[e.g.][]{Ballering2013}. Assuming this emission originates from blackbody-like grains would imply that they lie in the $\sim$10-100 au region of planetary systems.  At these distances and temperatures, belts are expected be volatile rich, and are therefore expected be populated by icy exocomets \citep[e.g.][]{Lebreton2012}; this is now corroborated by the ubiquity of CO gas in belts observed at sufficient sensitivity \citep[e.g.][]{Matra2019a}, whose origin lies in exocometary release for at least some (but not necessarily all) belts \citep[e.g.][]{Zuckerman2012, Matra2015, Matra2017b, Kral2017, Marino2020}. The majority of observed belts are therefore cold Kuiper belt analogues, although a number of systems also present warmer ($>120$ K) MIR emission that may originate from dust closer to the star and potentially produced within asteroid belt analogues at a few astronomical units %\LEt{***Please merge these two parentheses using a semicolon to separate their contents.***}
\citep[au; e.g.][]{Chen2014}. 

Early imaging confirmed the inference from unresolved photometry, locating belts at tens of au from the central star \citep{Smith1984, Koerner1998, Holland1998}. The advent of facilities with higher sensitivity and resolution (jointly key to imaging low-surface-brightness emission from planetesimal belts) led to a significant expansion of the number of imaged belts, with observations in optical/near-infrared(NIR) scattered light with the \textit{Hubble} Space Telescope \citep[HST; e.g.][]{Soummer2014, Schneider2014}, the Gemini Planet Imager \citep[GPI; e.g.][]{Esposito2020}, and the Spectro-Polarimetric High-contrast Exoplanet REsearch instrument \citep[SPHERE; e.g.][]{Dahlqvist2022}; in far-infrared (FIR) with the \textit{Herschel} Space Telescope \citep[e.g.][]{Booth2013, Morales2016, Marshall2021}; and at mm wavelengths with the \textit{James Clerk Maxwell} Telescope (JCMT), the Combined Array for Research in Millimeter-wave Astronomy (CARMA), the Submillimeter Array (SMA), and the Atacama Large Millimeter/submillimeter Array \citep[ALMA; e.g.][]{Holland2017,Steele2016,Lieman-Sifry2016,Matra2018b}. These surveys show that belts are typically detected at radii that are larger than inferred from unresolved photometry in the blackbody grain assumption by a (system-dependent) factor of up to a few \citep{Booth2013, Pawellek2014,Matra2018b}. 

The next step towards a comprehensive understanding of the planetesimal belt population, its origin, and its evolution is to resolve as many belts as possible. Such surveys should enable empirical constraints on belt evolution and should help us to understand how this evolution depends on stellar and belt properties. For example, collisional evolution models predict a dependence of collisional mass loss on belt radius \citep[e.g.][]{Wyatt2007a,Kenyon2008,Lohne2008,Kennedy2010} and dynamical excitation, which could be probed by vertically resolved observations \citep[][]{Matra2019b,Daley2019} or indirectly from their outer edges \citep{Marino2021}. When disentangled from collisional evolution and observational bias, resolved radial information could also yield crucial information on the birth location of planetesimal belts, informing planet and planetesimal formation processes \citep[]{Matra2018b}.

Motivated by the need for a larger sample of belts for population modelling studies, we present the REsolved ALMA and SMA Observations of Nearby Stars (REASONS) observing programme and archival reanalysis, presenting a uniform analysis of the planetesimal belts resolved so far using mm and sub-mm interferometry. This wavelength choice ensures that most of the emitting dust grains are not affected by radiation forces, and are therefore tracing the parent planetesimals. Additional benefits of this choice include the fact that stellar emission is faint or undetected in the majority of systems, leaving belt imaging unaffected (as opposed to shorter wavelength observations), and that resolution is sufficient to resolve belts across their width \citep[as opposed to \textit{Herschel}, whose limited resolution resolved mostly outer edges; e.g.][]{Kennedy2015, Moor2015, Marshall2021}.

Section \ref{sec:targetsel} introduces the REASONS sample, detailing aspects of observational bias and selection that should be considered in later analyses and future modelling studies. In Sect. \ref{sec:obs} we describe new ALMA and SMA observations, as well as archival observations reanalysed in this work. Section \ref{sec:contres} presents the gallery of resolved images and the uniform modelling of interferometric visibilities and multi-wavelength photometry carried out for the whole sample, results of which we release to the community. In Sect. \ref{sec:disc} we discuss certain trends and population properties of particular interest arising from the sample, before concluding with a summary of our findings in Sect. \ref{sec:summ}.

%--------------------------------------------------------------------
\section{Target selection and bias}%: the REASONS observing and archival programmes, and the REASONS sample} 
\label{sec:targetsel}

\subsection{The sample}
\label{sec:sample}
The REASONS observing programme observed 25 planetesimal belts interferometrically at mm wavelengths (1.27 mm) for the first time; 15 with ALMA (Sect. \ref{sec:almaobs}), and 10 with the SMA (Sect. \ref{sec:smaobs}). These observations, combined with archival observations, complete a resolved follow-up census of a flux-limited sample of sources detected at (sub-)mm wavelengths by the SCUBA-2 Observations of Nearby Stars (SONS) JCMT Legacy Survey \citep[detection threshold of $\gtrsim3$ mJy at 850 $\mu$m,][]{Holland2017}. Within the declination limits imposed by Mauna Kea observations \citep[-40$^{\circ}$ to +80$^{\circ}$ declination, with a few exceptions for bright targets; see][for details]{Holland2017}, the goal of the REASONS observing programme was to resolve all planetesimal belts previously detected at IR wavelengths and brighter than 3 mJy at 850 $\mu$m (or 1 mJy at 1.3 mm for a spectral slope $\alpha$ of 2.5). Of these 25 targets, 15 were resolved, and 10 (reported in Appendix \ref{sec:confused}) were either too low surface brightness for their spatial properties to be characterised, and/or contaminated.

In addition to the REASONS observing programme, we undertook an archival reanalysis effort (REASONS archival programme) to ensure uniformity of analysis and modelling for as large a population of mm-resolved belts as possible. As part of the archival programme, we 1) reanalysed SONS targets that had already been resolved interferometrically, and 2) analysed ALMA, SMA and/or CARMA archival data of planetesimal belts that became public before June 1 2020, or that became public more recently and have already been published in the literature. This broader sample includes belts that were not part of the SONS sample (mostly because they have a declination too southern for the JCMT) from a variety of programmes with different goals. We only report on archival observations of belts that were detected and resolved, as defined in the following paragraph; this is regardless of whether they would have been detected by the SONS JCMT survey or not. %Cases where our modelling results led to belts being formally undetected/unresolved, where this is in contrast to previous analysis of the same datasets, are reported in Appendix \ref{sec:discrepant}. 

Overall, from the joint REASONS archival and observing programmes, sources that were detected and resolved form a joint resolved sample of 74 belts, which we henceforth refer to as the REASONS sample. Formally, we defined a belt to be resolved if - upon fitting visibilities with a radially Gaussian belt model as described in Sect. \ref{sec:modelling} - there is a $\leq0.135$\% probability that the belt radius is equal to the lower boundary of our prior radius probability distribution. This lower prior boundary on the radius is always chosen to be much smaller than the smallest size scale (corresponding to the longest baseline) obtained by our observations; therefore, our criterion selects belts that are inconsistent with being point sources at the $\geq3\sigma$ level. 

\subsection{Observational bias and selection effects}
\label{sec:bias}
In population studies, considering selection effects is crucial to account for observational bias, and understand which belts would have ended up as part of the REASONS sample. The REASONS sample is a mix of different observing programmes with different goals, but two general selection criteria apply to all belts: 1) detectability at IR wavelengths (the discovery method), 2) detectability + resolvability at mm wavelengths. 
We direct the reader to \S3 of \citet[][]{Matra2018b} for a full description of the requirements for a belt to be selected. In summary, the first selection criterion is IR detection by \textit{Spitzer} at 24 or 70 $\mu$m, or by \textit{Herschel} at 100 or 160 $\mu$m. \textit{Herschel} detection is only considered for stars sufficiently nearby to have been included in the DUst Around NEarby Stars \citep[DUNES; e.g.][]{Eiroa2013} or Disc Emission via a Bias-free Reconnaissance in the Infrared/Submillimetre \citep[DEBRIS; e.g.][]{Phillips2010} survey samples. When evaluating detectability, we also have to consider whether belts may be resolved by these telescopes, effectively reducing the sensitivity to the flux density of the belt \citep[\S3.1.1 of][]{Matra2018b}.

The second selection effect is mm/sub-mm detectability + resolvability. Most belts in the REASONS sample broadly belong to two categories:
2A) belts detected by the SONS survey with the single dish JCMT telescope at 850 $\mu$m, all with flux densities $\gtrsim$3 mJy at 850 $\mu$m. Not all of the REASONS systems were observed as part of SONS. However, the majority of REASONS resolved targets (58/74) have 850 $\mu$m flux densities that meet the SONS 850 $\mu$m detection threshold. %As shown in \citet[][]{Matra2018b}, and confirmed in this work, any belt detected by SONS has sufficiently high surface brightness to be detected by ALMA in a reasonable amount of observing time. %All belts detectable by single dish telescopes so far have been resolvable by ALMA in at least one of its antenna configurations.

2B) All the 23 REASONS belts in the young Scorpius-Centaurus association (henceforth Sco-Cen) were observed directly with ALMA, avoiding the requirement for single-dish detectability. Of these, 13 have flux densities inferred to be $<3$ mJy at 850 $\mu$m (since they are $<$1 mJy at 1.3 mm for a spectral slope $\alpha$ of 2.5), implying they would not have been detected by the SONS survey. We note that all of the 23 Sco-Cen belts but two (HD95086 and HD36546) were first detected at mm wavelengths by either \citet[][]{Lieman-Sifry2016}, who selected them to have bright IR excesses at 70 $\mu$m ($>100$ times the stellar photospheric contribution), or by \citet[][]{Moor2017}, who selected cold ($T<$140 K), high fractional luminosity ($f>5\times10^{-4}$) belts around A-type stars. %We use the specific sensitivity and resolution of these surveys to determine whether belts belonging to Sco-Cen would have enough flux density and surface brightness to be detected, and whether they were observed at sufficient resolution to be resolved.

In summary, 71 out of 74 belts belong to one of the mm selection categories above (2A: SONS-detected/detectable; i.e. $\geq3$ mJy at 850 $\mu$m, or 2B: ALMA-detectable, belonging to Sco-Cen). The remaining three are HD38206 \citep{Booth2021b}, HD54341 \citep{Macgregor2022}, and HD216956C \citep[Fomalhaut C,][]{CroninColtsmann2021}, which, being below JCMT detectability, were detected and resolved directly by ALMA (but do not belong to the Sco-Cen association). 
In practice, this means that our sample is mostly flux density-limited by the sensitivity of the IR discovery observations, and by either the JCMT or ALMA mm detection thresholds.
With these selection criteria in hand, for our interpretation in Sect. \ref{sec:disc} and for future modelling studies, we can consider whether a system with a given set of belt and host star parameters could have made it into the REASONS sample.

\section{Observations} \label{sec:obs}
\subsection{New ALMA data} \label{sec:almaobs}

\begin{figure*}
\vspace{-0mm}
 \centering
 %\hspace{-20mm}
   \includegraphics[scale=0.38]{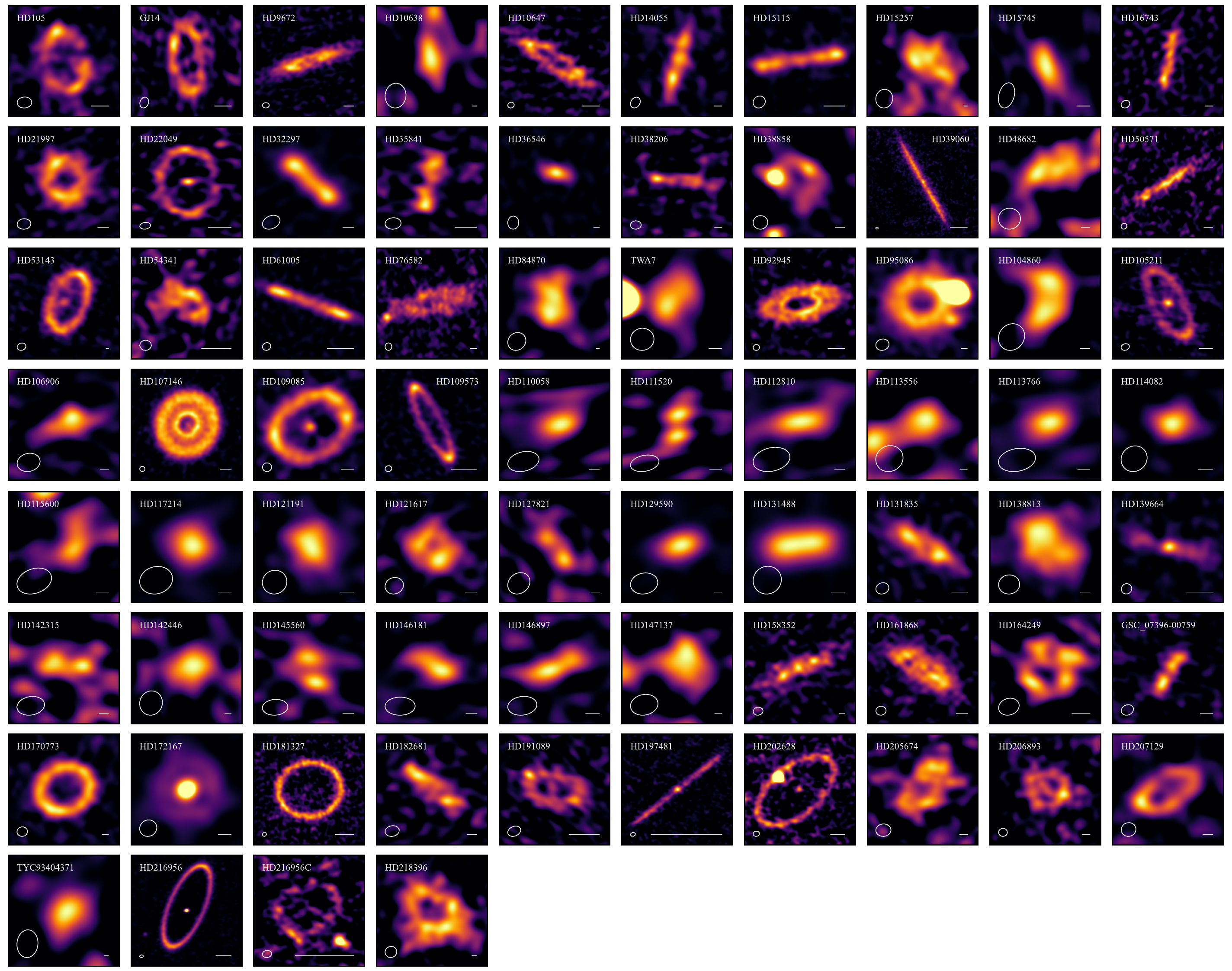}
%\vspace{-5mm}
\caption{Millimetre continuum images for the REASONS resolved sample of 74 belts, ordered by source RA. North is up and east is left. Bars indicate a physical scale of 50 au, and ellipses represent the synthesised beam of the observations. Images were obtained with the CLEAN algorithm as described in Sect. \ref{sec:obs}, with weighting parameters, resulting RMS noise levels, and beams listed in the observational log tables (available on \textsc{Zenodo}). All images are in a linear  scale, stretching from 0 (black) to the maximum intensity of the image, except in a few cases where the maximum was set to a lower value to highlight emission from a belt with respect to the star or a contaminating source. %Ciao%. 1D histograms represent probability distributions of each parameter marginalised over the other two, whereas contour maps represent 2D probability distributions of different pairs of parameters, marginalised over the third. Contours represent the central [68.3, 95.5, 99.73] \% of the distribution. Blue solid lines represent marginalised posterior probability distributions of the parameters given our observed sample, and should be compared with the model distributions.
} 
\label{fig:gallery}
\end{figure*}

%\begin{figure*}
%\vspace{-3mm}
% \centering
% %\hspace{3mm}
%   \includegraphics*[scale=0.3]{REASONS_comboplot_2.pdf}
%\vspace{-36mm}
%\caption{Continued from Fig. \ref{fig:gallery1}%Ciao%. 1D histograms represent probability distributions of each parameter marginalised over the other two, whereas contour maps represent 2D probability distributions of different pairs of parameters, marginalised over the third. Contours represent the central [68.3, 95.5, 99.73] \% of the distribution. Blue solid lines represent marginalised posterior probability distributions of the parameters given our observed sample, and should be compared with the model distributions.
%} 
%\label{fig:gallery2}
%\end{figure*}

We observed 15 systems with ALMA on Chajnantor, Chile during its Cycle 5. Fourteen targets were observed through project 2017.1.00200.S (PI: Matr\`a) and one (HD15745) through project 2017.1.00704.S (PI: Kral), due to project overlap given the similar resolution/sensitivity required. All observations were carried out using Band 6 receivers. Data were taken using the 12-m array with 43-50 antennas in a single configuration per target, varying for different targets.  Atacama Compact Array (ACA, 7-m antennas) observations were also obtained to recover flux on the shortest baselines (largest scales) for two of the targets, HD170773 and HD161868. For each target, observing dates, baseline ranges, on-source times, weather conditions, and number of antennas employed are listed in the table available on \textsc{Zenodo}\href{}{}. A single-pointing strategy was adopted, with observations centred at the proper motion corrected stellar position.

We adopted a uniform spectral setup for the correlator. This consisted of two 2 GHz-wide spectral windows centred at 243.1 and 245.1 GHz with a low spectral resolution (31.25 MHz), and two 1.875 GHz-wide windows centred at 227.2 and 230.1 GHz at higher spectral resolution (976.563 kHz, or twice the channel width of 488.281 kHz due to Hanning smoothing\footnote{\url{https://safe.nrao.edu/wiki/pub/Main/ALMAWindowFunctions/Note_on_Spectral_Response.pdf}}). The higher resolution spectral windows were set to cover the CN N=2-1 (J=5/2-3/2) and the CO J=2-1 transitions at 226.875 and 230.538 GHz, respectively. The corresponding velocity resolution for both lines is 1.29 km/s. The total bandwidth available for continuum was 7.75 GHz, with both polarisations combined.

Standard calibrations were applied to each visibility dataset by the ALMA observatory, using its pipeline. If available, and adding significantly to the sensitivity and/or resolution of the REASONS data, calibrated datasets from different dates and configurations were concatenated. This was done ensuring appropriate relative visibility weighting and/or correcting for pointing and phase center offsets (if comparable to the beam size of the observations).
For HD191089, we combined long baseline data from our project with more compact Band 6 observations from project 2017.1.00704.S (PI: Kral) and archival observations from project 2012.1.00437 (PI: Rodriguez). For HD158352, we combined our data with archival observations (at similar sensitivity and resolution) from project 2019.1.01517 (PI: Rebollido). %For project C, these were obtained with a correlator setup similar to the other projects, but with spectral windows centred at E, F, G, H GHz (project C), covering the CO but not the CN transition (CHECK).

All concatenated datasets were imaged in the Common Astronomy Software Applications (CASA) software v5.4.0 using the CLEAN algorithm implemented through the tclean task. %For continuum imaging, we averaged the visibility datasets along the frequency axis to as few channels per spectral window as possible while avoiding bandwidth smearing effects \footnote{see X for details}. This was to decrease the data size for faster imaging and modelling. 
%For line imaging, we extracted the spectral window covering the line frequency of interest from the visibility dataset, and subtracted the continuum (measured in a frequency region excluding channels within $\pm$50 km/s from the line frequency) using the \textit{uvcontsub} task.
The continuum imaging was carried out in multi-frequency synthesis mode with multiscale deconvolution \citep{Cornwell2008}. Different weighting schemes and u-v tapers were used for different targets to find an optimal balance between surface brightness sensitivity and resolution. The weighting choice is indicated, together with the achieved beam sizes, RMS noise levels, weather conditions, baseline lengths, dates, and time on source, in tables available on \textsc{Zenodo} (see Sect. \ref{sec:dataavail}). Typical continuum sensitivities, measured in a region of the images that is free of emission, are 12-68 $\mu$Jy for beam sizes ranging between 0.2$\arcsec$ and 3.1$\arcsec$. The flux calibration accuracy of all ALMA observations was conservatively assumed to be 10\%.

CO imaging was carried out after continuum subtraction from the visibility measurements (using the uvcontsub CASA task). We imaged a spectral region $\pm100$ km/s of the stellar barycentric velocity using the tclean task, with standard deconvolution. We chose to keep the native channel size of 488.281 kHz, and use natural weighting for all targets to maximise sensitivity. No clear CO detections are obtained for any of the targets; RMS noise levels in the cubes are reported in the rightmost column of tables available on \textsc{Zenodo} (see Sect. \ref{sec:dataavail}). We underline, however, that more detailed analysis (beyond the scope of this continuum-focused work) is needed to search for faint emission and extract CO gas mass upper limits from the data cubes. 

%Following continuum subtraction, we imaged the region within $\pm$50 km/s of the CO and CN line frequencies in the rest frame of each star (taking into account their heliocentric velocity, see Table X) at the native spectral resolution of 1.29 km/s, with natural weighting in order to achieve maximum sensitivity. %In addition, to search for extragalactic line emission (see Sect. X), we imaged the entire bandwidth with natural weights at a uniform spectral resolution of 62.5 MHz. 

\subsection{New SMA data} \label{sec:smaobs}
We observed 10 systems with the SMA (6-m antennas) on Mauna Kea (Hawaii, USA), between January 2018 and January 2019. We simultaneously used the 230 and 240 receivers, with between 5 and 8 antennas arranged in compact and/or subcompact configuration. Similarly to the ALMA data, we list observing dates, baseline ranges, on-source times, weather conditions, and number of antennas for each target in a table available on \textsc{Zenodo} (see Sect. \ref{sec:dataavail}). Once again, a single-pointing strategy was adopted, with observations centred at the proper motion-corrected stellar position.

The correlator was configured with 4 chunks per receiver per sideband, each providing $\sim$2 GHz of effective bandwidth, and centred near 224.5, 226.5, 228.5, 230.5 GHz (lower sideband) and 240.5, 242.5, 244.5, 246.5 GHz (upper sideband). The total bandwidth available for continuum was therefore $\sim$16 GHz per receiver, all at a spectral resolution of 140 kHz (corresponding to a velocity resolution of 0.18 km/s at the frequency of the CO J=2-1 line). The two receivers were set up to cover the same frequency range, yielding an overall $\sim\sqrt{2}$ improvement in sensitivity.

Observations typically included 30-60 minutes on a strong quasar used as bandpass calibrator, and 5-20 minutes on a Solar System planet or satellite used as flux calibrator (yielding typical absolute flux uncertainties of $\sim$20\%). Observations of the science target were interleaved with observations of two quasars as phase calibrators, typically with $\sim$2 minute integrations, repeated every $\sim$15 minutes. For daytime observations, we employed more rapid cycling through science target, which was dependent on weather conditions, in order to capture faster atmospheric phase variations. The two chosen quasars were located typically within a few to 20 degrees of the science target.
All calibrations were applied to the complex visibilities within the Millimeter Interferometer Reduction (MIR) package\footnote{\url{https://www.cfa.harvard.edu/~cqi/mircook.html}}, producing calibrated visibility datasets that were later exported to CASA v5.4.0 as Measurement Sets (MSs) for imaging.

After concatenation of observations from different dates, continuum and line imaging was carried out using the CASA tclean task in the same way as described in Sect. \ref{sec:almaobs} for the ALMA data. The weighting choice, achieved beam size and continuum RMS noise level of each observation are indicated in a table available on \textsc{Zenodo} (see Sect. \ref{sec:dataavail}). Continuum sensitivities achieved range between 100 and 290 $\mu$Jy for beam sizes in the 3$\arcsec$-6$\arcsec$ range.

\subsection{Archival observations} \label{sec:archivalobs}

We retrieved archival ALMA and SMA continuum observations of resolved belts that were made public before June 1 2020, or that became public more recently but have already been published in the literature. 
For the ALMA sources, where more than one project observed the same target, we analyse the project which produced the best combination of resolution and continuum sensitivity as listed in the ALMA archive. The few exceptions to this rule were sources where we deemed the additional sensitivity and/or baseline coverage to be beneficial. %(using visibility curves as in, e.g., Fig. \ref{fig:imagecombo1})
In these cases, observations were combined and jointly modelled as long as the local oscillator (LO) frequencies were within 20 GHz of one another.

For each project, and within it for each observation, we retrieved raw visibilities from the ALMA archive, and calibrated them using the provided pipeline or calibration scripts within the same version of CASA as done by the ALMA observatory.
For SMA and CARMA archival observations, we obtained calibrated, science-ready visibilities from PIs/co-Is of the respective projects, where similar calibration strategies as for the newly obtained REASONS targets (Sect. \ref{sec:smaobs}) were employed\footnote{D. Wilner, A.M. Hughes, private communication}.

Continuum imaging of the combined observations for each target was carried out using multiscale CLEAN deconvolution within CASA v5.4.0 in the same manner as for the new REASONS data (Sect. \ref{sec:almaobs} and Sect. \ref{sec:smaobs}), once again adapting the weightings and u-v tapers to observations of each belt. These choices, together with beam sizes and RMS levels achieved are listed in tables available on \textsc{Zenodo} (see Sect. \ref{sec:dataavail}).\\

\section{Results and modelling} \label{sec:contres}

\subsection{Image gallery}
\label{sec:imagegallery}

Figure \ref{fig:gallery} shows continuum images for the entire REASONS sample of 74 mm-resolved belts, ordered by right ascension (RA) left to right, and top to bottom. %, and including the 15 newly resolved belts from the REASONS observing programme (marked by an asterisk).
The belts are resolved at a wide variety of levels, from marginally resolved (e.g. HD110058) to resolved over a large number of beams (e.g. HD39060 - $\beta$ Pictoris). However, even belts that appear marginally resolved in the images of Fig. \ref{fig:gallery} are formally resolved by the longest baselines of the observations, and according to our formal definition of Sect. \ref{sec:targetsel}.

Of the 25 targets from the REASONS observing programme, 15 were detected and resolved, while 10 of them were not detected, often due to contamination by, or confusion with, likely background sources (see Appendix \ref{sec:confused} for details). Additionally, we note that 2 sources previously reported as detected and resolved in the literature, HD10700 \citep[$\tau$ Ceti,][]{Macgregor2016b} and HD115617 \citep[61 Vir,][]{Marino2017b}, were found not to be conclusively detected and/or resolved in our analysis, and are therefore not included in the REASONS sample. The likely reason for this discrepancy is that we did not use single dish data to constrain the total flux of the belt, which is largely resolved out in these specific interferometric ALMA datasets.

\subsection{Interferometric visibility modelling}
\label{sec:modelling}

\subsubsection{Data preparation}
Interferometric visibilities for all targets in the sample were imaged, modelled, and post-processed using a common software framework, available on GitHub as a package called MIAO: Modelling Interferometric Array Observations\footnote{\url{https://github.com/dlmatra/miao}}. For a given system and a given dataset (observing date), calibrated continuum visibility datasets were averaged in time and frequency using the CASA \textsc{mstransform} task to reduce the number of visibilities, and therefore the computing time needed for modelling. To avoid bandwidth and time smearing, for ALMA data we limited the averaging to at most 2 GHz in frequency, and at most 30s (or 60s) in time for 12m (or ACA) data. For stars observed over multiple dates, we note that the phase center was in most cases updated by ALMA or by the SMA PI to match the proper motion-corrected stellar position for every observing date. Therefore, datasets with identical spectral setups that were sufficiently close in time were merged before the averaging step.

For a given system, we imported all visibility datasets available from CASA into Python, and determined the pixel size and the number of pixels required in the model image for the combined datasets. The choice was analogous to the criteria described in \citet[][]{Tazzari2018} to ensure that the u-v plane covered by the data is appropriately sampled by the model visibilities to be produced. %The size of the image is then translated from angular to physical units (au) using the distance to the star in pc.

\begin{figure*}
\vspace{-0mm}
\hspace{-18mm}
\includegraphics[scale=0.28]{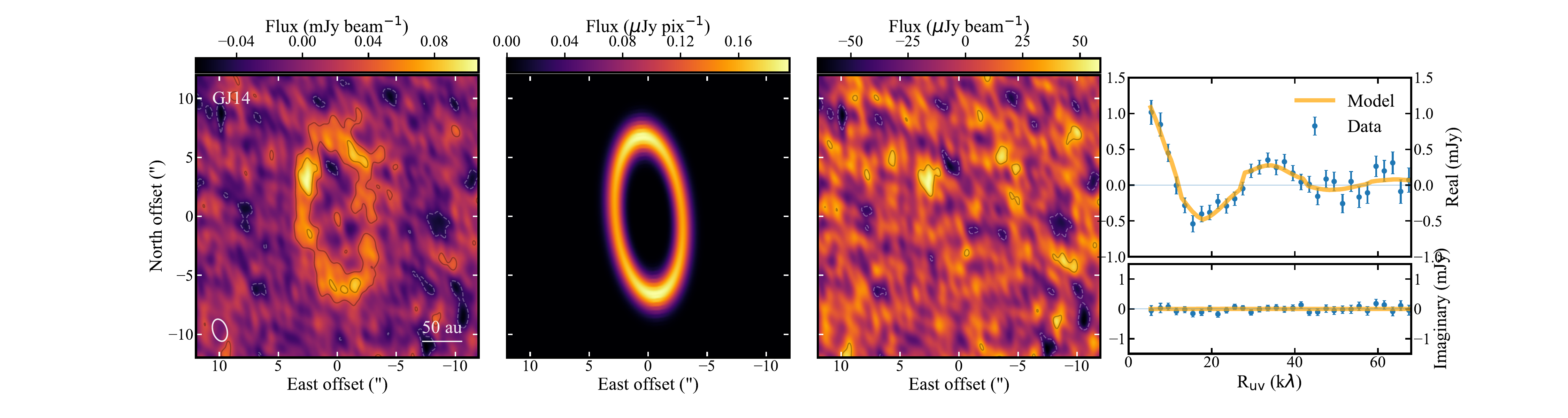}
%\vspace{-8mm}
\caption{Visuals used to support the modelling and fit evaluation process, carried out for each system, here shown for the GJ14 system as an example. \textit{Leftmost:} %\LEt{***Please provide a general title before introducing the different panels. Please make sure all figures have a general title.***}
ALMA continuum image of the GJ14 system (see imaging details in tables available on \textsc{Zenodo}). Contours are [2,4,..] $\times$ the RMS noise level. 
\textit{Center left:} full resolution best-fit belt model. \textit{Center right:} Residual image after subtraction of the best-fit visibilities from the data. Imaging parameters and contours are the same as the leftmost image. \textit{Rightmost:} Real and imaginary part of the azimuthally averaged de-projected complex visibility profiles, for both the data (blue points with uncertainties) and the best-fit model (orange lines). The de-projection was carried out using the best-fit inclination and PA from Table \ref{tab:resbelts}. }
\label{fig:evaluatemodel}
\end{figure*}

\subsubsection{The physical model}
The most general model comprises three components: the dust belt, the host star modelled as a point source, and background source(s). For any modelled system, we justified including the star and/or background sources by first inspecting the imaged data, and if necessary by inspecting the residuals after subtraction of a best-fit model including the belt only.

Each planetesimal belt was modelled as an axisymmetric ring of emission. The radial mass surface density distribution $\Sigma$ is Gaussian. While we acknowledge that at high resolution, most belts are unlikely to resemble this distribution \citep[as demonstrated by existing data, e.g.][]{Marino2018b, Faramaz2021}, we deemed a Gaussian a simple enough prescription to derive the centroid radius $R$ and width (FWHM) $\Delta R$ of the surface density distribution, which are of most interest to this study. Additionally, most of the belts were observed at moderate resolution, with at most a few beams across their widths, which resulted in a Gaussian producing a satisfactory fit for the vast majority of systems.

In the vertical direction, belts are modelled as a single Gaussian (mass) in number density. This is the expected vertical distribution for a Rayleigh distribution of particle inclinations \citep[][]{Matra2019b}, expected from gravitational perturbations between large stirrers and planetesimals in a thin disc \citep[e.g.][]{IdaMakino1992}. The full prescription of the particle mass number density distribution is therefore
\begin{equation}
\rho(r,z)=\Sigma_{\mathrm{dust, }r=r_{\rm c}}\ e^{-\frac{(r-r_{\rm c})^2}{2\sigma_r^2}}\frac{e^{-\frac{z^2}{2(hr)^2}}}{\sqrt{2\pi}hr},
\end{equation}
where symbols have the same meaning as for Eq. 1 in \citet[][]{Matra2019b,Matra2020}. The parameter describing the vertical thickness of the disc is the aspect ratio $h=\frac{H}{r}$, which we assume to be constant with radius. On the other hand, the parameter describing the radial width of the disc is $\sigma_r$, which is related to the FWHM $\Delta R$ of the Gaussian surface density distribution. We note that in an effort to minimise the number of free parameters in our modelling, we only include the aspect ratio as a free parameter in cases where the belt is clearly vertically resolved, or is observed at sufficiently high resolution and SNR that its vertical structure may be extracted from the observed azimuthal intensity profile \citep[as described in][]{Marino2016}. In other cases, we fix this value to $h=0.03$, motivated by the aspect ratio of the AU Mic disc \citep{Daley2019}.

We set the temperature distribution to have a $r^{-0.5}$ radial dependence, on the assumption that the large grains probed by millimetre observations are well approximated by blackbodies (an assumption that is not appropriate for smaller grains which dominate the belts' IR luminosity, as mentioned in Sect. \ref{sec:intro}). We note this radial dependence of the temperature distribution lead to a radial intensity distribution that is not exactly Gaussian. We then create a model image of the belt using the RADMC-3D radiative transfer code \footnote{\url{http://www.ita.uni-heidelberg.de/~dullemond/software/radmc-3d/}} \citep{Dullemond2012}. We initially centre the model belt at the origin of the image, and incline it from the plane of the sky by inclination angle $i$ (a free parameter, with $i=0^{\circ}$ indicating a face-on belt). We then rotate the belt in the plane of the sky so that the belt's sky-projected semimajor axis is at  a position angle PA (also a free parameter) compared to the declination direction, where this angle is measured East of North.

We renormalise the pixel values in the model image so that the integral of the pixel surface brightnesses (in Jy/pixel) over the entire image equals the belt's model flux density $F_{\nu_{\rm belt}}$ (Jy). Our visibility-based determination of the flux density is more accurate than a measurement obtained directly from the imaged data, as it does not depend on weighting schemes or suffer from imaging artifacts. However, it still assumes that our visibility data samples sufficiently short u-v distances; in cases where it does not \citep[e.g. Vega, see][]{Matra2020}, the total flux density measured is model-dependent and could change when considering non-Gaussian models.

\subsubsection{The fitting process}

The model image is then multiplied by the primary beam obtained during the CASA imaging process, to account for the response of the interferometer's antennas. For multi-pointing (mosaic) observations, we repeat this process for every pointing in the dataset being modelled; in practice, we treat different mosaic pointings as different datasets.

We then use the GALARIO\footnote{\url{https://github.com/mtazzari/galario/}} software package \citep[][]{Tazzari2018} to obtain a Fourier transform of the model image, and sample it at the same u-v locations as the data. Finally, we apply an RA and Dec offset (with each left as a free parameter) to the model belt as a phase shift in Fourier space. This allows us to account for astrometric offsets of the belt's centre from the phase centre of the observations.

To these belt-only model visibilities, we add the star as an additional point source component with flux density $F_{\nu_{\star}}$, and located exactly at the geometric centre of the belt; therefore the same astrometric offset applies to the star and the belt in the vast majority of systems. In some systems where the belt has been found to be significantly eccentric (HD53143, HD202628, HD216956), we model the eccentricity simply as an extra RA and Dec offset between the star and the belt's geometric centre.

In systems with one or more background sources, we model these sources initially as unresolved, point-like emission, with flux density $F_{\rm bkg}$, and offsets $\Delta$RA$_{\rm bkg}$, $\Delta$Dec$_{\rm bkg}$. In some cases, inspection of residuals shows that the sources are resolved, in which case we model them as 2D Gaussians with two extra free parameters being their FWHM along the sky-projected semimajor axis, and an inclination $i_{\rm bkg}$ and PA$_{\rm bkg}$ defined as for the planetesimal belt component. 

The uncertainty $\sigma$ on each visibility data point (real or imaginary part) is contained in a visibility weight $w = 1/\sigma^2$ delivered by each observatory. However, at least for ALMA it has been shown that the delivered visibility weights, while accurate relative to one another within a dataset, can be inaccurate in an absolute sense, and need rescaling by a factor common to all visibilities within any given dataset \citep[e.g.][]{Marino2018b,Matra2019b}. We therefore leave this rescaling factor as a free parameter in each of our modelled datasets.

For any given system, we fit the model visibilities to the data using the affine-invariant Markov Chain Monte Carlo (MCMC) ensemble sampler from \citet[][]{GoodmanWeare2010}, implemented through the \textsc{EMCEE} v3 software package \citep{Foreman-Mackey2013,Foreman-Mackey2019}. The likelihood function is proportional to $e^{-\chi^2/2}$. Where multiple datasets and or different pointings were fitted simultaneously for a given system, this $\chi^2$ was taken to be the sum of the $\chi^2$ of the individual datasets/pointings. 

We used uniform priors for all model parameters, with prior ranges chosen to allow the chains to explore a wide enough, yet physical region of parameter space. We note that to retain the Gaussian radial nature of the belt's surface density - in other words, to ensure there is an inner hole for the Gaussian ring - we ensure that the belt's radial peak is at least 2$\sigma_r$ away from the star. While again we acknowledge this Gaussian ring model is not necessarily an accurate description of every belt, we find that at the SNR and resolution of the data, it is sufficient to accurately capture the midpoint radius and width of the belts in our study.

We ran the MCMC to sample the posterior probability distribution of the parameters using a number of walkers equal to 10 times the number of free parameters (which is dependent on the system modelled), and for a number of steps $\geq$1000. This number of steps varied depending on the number of model components and free parameters, the number of datasets being fitted and the SNR of the emission for a given planetary system. In all cases, we ensured visual convergence of the MCMC chains.

\renewcommand{\arraystretch}{1.3}
\begin{table*}
\caption{REASONS newly observed and resolved belts} \label{tab:resbelts}
\centering
\begin{tabular}{ccccccccc}
\hline\hline
Target &
$\lambda$ &
$F_{\nu_{\star}}$ &
$F_{\nu_{\rm belt}}$ &
$R$ &
$\Delta R$ & 
$h$ & 
$i$ &
PA \\
 & mm & $\mu$Jy & mJy & au &
au &  & $^{\circ}$ & $^{\circ}$ \\
\hline
GJ14 & $1.27$ & $^{a}40^{+20}_{-20}$ & $1.8^{+0.2}_{-0.2}$ & $99^{+3}_{-3}$ & $33^{+7}_{-8}$ & $^{a}0.05^{+0.03}_{-0.03}$ & $64^{+1}_{-2}$ & $5^{+2}_{-2}$ \\
HD10638 & $1.27$ & - & $1.2^{+0.3}_{-0.3}$ & $160^{+80}_{-50}$ & $<400$ & - & $60^{+20}_{-30}$ & $10^{+27}_{-42}$ \\
HD14055 & $1.27$ & $<200$ & $3.4^{+0.2}_{-0.2}$ & $180^{+10}_{-10}$ & $160^{+30}_{-20}$ & $^{a}0.03^{+0.02}_{-0.02}$ & $81.1^{+0.8}_{-0.9}$ & $163.3^{+0.7}_{-0.7}$ \\
HD15257 & $1.27$ & - & $3.0^{+1.0}_{-0.7}$ & $270^{+60}_{-40}$ & $220^{+100}_{-110}$ & - & $^{d}40^{+20}_{-30}$ & $60^{+50}_{-20}$ \\
HD15745 & $1.27$ & - & $1.12^{+0.07}_{-0.07}$ & $65^{+6}_{-5}$ & $50^{+10}_{-20}$ & -$^{b}$ & $70^{+5}_{-5}$ & $29^{+4}_{-4}$ \\
HD35841 & $1.27$ & $<50$ & $0.62^{+0.04}_{-0.04}$ & $57^{+3}_{-3}$ & $^{a}15^{+9}_{-8}$ & $^{a}0.08^{+0.05}_{-0.05}$ & $^{c}84^{+4}_{-4}$ & $167^{+2}_{-2}$ \\
HD76582 & $1.27$ & $^{a}30^{+20}_{-20}$ & $3.2^{+0.2}_{-0.2}$ & $219^{+9}_{-8}$ & $210^{+20}_{-20}$ & $<0.1$ & $72^{+1}_{-1}$ & $103.7^{+0.9}_{-1.0}$ \\
HD84870 & $1.27$ & - & $1.9^{+0.5}_{-0.4}$ & $260^{+50}_{-50}$ & $260^{+60}_{-60}$ & - & $^{d}50^{+10}_{-30}$ & $10^{+20}_{-20}$ \\
HD127821 & $1.27$ & - & $1.9^{+0.4}_{-0.4}$ & $120^{+30}_{-20}$ & $<300$ & - & $^{c}78^{+8}_{-10}$ & $36^{+9}_{-8}$ \\
HD158352 & $1.30$ & $<90$ & $2.1^{+0.2}_{-0.1}$ & $270^{+20}_{-20}$ & $380^{+90}_{-60}$ & $0.17^{+0.03}_{-0.04}$ & $^{c}81^{+4}_{-2}$ & $114^{+1}_{-1}$ \\
HD161868 & $1.27$ & $^{a}50^{+20}_{-20}$ & $2.5^{+0.1}_{-0.1}$ & $124^{+6}_{-5}$ & $110^{+10}_{-10}$ & $^{a}0.13^{+0.04}_{-0.05}$ & $68^{+2}_{-2}$ & $57^{+2}_{-2}$ \\
HD170773 & $1.27$ & $^{a}30^{+20}_{-20}$ & $6.3^{+0.2}_{-0.2}$ & $194^{+2}_{-4}$ & $68^{+5}_{-5}$ & $<0.20$ & $33^{+2}_{-2}$ & $114^{+3}_{-3}$ \\
HD182681 & $1.27$ & $<70$ & $1.41^{+0.06}_{-0.06}$ & $143^{+4}_{-4}$ & $100^{+10}_{-10}$ & $<0.20$ & $76^{+1}_{-1}$ & $53.3^{+0.8}_{-0.8}$ \\
HD191089 & $1.27$ & $^{a}40^{+20}_{-20}$ & $1.83^{+0.03}_{-0.03}$ & $44.8^{+0.9}_{-0.9}$ & $16^{+3}_{-3}$ & $^{a}0.10^{+0.04}_{-0.05}$ & $60^{+1}_{-1}$ & $73^{+1}_{-1}$ \\
HD205674 & $1.27$ & - & $1.1^{+0.1}_{-0.1}$ & $160^{+10}_{-10}$ & $120^{+20}_{-20}$ & -$^{b}$ & $56^{+5}_{-6}$ & $138^{+6}_{-7}$ \\
\hline
\end{tabular}
\tablefoot{
\tablefoottext{a}{Marginally resolved or detected, i.e. having a posterior probability distribution with a non-zero peak but consistent with zero at the $3\sigma$ level.}
\tablefoottext{b}{Quantity unconstrained within prior boundaries.}
\tablefoottext{c}{Inclination consistent with 90$^{\circ}$ (perfectly edge-on) to within 3$\sigma$.}
\tablefoottext{d}{Inclination consistent with 0$^{\circ}$ (perfectly face-on) to within 3$\sigma$. }}
\end{table*}

\subsubsection{Modelling results}
Final posterior probability distributions were marginalised over parameters that were unrelated to the planetary system, such as those characterising background sources (if any), and visibility weight-rescaling factors. In Table \ref{tab:resbelts} and \ref{tab:resbeltsarc}, we present the 50$^{+34}_{-34}$th percentile values of the posterior probability distribution of each belt and stellar parameter, marginalised over all other parameters. It is important that these are interpreted as best-fit $\pm$1$\sigma$ uncertainties only in cases where the posterior probability distribution of a given parameter is single-peaked and approximately Gaussian in shape. Therefore, we make extensive use of footnotes in Table \ref{tab:resbelts} and \ref{tab:resbeltsarc} to highlight instances where this was not the case, and/or where parameters were not well constrained within the prior boundaries. Upper or lower limits reported in Table \ref{tab:resbelts} and \ref{tab:resbeltsarc} are at the $3\sigma$ level, and flux density uncertainties do not include absolute flux calibration systematics.

Figure \ref{fig:evaluatemodel} uses the GJ14 system as an example to illustrate how we evaluated the fit for each planetary system. First, we produced model images (centre-left panel in Fig. \ref{fig:evaluatemodel}) and visibilities using best-fit (median) parameters, and subtracted them from the data to produce residual visibilities. We then imaged the residual visibilities using the exact same imaging parameters as the data (leftmost panel in Fig. \ref{fig:evaluatemodel}), to produce residual maps and evaluate the goodness of fit (centre-right panel in Fig. \ref{fig:evaluatemodel}). To further confirm goodness of fit in visibility space, we also plotted the real and imaginary part of the complex (data and model) visibilities as a function of de-projected u-v distance from the phase centre (rightmost panel in Fig. \ref{fig:evaluatemodel}). To do so, we applied the de-projection method of \citet[][]{Hughes2007} and used the belt's best-fit $i$ and PA from the visibility fitting.

Based on the compatibility of residual images with pure noise, we find that 65/74 belts are fit well by our radially and vertically Gaussian model. This confirms that such a simple model is sufficient to capture the basic structure (centroid radius, width) of belts at the resolution and SNR of most of the data. Belts where our Gaussian model left significant residuals are marked by a $^{\star}$ in the leftmost (Target) column of Table \ref{tab:resbeltsarc}. In most cases this is due to substructure becoming apparent in data with higher resolution and/or SNR. One notable exception is HD36546, whose edge-on, highly centrally peaked emission morphology indicates the lack of a central hole interior to the belt. For this belt, ensuring a good, residual-free fit meant we had to relax the prior imposing the presence of an inner hole. This led to an artificially inflated belt FWHM, which in truth reflects the failure of the Gaussian ring model in accurately reproducing the observed emission. To avoid biasing the population of belt widths, we exclude this system from our discussion of belt widths in Sect. \ref{sec:widths}.

\subsection{SED modelling}
\label{sec:sed}

\begin{figure}
\vspace{-0mm}
\hspace{-3mm}
\includegraphics[scale=0.65]{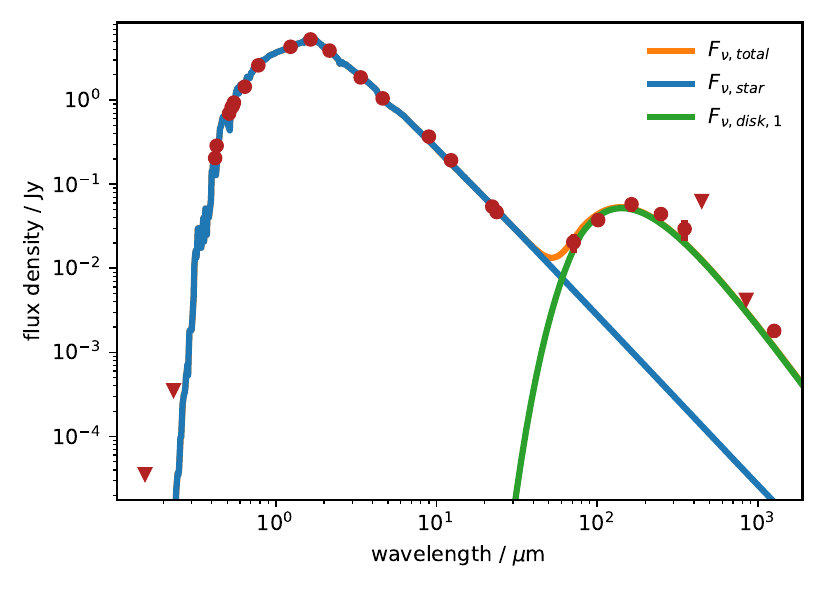}
%\vspace{-5mm}
\caption{Example of multi-wavelength photometry gathered for the GJ 14 system (brown circles for detections, and downward-pointing triangles for upper limits), and best-fit star (blue) and single-component modified blackbody belt model (green) obtained following the method of \citet[][]{Yelverton2019}. Best-fit parameters for this and other systems are listed in Table \ref{tab:sdbprops}.} 
\label{fig:evaluatesedmodel}
\end{figure}

For each star in our REASONS sample, we derive stellar and belt properties by fitting multi-wavelength photometry. We gather photometry (in addition to mm flux densities reported in this work) and fit it with a star + modified blackbody model, following the method of \citet[][]{Yelverton2019}. Figure \ref{fig:evaluatesedmodel} uses the GJ14 system once again as an example to illustrate a typical fit as carried out for each planetary system. 
Stellar and dust properties of interest derived are listed in Table \ref{tab:sdbprops}, with parameters having the same meaning as in \citet[][]{Yelverton2019}. In some cases (flagged as `Warm dust' systems in Table \ref{tab:sdbprops}), an additional modified blackbody representing a warmer dust population was necessary to fit a system's mid-IR photometry, which was otherwise found to be underestimated by a single modified blackbody fit. In these cases, we report dust properties (fractional luminosity L$_{\rm dust}$/L$_{\ast}$, temperature T=T$_{\rm cold}$, $\lambda_{0}$ and $\beta$) for the colder dust population only, which dominates the mm-wavelength emission in all cases.

\section{Discussion} \label{sec:disc}
In previous sections we presented the REASONS sample including the vast majority of planetesimal belts resolved at mm wavelengths to date. We undertook a uniform interferometric visibility modelling analysis for all systems to construct a sample of 74 planetesimal belts with spatially resolved properties. Combined with modelling of multi-wavelength photometry, the final product is a N-dimensional dataset of star and belt properties (N being all the properties listed in Tables \ref{tab:resbelts}, \ref{tab:resbeltsarc}, and \ref{tab:sdbprops}) for the whole REASONS sample. As described in Sect. \ref{sec:dataavail}, all processed data and results are available to the reader and can be readily explored online. %\LEt{Single-sentence paragraphs are not permitted and must be joined to one or other of the surrounding paragraphs.}
Using this new N-dimensional REASONS dataset, in this section we discuss emerging population properties and trends by projecting this multi-dimensional dataset onto 2D parameter spaces.

\subsection{The distribution of planetesimal belt radii: an observed dearth of small belts}
\label{sec:rvsLstar}

\begin{figure}
%\vspace{-2mm}
  %\centering 
\hspace{-0mm}
\includegraphics[scale=0.35]{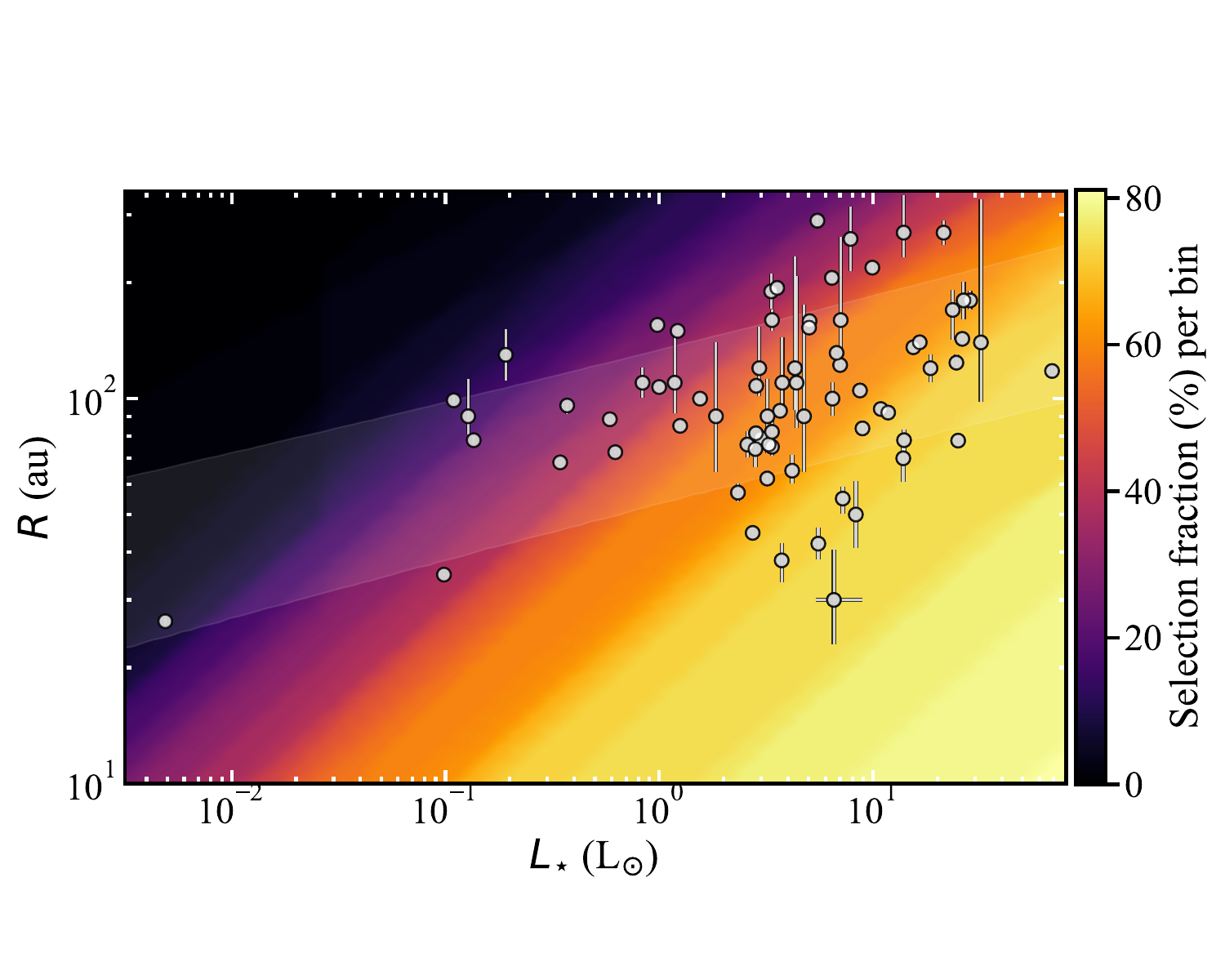}
%\vspace{-13mm}
  \caption{Radius of observed planetesimal belts as a function of host star luminosity (black and white points with error bars). The white shaded region represents the $\pm 1\sigma$ range of power laws (about the best fit) allowed by the data, including the intrinsic scatter as well as the uncertainty in the observed radii. The background colour map represents the selection probability (\%), or percentage of belts that would pass the selection effects at a given [R-L$\star$] location, assuming unobserved belts have the same distribution of parameters [$d$,R$_{\rm BB}$/R, M, $\lambda_0$, $\beta$] as the observed population.}
 \label{fig:rvsl}
\end{figure}

Figure \ref{fig:rvsl} shows the distribution of planetesimal belt radii as a function of their host star luminosity. Before consideration of selection bias, we find the same positive-sloping, shallow trend noticed by \citet[][]{Matra2018b} and \citet{Marshall2021}, though with much larger scatter and consequently lower degree of correlation. A fit as described in \S2 of \citet[][]{Matra2018b} leads to a slope $\alpha=0.14^{+0.03}_{-0.03}$, a vertical offset $R_{\rm 1 L_{\odot}}=92^{+6}_{-6}$ au, and an intrinsic scatter $f=0.44^{+0.05}_{-0.04}$, where the latter describes the vertical scatter of the distribution measured as a fraction of radius. While the stellar luminosity dependence remains consistent, albeit slightly shallower compared to earlier results, we find REASONS radii to be on average larger, and to display a significantly larger intrinsic scatter ($0.44^{+0.05}_{-0.04}$) compared to the previous inference from a smaller sample ($0.17^{+0.07}_{-0.08}$). In other words, the REASONS sample shows a broader range of radii $R$ at any host star luminosity $L_{\star}$. This is evident for belts around F- to late-A type stars (2-10 L$_{\odot}$), where a number of smaller ($R\sim$20-60 au) belts have been newly resolved.

We then consider selection effects through a method that can be employed to 2D plots of any 2 belt parameters X and Y amongst the N parameters reported in the REASONS dataset. We create a synthetic population of 1000 belts per log-uniform log$_{10}$(X)-log$_{10}$(Y) bins across the 2D parameter space displayed in Fig. \ref{fig:rvsl}.
To pass each belt through the selection effects described in Sect. \ref{sec:bias} we need to calculate the detectability and thus the flux density of a belt at several wavelengths. This in turn requires assuming a set of (N-2) star and belt parameters (2 representing the X and Y parameters considered in the 2D plot). This is because overall, N parameters are needed to calculate the flux densities of the star and the belt, namely [$L_{\star}$, $T_{\star}$, $R_{\star}$, $d$, $R$, $R/R_{\rm BB}$, $\sigma_{\rm tot}$, $\lambda_0$, $\beta$]. In order, these represent the star's luminosity, effective temperature, radius and distance from Earth, the belt's true radius, the ratio between the true radius and blackbody radius (determining the temperature of the grains that dominate the emission in the belt's spectrum), the belt's total cross sectional area in dust grains $\sigma_{\rm tot}$, and the modified blackbody parameters $\lambda_0$ and $\beta$, describing the long-wavelength falloff in the emission spectrum. For simplicity, we ignore the effect of belt width and assume all grains are located at the midpoint radius derived in our modelling (Sect. \ref{sec:modelling}). 

To choose these N-2 parameters for each of the 1000 synthetic belts, we randomly draw one of the 74 belts in REASONS, take its N-2 parameters and assign them to this synthetic belt. This approach ensures that we retain the same (N-2) dimensional distribution of parameters as the observed REASONS sample, including correlations between any of the N-2 parameters. On the other hand, this approach does not retain correlations between quantities X or Y and any of the other N-2 parameters. Then, if either X or Y is a stellar parameter amongst [$L_{\star}$,$T_{\star}$ or $R_{\star}$], we derive the other two stellar parameters assuming the star has reached the main sequence, interpolating from tabulated values from \citet[][]{PecautMamajek2013}\footnote{\url{https://www.pas.rochester.edu/~emamajek/EEM_dwarf_UBVIJHK_colors_Teff.txt}}. We then pass the 1000 belts through our selection effects to obtain a selection fraction per bin, which represents the fraction of belts (out of 1000) that we could have detected and resolved. We will henceforth call this a `bias map'. We note that because we are drawing the N-2 parameters behind every 2D plot from the observed distribution, the question we are asking with our bias maps is `What fraction of belts at this [X,Y] location would have ended up in REASONS if they existed, assuming they had the same joint distribution of N-2 other parameters as the observed REASONS population?'. 

In [$R$-$L_{\star}$] space, the bias map (colour map in Fig. \ref{fig:rvsl}) shows that the detectability of belts decreases as we go to larger belts and less luminous stars \citep{Luppe2020}, simply because these are colder and thus harder to detect. Indeed, the slope in the observed bias map largely follows $R\propto \sqrt{L_{\star}}$, as expected for belts observed with a fixed flux sensitivity at any wavelength (both on the Rayleigh-Jeans side of the dust's spectrum, where $B_{\nu}(T)\propto T$, and on the Wien side where $B_{\nu}(T)\propto e^{-h\nu/kT}$), and for a fixed set of N-2 parameters. This selection effect explains the absence of large ($\gg$100 au) belts around low luminosity stars, and accounting for it would make the weakly positive $R$-$L_{\star}$ trend even shallower.

However, selection effects cannot explain the lack of easily detectable belts smaller than 10 to a few tens of au observed by \citet[][]{Matra2018b}, which is confirmed in the REASONS sample. This observed dearth of belts could imply either that smaller belts are truly rarer, for example, if belts preferentially formed at larger radii, or that they are preferentially less massive than larger belts because they were born or evolved that way \citep[][see further discussion in Sect. \ref{sec:mvsr}]{Matra2018b}.

\subsection{The width of planetesimal belts}
\label{sec:widths}

Scattered light observations of debris discs show a wide range of widths from the narrow belts of  HR~4796 \citep{Schneider1999} and Fomalhaut \citep{Kalas2005} to the broad discs of $\beta$~Pic \citep{Smith1984, KalasJewitt1995} and AU Mic \citep{Kalas2004}. \citet{Strubbe2006} developed a ``birth ring'' model, which showed that the AU Mic observations could be explained by a narrow belt of parent planetesimals that produce dust through collisions, which is then spread out by transport processes. They proposed that this birth ring model could be prevalent amongst debris discs and it has been commonly used to model other systems. However, infrared observations, which are less affected by transport forces, showed that some systems were harder to explain with the narrow birth ring model \citep[e.g.][]{Su2009,Booth2013}. With ALMA, we are observing at a wavelength long enough that the observations are dominated by dust grains that are too big to be affected by transport forces and we have a resolution necessary for us to clearly determine the radial distribution of the large, gravitationally bound grains. The width of the parent planetesimal belts can therefore be determined \citep[e.g.][]{Matra2018b}, but until now the number of resolved discs was still too low to draw definitive conclusions about the distribution of widths.

\begin{figure}
  \centering 
 \includegraphics[trim=0.0cm 0.0cm 0.0cm 0.0cm,
    clip=true, width=0.85\columnwidth]{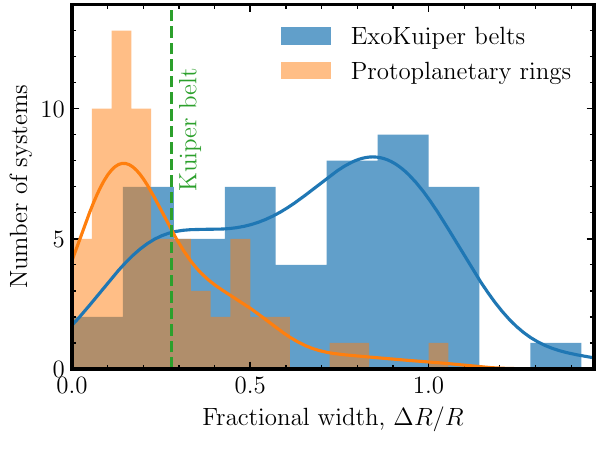}
\includegraphics[trim=0.0cm 0.0cm 0.0cm 0.0cm,
    clip=true, width=0.85\columnwidth]{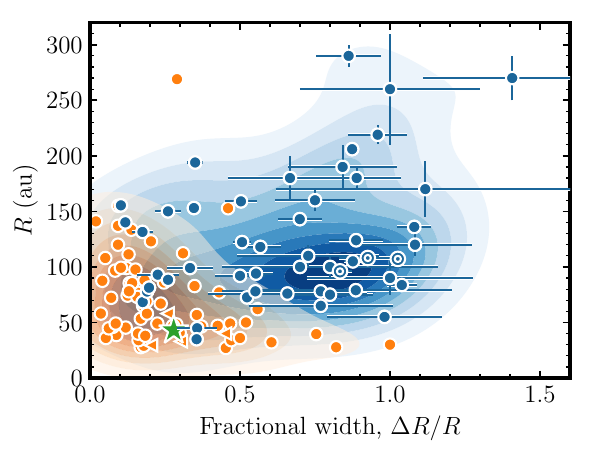}
    %\vspace{-3mm}
  \caption{Distribution of fractional widths ($\Delta R/R$) for exo-Kuiper belts (blue) and protoplanetary rings (orange). The top panel shows the histogram of widths while the bottom panel shows the 2D distribution of fractional widths and central radii. The solid lines (top panel) and filled contours (bottom panel) represent kernel density estimations using a Gaussian kernel and a bandwidth chosen following Scott's rule \citep{Scott2015}. The green dashed line and green circle represent the Kuiper belt fractional width and central radius. The belts with gaps around HD~92945, HD~107146 and HD~206893 are represented by $\circledcirc$ symbols.}
 \label{fig:frac_width}
\end{figure}

From the 74 discs analysed here, we now focus on those that have good estimates of their fractional widths, defined as the ratio between the FWHM and central radius ($\Delta R/R$). We use a threshold value of 50\% in the fractional error of the fractional widths. Using this threshold we obtain a subsample of 50 discs, excluding HD36546 as justified in Sect. \ref{sec:modelling}. Figure~\ref{fig:frac_width} shows in blue the distribution of fractional widths for this subsample (top) and the distribution of fractional widths against the central radius (bottom). We find that the distribution of fractional widths is wide and there is not a strong peak. Nevertheless, we find that roughly 70\% of discs are wide ($\Delta R/R>0.5$), with a median fractional width of 0.71. These numbers do not change significantly if we lower or increase the threshold defined above. This leads us to our first conclusion that very narrow rings such as HR~4796, Fomalhaut and HD~202628 are rare amongst detectable (and hence relatively massive) belts and thus should not be used as good references for the larger population of observable planetesimal belts. This conclusion is unlikely to be biased by narrower belts being generally fainter than broad discs and thus harder to detect. When examining the belt fluxes as a function of fractional widths we do not find any strong correlation. 

\begin{figure}
  \centering 
 \includegraphics[trim=0.0cm 0.0cm 0.0cm 0.0cm,
    clip=true, width=0.85\columnwidth]{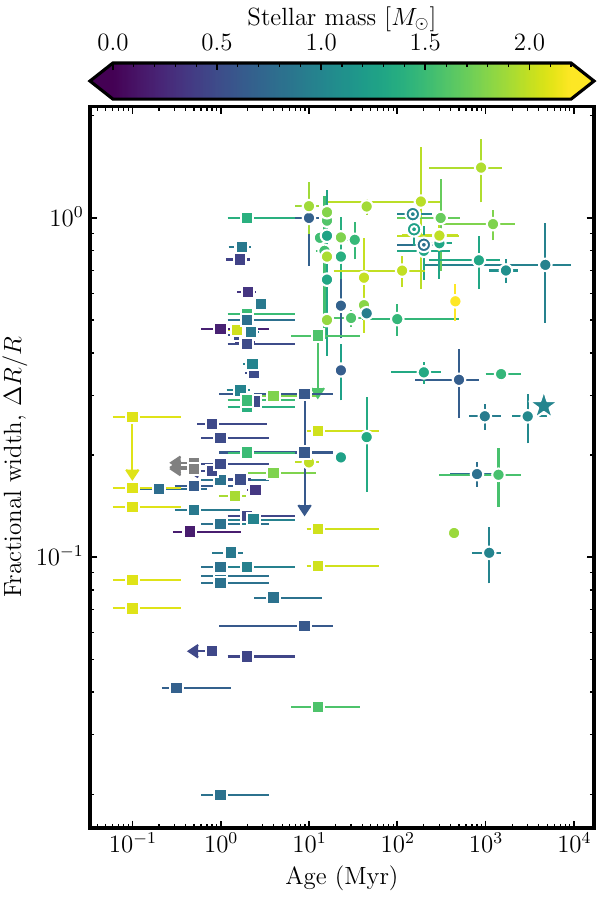}
    %\vspace{-3mm}
  \caption{Estimated ages and fractional widths ($\Delta R/R$) for exo-Kuiper belts (circles), protoplanetary rings (squares), and the Kuiper belt (star).  The belts with gaps around HD~92945, HD~107146, and HD~206893 are represented by $\circledcirc$ symbols. Ages and uncertainties for systems in the DSHARP survey were taken from \cite{Andrews2018}. For systems in Taurus, we randomised their ages with a mean 2~Myr and a standard deviation of 0.5~Myr. For systems in Ophiuchus we assume use the ages reported by \cite{Cieza2021} and assume an age uncertainty of 0.4~dex.}
 \label{fig:age_vs_fwidth}
\end{figure}

We are also interested in comparing this distribution to the fractional widths of rings in protoplanetary discs since those are ideal places for planetesimal formation via streaming instability \citep[e.g.][]{Stammler2019}, and cover a similar range of radii as exoKuiper belts. We compile a sample of 65 protoplanetary rings from three ALMA surveys: DSHARP \citep[Table 1 in ][including HL~Tau and TW~Hya]{Huang2018}, the Taurus star-forming region survey \citep[Table 4 in][]{Long2018}, and ODISEA that target the Ophiuchus star-forming region \citep[Table 6 in][]{Cieza2021}. We note that the widths of rings reported by \cite{Huang2018} and \cite{Cieza2021} for DSHARP and ODISEA are equivalent to a FWHM and are measured from CLEAN images, and thus the width values could be overestimated due to the beam convolution. The widths reported by \cite{Long2018} are derived from visibility modelling assuming Gaussian profiles and are defined as twice the standard deviation (F. Long, private communication). Hence these are deconvolved widths and we convert them to FWHM's by multiplying by a factor of 1.2 (FWHM/(2$\sigma$)). Finally, the widths of four rings in the DSHARP sample are only constrained by upper limits, and here we take them as conservative estimates of their widths.

In orange colour, Fig.~\ref{fig:frac_width} shows the distribution of the 65 rings in our sample of protoplanetary discs. We find that protoplanetary rings tend to be narrower than debris discs, with a median fractional width of 0.18 and only 9\% having values above 0.5. A simple Kolmogorov-Smirnov test indicates a probability below $10^{-9}$ of both fractional width distributions being drawn from the same distribution. We note that both distributions are biased and the test does not take into account the uncertainties, meaning that this comparison is not strictly valid. Nevertheless, they show that the observed distributions are not consistent with each other. 

Figure~\ref{fig:age_vs_fwidth} shows the fractional widths as a function of system age, and coloured by their estimated stellar mass. We find no correlation between the age of systems and their fractional widths. This figure shows, however, that our sample of protoplanetary discs is dominated by low-mass stars ($<1\ M_{\odot}$) whereas REASONS is biased towards intermediate-mass stars ($>1\ M_{\odot}$).

The green vertical dashed line and green star symbol in the top and bottom panels of Figs.~\ref{fig:frac_width} and \ref{fig:age_vs_fwidth} represent the location fractional width and age of the Kuiper belt. These values are estimated from the L7  synthetic model of the inner, main and outer Kuiper belt \citep{Kavelaars2009, Petit2011}\footnote{\url{http://www.cfeps.net/?page\_id=105}}. This synthetic and de-biased model includes the classical, scattered, detached and resonant populations in the Kuiper belt with relative weights set to match the observed populations. We fit a Gaussian profile to this synthetic population and estimate a central radius of 43~au and a FWHM of 12~au. We note that the distribution is wide due to the scattered, detached and resonant components, but it is still heavily peaked around 43~au where the classical belt is located. This synthetic population is an approximation of what the Kuiper belt would look like if detectable and observed by ALMA around another system. Its inferred fractional width of 0.28 makes it closer to the minority of narrow exoKuiper belts and the typical width of protoplanetary rings (0.18). Thus, despite the Kuiper belt having extended components, it would appear narrower than most of the observed exoKuiper belts.

When examining the two-dimensional distribution of fractional widths and radii, we see no strong correlations for exoKuiper belts. Broad and narrow belts are found in both small and large belts, although the five largest belts ($r>200$~au) are all broad belts ($\Delta R/R>0.8$), but these are still low number statistics. On the other hand, the seven widest protoplanetary rings ($\Delta R/R\geq0.5$) are all at a relatively small radius ($R<70$~au). It is possible that these wide protoplanetary rings could be split into multiple narrower rings that are unresolved  \citep[as has been found in some large protoplanetary rings; e.g.][]{Perez2020}, pushing the distribution of protoplanetary rings towards smaller fractional widths.

If the planetesimal population in exoKuiper belts is truly formed in these protoplanetary rings, we can conclude that the planetesimals do not simply inherit the observed dust distribution. We identify three mechanisms that could explain the observed differences. 

\textit{Wide exoKuiper belts have unresolved substructure} - Wide belts could be hiding substructures such as gaps, splitting these wide belts into narrower multiple belts. Such is the case of the wide belts HD~107146 \citep{Marino2018b}, HD~92945 \citep{Marino2019} and HD~206893 \citep{Marino2020, Nederlander2021}, which are represented as double circles in Fig.~\ref{fig:frac_width}. If we considered these wide belts as double, each component would have a fractional width close to $\sim0.4$. Therefore many of these wide belts may be made of multiple narrower belts. Nevertheless, some wide belts such as the ones around q$^1$~Eri and HR~8799 are wide and have been well resolved with multiple beams across showing no evidence of gaps \citep{Lovell2021, Faramaz2021}. Therefore, we conclude that substructures could make some double belts appear as wide single belts, but it is unlikely to explain the whole population of wide belts. The ongoing ALMA large programme ARKS is studying several of these wide belts to determine if they are made of multiple narrow components or not (Marino et al. in prep).
 
 \textit{Protoplanetary rings are not stationary} - If the location of dust-rich rings in protoplanetary rings evolves in time, then planetesimal formation will occur in a wider range of radii compared with the widths of protoplanetary rings. These rings could appear and disappear at different locations \citep[e.g.][]{Dittrich2013, Lenz2019}, or continuously move in time if caused by a planet that is migrating in or other processes \citep[][]{Meru2019, Shibaike2020, Miller2021, Jiang2021}. In particular, \cite{Miller2021} used numerical simulations of dust evolution in protoplanetary discs to show that moving rings could form wide planetesimal belts at tens of au that can explain the large widths found in this sample. This requires a high disc viscosity to enable a fast ring migration. It is still uncertain if the viscosity in protoplanetary discs is high enough for the migrating rings scenario to work. There is, however, tentative evidence that the dust component in protoplanetary discs becomes smaller with time which would agree with this scenario \citep{Hendler2020}.

 \textit{Planetesimal belts widen with time} - It is also possible that planetesimal belts are born narrow and widen due to (i) dynamical instabilities or (ii) viscous spreading. In (i), an initially narrow belt could be disrupted shortly after the protoplanetary disc dispersal if inner planets went through an instability as in the Nice model \citep{Gomes2005}. If so, we would expect the widest belts to be less massive since much mass is lost shortly after the instability \citep{Booth2009}. While these and other trends should be searched for and examined in more detail in dedicated follow-up work, we preliminarily do not find any correlation between the fractional width and fractional luminosity or dust mass in this sample. These and other trends between physical parameters of interest should be addressed in more detail in future, dedicated works.
 Moreover, even a highly disrupted belt like the Kuiper belt is still narrower than most of the observed population. Hence it is unclear whether this scenario could significantly widen a narrow disc up to fractional widths above 0.7. Even the broad disc around HR~8799 which shows evidence of having a scattered disc, still requires a dynamically cold and broad belt to explain the observations \citep{Geiler2019}. In (ii), planetesimal discs could slowly widen due to scattering and collisions \citep{Heng2010}. However, we do not find a width vs age correlation in our sample as shown in Fig.~\ref{fig:age_vs_fwidth}. For example, the three narrowest belts (HR~4796, Fomalhaut, HD~202628) have estimated ages of 10~Myr, 440~Myr and 1.1~Gyr, a distribution that is not particularly young when compared with the wider belts. Therefore, we conclude that it is unlikely that dynamical processes or viscous spreading alone could explain the large width of exoKuiper belts. 
  
All these mechanisms may play a role in the observed population. Further higher-resolution observations of wide belts could answer whether these are composed of multiple narrow belts or not, confirming or ruling out these hypotheses. Similarly, those observations could also reveal if the edges of the wide belts are smooth as expected if they broaden with time. This has only been done for a limited sample of well-resolved belts, showing that exoKuiper belts can display both sharp and smooth edges \citep{Marino2021, Imaz-Blanco2023}. Further modelling and simulations are also crucial to compare dynamical scenarios that could broaden belts with these and future higher-resolution observations.

An important caveat in this comparison is that the protoplanetary discs in this sample include a much larger fraction of low-mass stars compared to REASONS: 71\% of protoplanetary discs in this sample have stellar masses below 1~$M_{\odot}$ whereas this fraction is only 22\% for the well-resolved belts. If rings in protoplanetary discs around more massive stars tended to be much wider, this could solve this discrepancy. However, when examining only the protoplanetary discs around stars more massive than 1~$M_{\odot}$ we find a similar distribution with a median fractional width of 0.2. \cite{Pinilla2018} studied the radius and width of transition discs and found no strong correlation between the fractional width and stellar mass. Dust evolution models predict that rings are narrower for low-mass stars due to more efficient drift, however, the expected correlation is small and likely hidden by the resolution of those observations.

\subsection{Distribution of planetesimal belt masses: Evidence for collisional evolution}
\label{sec:mvsr}

\begin{figure*}
%\vspace{-5mm}
  %\centering 
\hspace{-3mm}
\includegraphics[scale=0.63]{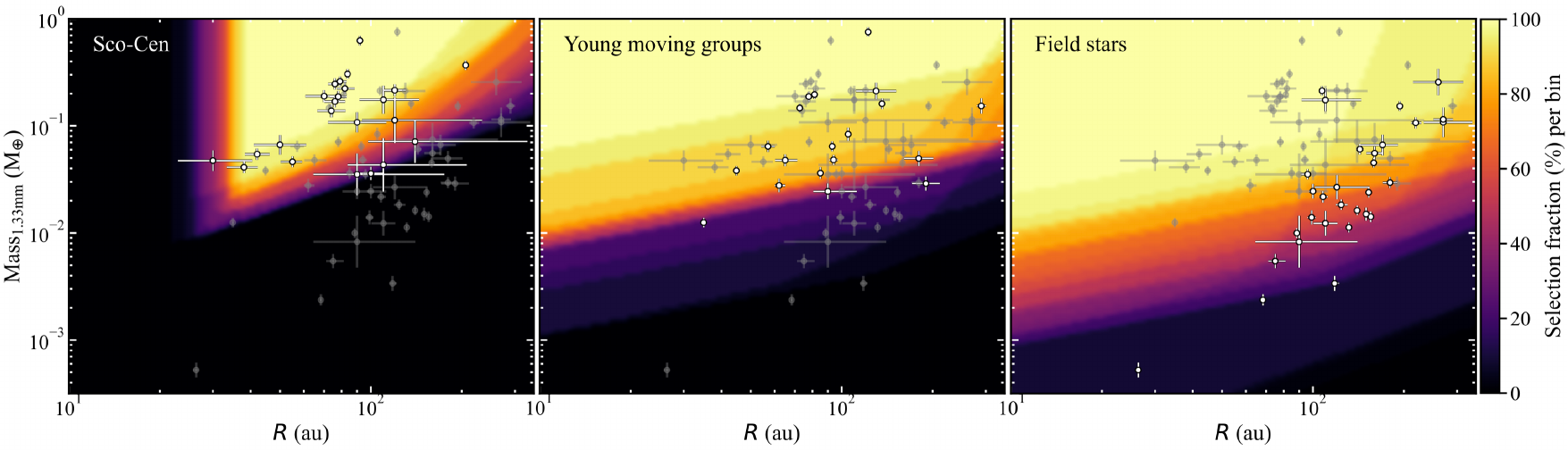}
%\vspace{-7mm}
  \caption{Mass of grains as measured from belt flux densities at 1.33 mm as a function of the radius of observed planetesimal belts (all points with error bars). Each panel focuses on a different subgroup of REASONS belts (shown as the white points, rest of the population in grey), those orbiting stars in the Sco-Cen association (left), young moving groups (middle) and field stars (right). In all panels, the background colour map represents the selection probability (\%), or percentage of belts in that subgroup that would pass the selection effects at a given [Mass$_{\rm 1.33 mm}$-$R$] location, assuming unobserved belts have the same distribution of parameters [$d$,R$_{\rm BB}$/R, L$_{\star}$, $\lambda_0$, $\beta$] as the observed population. The changing colour maps in each panel show how selection biases vary depending on stellar properties and observing strategies employed for different subgroups.}
 \label{fig:mvsr}
\end{figure*}

Figure \ref{fig:mvsr} shows the distribution of planetesimal belt masses derived from mm-wavelength measurements, as a function of the belts' true (resolved) radii. For each belt, dust masses were derived by first extrapolating flux densities from their measured wavelength (Table \ref{tab:resbelts} and \ref{tab:resbeltsarc}) to a common wavelength of 1.33 mm, using best-fit mm slope values $\beta$ from spectral modelling of the cold dust component (Table \ref{tab:sdbprops}). Then, we estimate masses of grains dominating the emission at 1.33 mm using 
\begin{equation} \label{eq:dustmass}
\mathrm{Mass}_{\rm 1.33mm} = \frac{F_{\nu_{\rm belt}} d^2}{\kappa_\nu B_\nu (T(R))},
\end{equation}
where $d$ is the distance to the star from Earth in m, $F_{\nu_{\rm belt}}$ is the flux density of the belt in Wm$^{-2}$Hz$^{-1}$, and $\kappa_\nu$ is the dust opacity, assumed to be 0.23 m$^2$ kg$^{-1}$, by scaling down 1 m$^2$ kg$^{-1}$ at 1000 GHz linearly to the frequency of our observations \citep[][]{Beckwith1990}. $B_{\nu}(T(R))$ is the Planck function, and $T(R)$ is the temperature of the grains (in K) dominating the emission at 1.33 mm. We assume these large grains to behave similar to blackbodies leading to
\begin{equation} \label{eq:dusttemp}
T(R) = 278.3 L_{\star}^{0.25} R^{-0.5},
\end{equation}
where $L_{\star}$ is the stellar luminosity in Solar luminosities, and R is the belt midpoint radius in au.

As noticed in Sect. \ref{sec:rvsLstar}, the majority of belts in our sample have large radii; only 9/74 belts are smaller than $60$ au. In Fig. \ref{fig:mvsr}, we construct bias maps as described in Sect. \ref{sec:rvsLstar} for each of three subgroups of the REASONS sample: belts around stars in Sco-Cen, in young moving groups, or around field stars.

For young moving groups and field stars, we find that selection effects (1+2A from Sect. \ref{sec:bias}) generally favour belts that are smaller (lower $R$) and/or more massive, once again because they are easier to detect (colour scale brighter towards the top left). 
The shape of the lower envelope of selection in the bias map can have two regimes. For most radii considered, selection is limited by detection at wavelengths on the Rayleigh-Jeans side of the blackbody function, leading to a trend following Mass$_{\rm 1.33mm}\propto \sqrt{R}$. At the largest radii displayed in Fig. \ref{fig:mvsr}, the slope of the trend steepens; this is because the dust becomes cold enough that selection is now limited by detection at wavelengths where the Rayleigh-Jeans approximation breaks down, leading to Mass$_{\rm 1.33mm}\propto e^{CR^{0.5}}$ (where C is a constant dependent on $L_{\star}$) on the Wien side of the blackbody function.  For stars in Sco-Cen, we find a similar trend with selection effects (1+2B from Sect. \ref{sec:bias}) favouring small and/or massive belts. However, the two key differences are that 1) there is a small radius cut-off set at $R=0.25\arcsec$, below which belts could not be resolved even interferometrically given the beam FWHM of $\sim1\arcsec$ at which most Sco-Cen observations were carried out \citep{Lieman-Sifry2016}; and 2) the lower envelope of detectability which is set again for most stars by the survey sample design of \citet{Lieman-Sifry2016}, selecting only belts with a 70 $\mu$m fractional excess of $>100$.

We also underline that a significant factor moving the lower envelope of selection vertically in any panel and between panels is the distance of a system from Earth, with the median distance of stars being 127 pc in the Sco-Cen subsample, 49 pc in the moving group subsample, and 24 pc for the field star subsample. These selection effects are apparent within subsamples as well as in the overall REASONS sample, which does show a trend where lower mass belts (towards the bottom of Fig. \ref{fig:mvsr}) tend to be the ones closest to Earth (e.g. in the field subsample), simply because they would not have been detectable had they been located further away (e.g. around Sco-Cen stars).

%Having considered our observational bias, it becomes clear that there is a true lack of small, massive belts ($R\lesssim60$ au and Mass$_{\rm 1.33mm}\gtrsim0.1$ M$_{\oplus}$). We also observe a lack of small ($R\lesssim60$ au) and very large ($R\gtrsim150$ au) belts in the lowest mass range (Mass$_{\rm 1.33mm}\lesssim2\times10^{-2}$ M$_{\oplus}$), particularly when neglecting AU Mic and Fomalhaut C as being very different from other systems in terms of stellar host properties. This is only applicable to field stars, closest to Earth, where low mass belts could have been detected, and likely somewhat affected by the difficulty in detecting very large, cold low-mass belts.

\begin{figure}
%%\vspace{-20mm}
  %\centering 
\hspace{-5mm}
\includegraphics[scale=0.36]{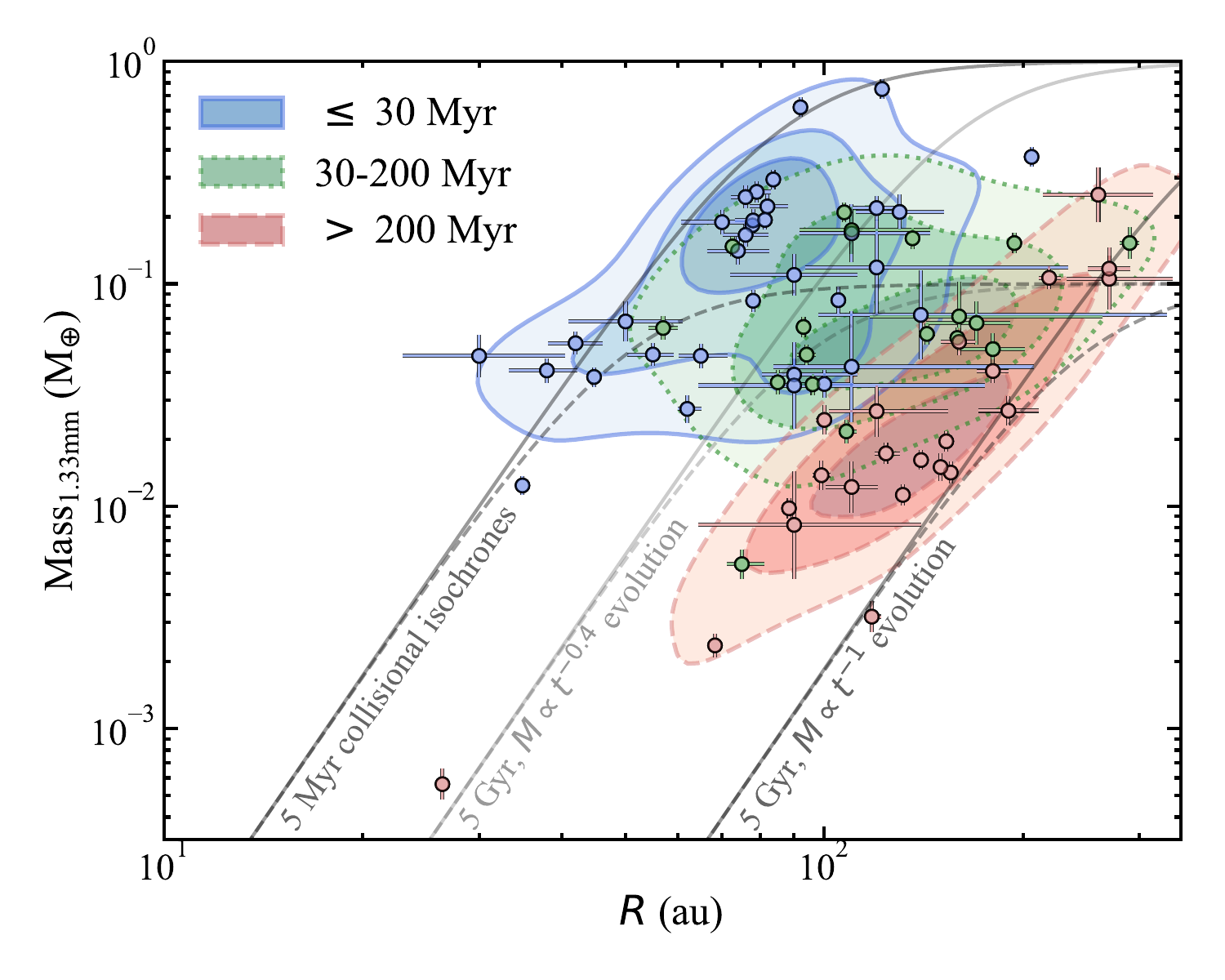}
%\vspace{-5mm}
  \caption{Belt dust mass (as measured at 1.33 mm) as a function of resolved belt radius for the REASONS sample (coloured points with error bars). The filled contours represent the 2D density distributions of belts $\leq30$ (blue solid), 30-200 (green dotted), and $>200$ (red dashed) Myr-old. Contour lines are defined to contain 20, 50, and 80\% of the population for each age range. Black lines represent predictions from a simple collisional evolution model as described in Sect. \ref{sec:mvsr}, for collisional ages of 5 Myr (left) and 5 Gyr (right), an initial dust mass of 1 M$_{\oplus}$ (solid) or 0.1 M$_{\oplus}$ (dashed), and dust masses evolving with time as $t^{-0.4}$ or $t^{-1}$ between 5 Myr and 5 Gyr.}
 \label{fig:mvsrvsage}
\end{figure}

Having considered our observational bias, in Fig. \ref{fig:mvsrvsage} we explore [Mass$_{\rm 1.33mm}$-$R$] trends by separating belts in three simple, empirical evolutionary groups. While we acknowledge that ages (Table \ref{tab:sdbprops}) are uncertain, we divide stars by age as $\leq$30 Myr (youngest, blue points in Fig. \ref{fig:mvsrvsage}), 30-200 Myr (intermediate, green points), and $>$200 Myr (oldest, red points). For each group, we create a 2D density distribution using kernel density estimation, employing a Gaussian kernel with bandwidth according to Scott's rule \citep[][]{Scott2015}. The 2D distributions are shown as the filled blue solid ($\leq$30 Myr), green dotted (30-200 Myr), and red dashed ($>$200 Myr) contours in Fig. \ref{fig:mvsrvsage}.

We observe a clear trend with belts around old field stars being on average significantly less massive and at the same time larger than belts around the youngest moving group, and slightly older stars. While keeping in mind (Fig. \ref{fig:mvsr}) that young belts less massive than observed would not have been detectable at the distance of Sco-Cen, we conclude that belts that are both as massive and as small as those observed in the Sco-Cen sample (moving towards the top left in the plot) must be rare around field stars. This would imply that either belts around field stars were born with different properties compared to belts around stars that are currently young, which we deem unlikely, or that belts evolve to lower masses and/or larger sizes with time. 

To explain this trend, we consider a simple evolutionary model inspired by the analytical collisional evolution model of \citet[][]{Wyatt2007a}. In this model Mass$_{\rm 1.33mm}$($t$)=Mass$_{\rm 1.33mm}$($t_0$)$/(1+[(t-t_0)^{\epsilon}]/t_c)$, where $t$ is the age of the system and $t_0$ is the time at which collisional evolution begins, assumed to be $t_0=10$ Myr for simplicity. Mass$_{\rm 1.33mm}$($t_0$) is the belt mass in mm grains at birth, assumed to be independent of stellar and belt properties. $t_c$ is the collisional timescale of the largest planetesimals in the belt, assumed to be $t_c=(D/$Mass$_{\rm 1.33mm}$($t_0$))$R^{\delta}$, where $D$ is a constant, incorporating the dependence of the collisional timescale on other stellar and belt properties \citep[e.g. see Eq. 16 in][]{Wyatt2008}. For a given system age $t$, time evolution exponent $\epsilon$, radial dependence exponent $\delta$, and constants Mass$_{\rm 1.33mm}$($t_0$), and $D$, we can draw a collisional isochrone (grey lines) representing the expected locus of belts in Fig. \ref{fig:mvsrvsage} at a given age.

While a detailed fit is beyond the scope of this paper, we assume $\epsilon=1$, $\delta=13/3$ \citep[as in the simple model of][]{Wyatt2007a}, and draw collisional isochrones in Fig. \ref{fig:mvsrvsage} (dark grey lines) for ages $t=15$ Myr (roughly representing most of the young stars in our sample, belonging to the Sco-Cen association), and $t=5$ Gyr (representing the oldest field stars in our sample). We assume an initial belt mass of mm grains Mass$_{\rm 1.33mm}$($t_0$) of 1 M$_{\oplus}$ (solid lines) or 0.1 M$_{\oplus}$ (dashed lines); this sets the vertical location of the horizontal regime of the collisional isochrones, along which belts are yet to reach collisional equilibrium ($t<t_c$). On the other hand, the factor $D$ affects the horizontal location of the diagonal part of the isochrones, representing belts that have reached collisional equilibrium ($t>t_c$). For example, increasing $D$ by an order of magnitude would make the collisional timescale $t_c$ 10 times longer, which means *both* the 5 Myr and 5 Gyr isochrones would shift to the left in the plot. This is because belts at a given radius would retain more mass at the same collisional age. We find that a good qualitative fit to the data can be found by setting $D\sim2\times10^{-8}$ Myr M$_{\oplus}$ au$^{-13/3}$.

Overall, at any given age $t>t_c$, we should expect a diagonal locus in [Mass$_{\rm 1.33mm}$-$R$] representing belts that have reached collisional equilibrium. This is in clear agreement with the older field population (red in Fig. \ref{fig:mvsrvsage}), where this locus lies along a slope roughly consistent (within the uncertainties) to the Mass$_{\rm 1.33mm}\propto R^{13/3}$ expected from analytical collisional evolution models.

Moreover, for belts along this diagonal locus (i.e. in collisional equilibrium), the rate of mass depletion represented by the exponent $\epsilon$ in Mass$_{\rm 1.33mm}(t)\propto (t-t_0)^{-\epsilon}$ determines the mass ratio between belt populations of different ages. This ratio corresponds to a vertical offset in the log-log plot of Fig. \ref{fig:mvsrvsage}, formally written as
\begin{equation}
\epsilon=\frac{\mathrm{log}[\mathrm{Mass}_{\rm 1.33mm}(t_{\rm young})]-\mathrm{log}[\mathrm{Mass}_{\rm 1.33mm}(t_{\rm field})]}{\mathrm{log}(t_{\rm field}-t_0)-\mathrm{log}(t_{\rm young}-t_0)}.
\end{equation}

Figure \ref{fig:mvsrvsage} shows a $\sim$2 dex mass depletion (vertical offset in [Mass$_{\rm 1.33mm}$-$R$]) in $\sim$2 dex of collisional age (between $\sim$10 Myr and $\sim$1 Gyr). This would imply that mass collisionally depletes linearly with age ($\epsilon\sim1$, so Mass$_{\rm 1.33mm}\propto (t-t_0)^{-1}$), in line with the expectation from the simple collisional evolution model of \citet{Wyatt2007a}, and assuming the observed field and young populations of belts share the same stellar and collisional properties. For comparison, we also test models with $\epsilon=0.4$ \citep[as predicted by other models e.g.][]{Lohne2008, Kral2013}, adjusting $D$ such that the 5 Myr isochrone matches with the one from the $\epsilon=1$ model. Then, we clearly see that the 5 Gyr isochrone of this slower evolution model (lighter grey line in Fig. \ref{fig:mvsrvsage}) significantly overestimates the mass of belts in the old field population.

In conclusion, the evolutionary trends observed in [Mass$_{\rm 1.33mm}$-$R$] space provide strong evidence for (radius-dependent) collisional evolution of planetesimal belts, with a mass depletion that is consistent with being linear with time after reaching collisional equilibrium. %Additionally, the population data strongly supports a collisional timescale that varies steeply with radius, similar to the $R^{13/3}$ dependence expected from simple collisional evolution models. In the future, these results should be further tested by forward modelling of the population including effects of parameters neglected in this basic analysis, for example differing stellar properties, planetesimal sizes, strengths and resolved radial widths.
%\subsection{The distribution of planetesimal belt masses: selection effects explain the lack of belts around late-type stars}
%\label{sec:mvsLstar}

\subsection{Distribution of belt vertical aspect ratios: No evolutionary trend}
\label{sec:h}
\begin{figure} 
\hspace{-3mm}
\includegraphics[scale=0.36]{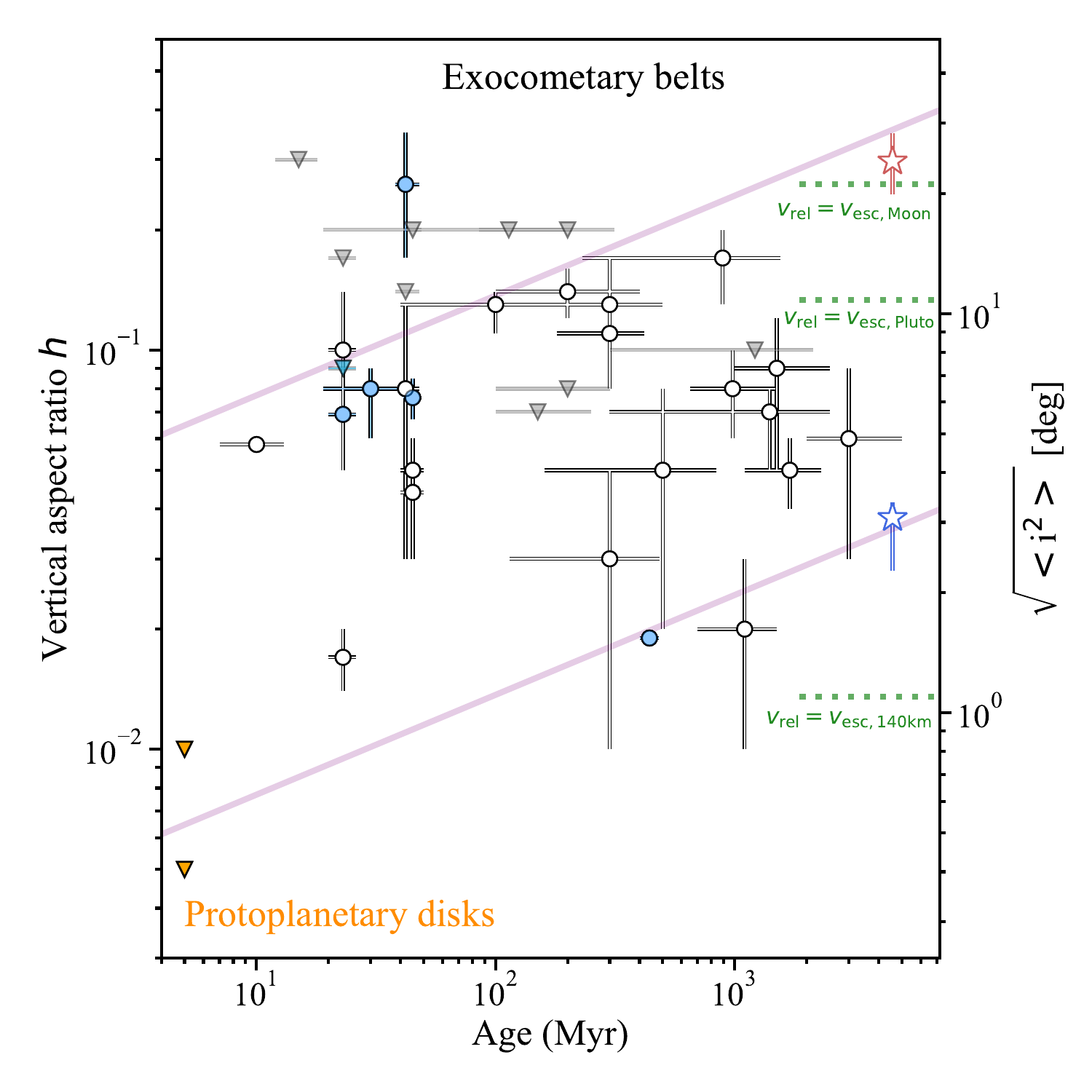}
%\vspace{-5mm}
  \caption{Vertical aspect ratios $h$ as a function of host-star age for REASONS planetesimal belts (left y axis), a measure of the RMS inclination of dust grains (right y axis) assuming a Rayleigh distribution of inclinations. Black and white points with errors are measured values. The red and blue star symbols are the observed RMS inclinations of the hot and cold classical populations of Kuiper belt objects \citep[][]{Brown2001}. Purple lines represent the $\sqrt{\langle i^2\rangle}\propto t^{\frac{1}{4}}$ increase with age expected for planetesimal belts stirred by large bodies, with mass $\times$ surface density $\Sigma M=10^{-2}$ M$_{\oplus}^2$ au$^{-2}$ (top) and $\Sigma M=10^{-6}$ M$_{\oplus}^2$ au$^{-2}$ (bottom), from \citet{IdaMakino1993}. Green dotted lines represent the maximum aspect ratios/RMS inclinations expected for belts with Moon-like, Pluto-like and 140 km-sized stirring bodies (assuming the median stellar host mass and belt radius of the REASONS sample).}
 \label{fig:hvsage}
\end{figure}
The REASONS survey also allows us to look at the vertical aspect ratio $h$ of belts as a population, for the first time. We find the aspect ratio to be meaningfully constrained (i.e. to have a non-zero peak in its posterior probability distribution) in 24 belts. However, we caution the reader that this does not necessarily mean that these belts were all vertically resolved, as height information may also have been extracted from a belt's on-sky projected azimuthal intensity profile. This is because assuming the belt to be intrinsically azimuthally symmetric, a higher intensity contrast between the sky-projected semimajor and semiminor axis implies a larger scale height compared to a lower intensity contrast \citep{Marino2016}. Additionally, we remind the reader that the aspect ratio was assumed to be constant with radius, though that is not necessarily the case \citep[e.g.][]{Matra2019b}, which could lead to biasing of this value, especially for broader belts.

\citet{Matra2019b} demonstrated how a belt's vertical structure may be linked to the distribution of orbital inclinations of particles, based on a similar analysis for the Solar System's Kuiper Belt \citep{Brown2001}. Assuming a Rayleigh distribution of particle inclinations, as expected from gravitational perturbations of planetesimals in a thin disc \citep{IdaMakino1992}, the expected vertical (latitudinal) distribution of particles is Gaussian, with the aspect ratio $h$ being related to the RMS $\sqrt{\langle i^2\rangle}$ of the inclination distribution through $\sqrt{\langle i^2\rangle}=\sqrt{2} h$.

Particle inclinations may be inherited from the belt formation process in protoplanetary discs, but may also be, over time, imparted by gravitational interaction with large bodies interior to, or exterior to, the belt. In the case of a large body (or bodies) within a belt, viscous stirring would continuously act to increase the velocity dispersion $v_{\mathrm{rel}}$ of planetesimals \citep{KokuboIda2012}, producing a Rayleigh distribution of inclinations with $\sqrt{\langle i^2\rangle}=v_{\mathrm{rel}}/(\sqrt{6}v_{\mathrm{Kep}})$. In this scenario, $v_{\mathrm{rel}}$ and therefore the observed $h$ will be dependent on the product between the mass and surface density of large bodies, the stellar mass, the belt radius, and the time since the stirring initiated \citep[][]{IdaMakino1993}. Over time, $v_{\mathrm{rel}}$ will increase until collisions become important, effectively destroying stirred-particles, causing $v_{\mathrm{rel}}$ to stop increasing. This maximum $v_{\mathrm{rel}}$ can be shown to be of order the escape velocity $v_{\mathrm{esc}}$ of the largest stirring bodies \citep[]{Matra2019b}. In this scenario, setting $v_{\mathrm{rel}}=v_{\mathrm{esc}}$ allows us to link the measured aspect ratio $h$ to a lower limit to the size and mass of the large bodies stirring the planetesimal belt from within.

Figure \ref{fig:hvsage} (circles  for detections and upside down triangles for upper limits) shows the aspect ratios measured for our REASONS population. We find well-constrained aspect ratios in the range between $\sim$0.01 and 0.3, corresponding to $\sqrt{\langle i^2\rangle}$ values between $\sim1-20^{\circ}$, and relative velocities between $\sim0.1-4$ km/s \citep[using Eq. 10 from][and using stellar masses derived from the observed luminosities assuming all stars are on the main sequence]{Matra2019b}. 
With the caveats mentioned above, this implies that relative velocities larger than this (dynamically hotter belts) are rare, as we would have been able to detect/resolve belts with larger scale heights. On the other hand, we cannot exclude that dynamically cold belts (aspect ratios thinner than $h\sim0.01$) exist, as we might not have been sensitive to/able to resolve such small aspect ratios.
In the above scenario of a large body stirring the planetesimals and observable grains within the belt, the measured aspect ratios would imply minimum sizes for the belt's largest bodies between $\sim$140~km and the size of the Moon (green lines in Fig. \ref{fig:hvsage}, assuming the density of Pluto and the Moon respectively, and assuming the median host mass and belt radius of the REASONS sample).

While we might expect a trend of increasing $h$ with age in the viscous stirring picture drawn above \citep[following  $\sqrt{\langle i^2\rangle}\propto t^{\frac{1}{4}}$, purple lines in Fig. \ref{fig:hvsage},][]{IdaMakino1993}, we find no such trend in our data. This could be explained in one of the following two ways. In the first one, planetesimals and/or mm grains are viscously stirred by large bodies within the belt to the observed inclination distributions early on, within the first $\sim$10-20 Myr. In this case, the observed $\sqrt{\langle i^2\rangle}$ would be a genuine tracer of the escape velocity of the largest bodies within the belt. We note that mm grains cannot be born with the observed inclination distributions within protoplanetary discs, because ALMA observations of Class II discs so far show that mm grains are heavily settled, with upper limits of $h\lesssim0.005-0.01$ at 100 au \citep[e.g.][and references therein]{Villenave2023}. Indeed, we report a sharp transition in the evolution of vertical aspect ratios of mm grains between protoplanetary discs and exocometary belts (Fig. \ref{fig:hvsage}). This is likely produced by the presence of large amounts of gas in the former compared to the latter, favouring strong settling of these grains to the disc midplane in protoplanetary discs. Then, the observed sharp transitions would indicate that either planetesimals and mm grains are stirred very quickly once the protoplanetary disc is dissipated, or that (unobservable) planetesimals are born stirred and mm grains are quickly collisionally replenished on orbits tracing the stirred planetesimals. Either way, our results indicate that for observable belts, stirring is already in place by the time a belt is $\sim10-50$ Myrs old.

Another possibility is that the diversity of belt aspect ratios observed is produced by different and/or multiple evolutionary pathways with processes like viscous stirring, collisional damping \citep{PanSchlichting2012}, and/or gravitational perturbations by bodies exterior to \citep[e.g.][]{Mustill2009} or migrating into the belt \citep[as is the case for the Solar System; e.g.][]{Nesvorny2015} playing a role. We note that no clear trends are found for the aspect ratios as a function of stellar properties, belt mass in mm grains, or belt radius. Interestingly, no trend is found as a function of presence of gas in a given system either; this implies that gas-bearing systems (blue data points\footnote{Gas-bearing discs considered: HD21997, HD32297, HD39060, HD9672, HD181327, HD216956} in Fig. \ref{fig:hvsage}),  have aspect ratios that are no smaller than other systems where gas has not been detected. Therefore, there is no evidence in the REASONS sample that $\sim$mm-sized grains settle to the midplane as expected in the presence of high gas densities \citep[][]{Olofsson2022}. This in turn suggests that the gas densities in gas-bearing debris discs are very different when compared to younger protoplanetary discs, and might not be sufficient to affect the dynamics of mm-sized grains.

Finally, we find that our Kuiper belt's dynamically cold classical population falls right within the range of extrasolar RMS inclinations measured in extrasolar belts, whereas only one belt (HD21997) has a value consistent with our belt's dynamically hot population. As found for $\beta$ Pictoris, it is possible that extrasolar belts also host multiple dynamical populations with different levels of excitation like our Kuiper belt, but follow-up ALMA observations at higher angular resolution and sensitivity will be needed to test this for the majority of the belts in Fig. \ref{fig:hvsage}.

\section{Summary} \label{sec:summ}
In this work, we present the results of the REASONS survey and archival reanalysis programme, leading to a joint sample of 74 planetesimal belts interferometrically resolved in dust continuum emission at mm wavelengths with ALMA and the SMA. We imaged, analysed, and modelled all visibility datasets uniformly to derive the spatial properties of all belts, which we combined with fits to multi-wavelength photometry to obtain host-star and dust emission properties. This results in an N-dimensional table of system properties for the whole REASONS sample, which we release to the community (links at the beginning of Sect. \ref{sec:disc}) together with the mm-wave-calibrated data products and fit results.

A first analysis of this legacy population dataset led us to the following findings:
\begin{itemize}
    \item In agreement with the literature, we find a shallow trend in belt radius with stellar luminosity in the observed (biased) population, although we obtain a higher intrinsic scatter and consequently a lower degree of correlation compared to previous findings. We confirm a general lack of belts smaller than a few tens of au%\LEt{*** The meaning of "smaller than (a range)" is unclear.***}
    , which would have been easily detected and resolved if they were as massive as other belts in the observed population. At the same time, we find a joint trend in belt mass versus radius, with younger belts appearing on average smaller and more massive than older belts around field stars. We attribute this to collisional evolution, depleting smaller belts faster than larger ones. Coeval belts at collisional equilibrium should follow a locus in mass--radius space, which is clearly observed for the older field population. The spread in mass--radius space between young and old belts %\LEt{*** is there not redundancy in the terminology here? Or, please consider expanding "across ages" or rewording as the meaning is not explicitly clear.***}
    is consistent with a roughly $t^{-1}$ rate of mass depletion, as expected from simple collisional evolution models.
    \item For the 50 belts with well-constrained widths, we find that $\sim$70\% of them are broad ($\Delta R/R>$0.5), with a median fractional width of 0.71 for the REASONS sample. This implies that well-known narrow rings such as that found around Fomalhaut are rare amongst the observed population. The distribution of observed widths is inconsistent with the distribution of protoplanetary ring widths observed by ALMA. We attribute the high fraction of broad belts to either unresolved substructure, or moving protoplanetary rings. The lack of correlation between belt widths and system age disfavours a scenario where belts start narrow and broaden with time due to outward scattering by inner planets.
    \item For the 24 belts with well-constrained vertical aspect ratios $h$, we find values of $\sim$0.01-0.3, indicating RMS orbital inclinations of $\sim$1-20$^{\circ}$. We find that approximately mm-sized grains in planetesimal belts have much broader inclination distributions than observed in protoplanetary discs, which is likely because they are not affected by settling to the midplane in the presence of large amounts of gas. Additionally, gas-bearing belts do not appear significantly vertically thinner than the rest of the belt population, suggesting that they host much lower gas densities than protoplanetary discs, and that gas does not play a significant role in the dynamics of mm grains. The observed aspect ratios of planetesimal belts show no correlation with system age, as would be expected in a scenario where they are stirred by large bodies within them. This indicates that mm grains are either stirred early ---within the first 10 Myr of evolution--- by large bodies with sizes constrained to at least $\sim$140 km (and up to at least the size of the Moon in some cases), or are rapidly collisionally produced with high inclinations by larger grains and/or planetesimals that are themselves stirred very early%\LEt{***Please check that I have retained your intended meaning.***}. 
    A diversity of evolutionary pathways may also explain the lack of an observed age trend.
\end{itemize}

Overall, the REASONS survey presented here shows the power of samples of resolved planetesimal belts, motivating follow-up at other wavelengths, and at even higher resolution with ALMA, which will provide an invaluable dataset for upcoming population modelling studies. Population modelling is necessary to gain a comprehensive picture of the evolution of planetesimal belts, and to infer their origin in protoplanetary discs. The population trends highlighted in this work showcase the potential information to be gleaned from this resolved sample, which we expect the community will fully explore  in the coming years. 

We remain conscious of the biases and limited (although much improved) size of our sample, which complicate interpretations of population properties. These are largely due to surface brightness sensitivity at mm wavelengths, which prevents intrinsically fainter and/or more distant belts from being imaged. To improve on the current sensitivity and enable even larger samples, we look forward to advances in bandwidth and thus continuum sensitivity at mm wavelengths \citep[such as foreseen by the ALMA Development Roadmap;][]{Carpenter2020}. In the long term, we encourage efforts to develop subarcsecond imaging capabilities from space at FIR wavelengths where belts are brightest (Matthews et al. in prep), and/or efforts to significantly increase the collecting area of ground-based interferometers to enable the next big leap in our understanding of the solar neighbourhood's population of planetesimal belts.

\section{Data Availability}
\label{sec:dataavail}
Observational log tables for the REASONS ALMA observing programme, the REASONS SMA observing programme and for the archival programme are available on \textsc{Zenodo} \href{https://doi.org/10.5281/zenodo.13959902}{here}.

The N-dimensional sample of belt properties is available as a \textsc{Pandas} dataframe \citep[][]{pandas2020, Mckinney2010} \href{https://github.com/dlmatra/REASONSbokehbinder/blob/master/bokehplots/static/REASONS_DataFrame_withsdbinfo}{here}, and can be explored through interactive plots \href{https://mybinder.org/v2/gh/dlmatra/REASONSbokehbinder/master?urlpath=/proxy/5006/bokehplots}{here}. 

For each system, the mm calibrated visibility continuum data used for the analysis, the CLEAN images shown in Fig. \ref{fig:gallery} (in FITS and PDF format), a corner plot with the results of the interferometric modelling, a goodness-of-fit check multi-panel image such as Fig. \ref{fig:evaluatemodel} and an SED with best-fit model overlaid such as Fig. \ref{fig:evaluatesedmodel} are available in compressed format in a dedicated \textsc{Zenodo} repository \href{https://doi.org/10.5281/zenodo.12583396}{here} \citep[][]{zenodoreasons}.

\begin{acknowledgements}
LM acknowledges funding from the Irish Research Council (IRC) under grant number IRCLA-2022-3788, from the European Union's Horizon 2020 research and innovation programme under the Marie Sklodowska-Curie grant agreement No 101031685, and by the Smithsonian Institution as a Submillimeter Array (SMA) Fellow. SM is supported by the Royal Society through a University Research Fellowship. JPM acknowledges research support from the National Science and Technology Council of Taiwan under grants NSTC109- 2112-M-001-036-MY3 and NSTC112-2112-M-001-032-MY3. CdB acknowledges support from a Beatriz Galindo senior fellowship (BG22/00166) from the Spanish Ministry of Science, Innovation and Universities. AGS acknowledges support by the National Science Foundation Graduate Research Fellowship Program under Grant No. 1842402. AMH is supported by NSF grant AST-2307920. MB received funding from the European Union's Horizon 2020 research and innovation program under grant agreement No. 951815 (AtLAST). MB and AVK were supported by DFG projects Kr 2164/13-2 and Kr 2164/14-2.

The Submillimeter Array is a joint project between the Smithsonian
Astrophysical Observatory and the Academia Sinica Institute of Astronomy and Astrophysics and is funded by the Smithsonian Institution and the Academia Sinica.
This paper makes use of ALMA data ADS/JAO.ALMA\#2017.1.00200.S. ALMA is a partnership of ESO (representing its member states), NSF (USA) and NINS (Japan), together with NRC (Canada), NSC and ASIAA (Taiwan), and KASI (Republic of Korea), in cooperation with the Republic of Chile. The Joint ALMA Observatory is operated by ESO, AUI/NRAO and NAOJ. 
\end{acknowledgements}

% WARNING
%-------------------------------------------------------------------
% Please note that we have included the references to the file aa.dem in
% order to compile it, but we ask you to:
%
% - use BibTeX with the regular commands:
\bibliographystyle{aa} % style aa.bst
\bibliography{lib} % your references Yourfile.bib
%
% - join the .bib files when you upload your source files
%-------------------------------------------------------------------

\onecolumn

\begin{appendix} %First appendix

\section{Additional Tables}

\begin{longtable}{ccccccccc}
\caption{REASONS Archival Resolved Belts}
\label{tab:resbeltsarc} \\
\hline\hline
Target & $\lambda$ & $F_{\nu_{\star}}$ & $F_{\nu_{\rm belt}}$ & $R$ & $\Delta R$ & $h$ & $i$ & PA \\ 
 & mm & $\mu$Jy & mJy & au & au &  & $^{\circ}$ & $^{\circ}$ \\
\hline
HD105 & $1.35$ & $-$ & $1.36^{+0.10}_{-0.10}$ & $85^{+2}_{-2}$ & $<30$ & $<0.2$ & $49^{+3}_{-2}$ & $14^{+3}_{-3}$ \\
HD9672 & $0.855$ & $<100$ & $16.3^{+0.8}_{-0.8}$ & $136^{+2}_{-2}$ & $147^{+8}_{-7}$ & $0.076^{+0.009}_{-0.009}$ & $79.1^{+0.4}_{-0.4}$ & $107.4^{+0.5}_{-0.4}$ \\
HD10647 & $1.26$ & $^{a}40^{+10}_{-10}$ & $5.2^{+0.6}_{-0.6}$ & $100^{+2}_{-3}$ & $70^{+5}_{-6}$ & $^{a}0.05^{+0.01}_{-0.01}$ & $77.2^{+0.5}_{-0.6}$ & $56.8^{+0.5}_{-0.5}$ \\
HD15115 & $1.34$ & $^{a}40^{+20}_{-20}$ & $2.02^{+0.06}_{-0.06}$ & $93.0^{+1.0}_{-1.0}$ & $^{a}21^{+6}_{-7}$ & $^{a}0.05^{+0.01}_{-0.02}$ & $^{b}88^{+1}_{-1}$ & $98.5^{+0.3}_{-0.3}$ \\
HD16743 & $1.27$ & $^{a}20^{+10}_{-10}$ & $1.24^{+0.04}_{-0.04}$ & $159^{+2}_{-2}$ & $80^{+8}_{-9}$ & $0.13^{+0.01}_{-0.02}$ & $>80$ & $168.4^{+0.6}_{-0.5}$ \\
HD21997 & $0.895$ & $<70$ & $3.5^{+0.1}_{-0.1}$ & $94^{+3}_{-3}$ & $52^{+5}_{-5}$ & $0.26^{+0.09}_{-0.09}$ & $34^{+3}_{-4}$ & $27^{+6}_{-6}$ \\
HD22049 & $1.26$ & $890^{+70}_{-60}$ & $12.0^{+0.8}_{-0.8}$ & $68.3^{+0.5}_{-0.5}$ & $12^{+1}_{-1}$ & $-$ & $30^{+2}_{-2}$ & $178^{+4}_{-4}$ \\
HD32297 & $1.34$ & $<80$ & $3.46^{+0.03}_{-0.03}$ & $122.3^{+0.7}_{-0.7}$ & $62^{+4}_{-3}$ & $^{a}0.08^{+0.01}_{-0.02}$ & $^{b}87^{+2}_{-2}$ & $47.8^{+0.3}_{-0.3}$ \\
HD36546 & $1.33$ & $-$ & $2.45^{+0.04}_{-0.04}$ & $70^{+10}_{-10}$ & $150^{+20}_{-20}$ & $-$ & $>80$ & $257.3^{+0.9}_{-0.9}$ \\
HD38206 & $1.35$ & $-$ & $0.9^{+0.1}_{-0.1}$ & $180^{+20}_{-20}$ & $120^{+30}_{-40}$ & $<0.14$ & $^{b}84^{+2}_{-2}$ & $84^{+2}_{-1}$ \\
HD38858$^{\star}$ & $1.26$ & $-$ & $2.7^{+0.7}_{-0.6}$ & $110^{+10}_{-10}$ & $80^{+30}_{-20}$ & $-$ & $^{c}40^{+10}_{-20}$ & $60^{+20}_{-20}$ \\
HD39060$^{\star}$ & $1.33$ & $^{d}80^{+20}_{-20}$ & $20^{+2}_{-2}$ & $105^{+1}_{-1}$ & $92^{+3}_{-3}$ & $0.069^{+0.003}_{-0.003}$ & $86.6^{+0.4}_{-0.3}$ & $30.0^{+0.1}_{-0.1}$ \\
HD48682 & $1.31$ & $-$ & $2^{+1}_{-1}$ & $90^{+40}_{-30}$ & $<200$ & $-$ & $^{e}60^{+30}_{-60}$ & $^{f}110^{+70}_{-110}$ \\
HD50571 & $0.873$ & $^{a}40^{+30}_{-30}$ & $3.8^{+0.4}_{-0.3}$& $190^{+20}_{-20}$ & $160^{+30}_{-30}$ & $0.11^{+0.01}_{-0.02}$ & $>80$ & $121.9^{+1.0}_{-0.9}$ \\
HD53143$^{\star}$ & $1.26$ & $53^{+5}_{-5}$ & $1.59^{+0.05}_{-0.05}$ & $88.4^{+0.7}_{-0.3}$ & $23^{+2}_{-2}$ & $^{a}0.08^{+0.02}_{-0.02}$ & $56.1^{+0.6}_{-0.5}$ & $157.5^{+0.4}_{-0.4}$ \\
HD54341 & $1.34$ & $-$ & $0.58^{+0.12}_{-0.09}$ & $170^{+20}_{-30}$ & $190^{+110}_{-50}$ & $-$ & $^{c}50^{+10}_{-10}$ & $60^{+20}_{-10}$ \\
HD61005$^{\star}$ & $1.29$ & $<50$ & $6.3^{+0.06}_{-0.06}$& $72.6^{+0.6}_{-0.4}$ & $38^{+1}_{-1}$ & $0.044^{+0.003}_{-0.004}$ & $85.7^{+0.2}_{-0.2}$ & $70.33^{+0.08}_{-0.09}$ \\
TWA7 & $0.878$ & $-$ & $2.8^{+0.09}_{-0.08}$ & $90^{+20}_{-10}$ & $90^{+20}_{-20}$ & $-$ & $<70$ & $-^{g}$ \\
HD92945$^{\star}$ & $0.856$ & $30^{+10}_{-10}$ & $10.8^{+0.5}_{-0.4}$ & $96^{+1}_{-5}$ & $80^{+3}_{-3}$ & $<0.08$ & $66.8^{+1.0}_{-1.3}$ & $100^{+2}_{-1}$ \\
HD95086 & $1.30$ & $<9$ & $3.04^{+0.04}_{-0.04}$ & $206^{+2}_{-1}$ & $180^{+4}_{-3}$ & $-$ & $29^{+1}_{-1}$ & $93^{+2}_{-2}$ \\
HD104860 & $1.36$ & $-$ & $4.0^{+0.7}_{-0.7}$ & $110^{+30}_{-20}$ & $<200$ & $-$ & $^{c}50^{+10}_{-30}$ & $170^{+30}_{-20}$ \\
HD105211$^{\star}$ & $1.29$ & $180^{+10}_{-10}$ & $2.4^{+0.1}_{-0.1}$ & $131.5^{+0.9}_{-0.8}$ & $23^{+4}_{-5}$ & $0.07^{+0.01}_{-0.02}$ & $66.4^{+0.8}_{-0.7}$ & $28.2^{+0.3}_{-0.5}$ \\
HD106906 & $1.26$ & $-$ & $0.36^{+0.03}_{-0.03}$ & $100^{+10}_{-10}$ & $^{a}80^{+30}_{-40}$ & $<0.3$ & $>60$ & $111^{+5}_{-5}$ \\
HD107146$^{\star}$ & $1.14$ & $^{a}17^{+6}_{-6}$ & $20.6^{+0.2}_{-0.2}$ & $107.2^{+0.4}_{-0.6}$ & $110.0^{+0.8}_{-0.8}$ & $<0.07$ & $22.2^{+0.6}_{-0.9}$ & $149^{+2}_{-2}$ \\
HD109085 & $0.880$ & $320^{+20}_{-20}$ & $16.8^{+0.7}_{-0.7}$ & $153^{+2}_{-1}$ & $53^{+2}_{-3}$ & $^{a}0.09^{+0.03}_{-0.04}$ & $39^{+1}_{-1}$ & $120^{+1}_{-1}$ \\
HD109573 & $0.880$ & $70^{+30}_{-30}$ & $15.4^{+0.2}_{-0.2}$ & $77.8^{+0.4}_{-0.2}$ & $14.8^{+0.6}_{-0.6}$ & $0.058^{+0.002}_{-0.002}$ & $76.5^{+0.2}_{-0.2}$ & $26.7^{+0.1}_{-0.1}$ \\
HD110058 & $1.25$ & $-$ & $0.64^{+0.10}_{-0.09}$ & $50^{+10}_{-10}$ & $<100$ & $-$ & $-^{h}$ & $-^{h}$ \\
HD111520 & $1.25$ & $-$ & $1.33^{+0.08}_{-0.08}$ & $76^{+6}_{-6}$ & $^{a}50^{+20}_{-30}$ & $-$ & $^{b}84^{+4}_{-5}$ & $168^{+4}_{-4}$ \\
HD112810 & $1.25$ & $-$ & $0.56^{+0.08}_{-0.08}$ & $90^{+20}_{-20}$ & $<200$ & $-$ & $>60$ & $97^{+8}_{-9}$ \\
HD113556 & $1.25$ & $-$ & $^{a}0.4^{+0.2}_{-0.2}$ & $110^{+70}_{-30}$ & $<400$ & $-$ & $-^{h}$ & $^{f}101^{+34}_{-51}$ \\
HD113766 & $1.25$ & $-$ & $0.65^{+0.07}_{-0.07}$ & $30^{+9}_{-8}$ & $<60$ & $-$ & $-^{h}$ & $-^{h}$ \\
HD114082 & $1.26$ & $-$ & $0.69^{+0.03}_{-0.03}$ & $38^{+4}_{-5}$ & $<40$ & $-$ & $>40$ & $110^{+9}_{-8}$ \\
HD115600 & $1.25$ & $-$ & $^{a}0.3^{+0.1}_{-0.1}$ & $^{a}90^{+60}_{-30}$ & $<200$ & $-$ & $-^{h}$ & $^{f}160^{+50}_{-40}$ \\
HD117214 & $1.26$ & $-$ & $0.75^{+0.03}_{-0.03}$ & $42^{+4}_{-4}$ & $<50$ & $-$ & $^{e}40^{+20}_{-20}$ & $^{f}10^{+30}_{-20}$ \\
HD121191 & $1.26$ & $-$ & $0.41^{+0.02}_{-0.02}$ & $55^{+4}_{-5}$ & $54^{+8}_{-11}$ & $-$ & $^{c}40^{+10}_{-10}$ & $30^{+10}_{-20}$ \\
HD121617 & $1.33$ & $-$ & $1.7^{+0.1}_{-0.1}$ & $78^{+5}_{-5}$ & $60^{+10}_{-10}$ & $-$ & $^{c}37^{+7}_{-10}$ & $60^{+10}_{-10}$ \\
HD129590 & $1.26$ & $-$ & $1.30^{+0.03}_{-0.03}$ & $79^{+4}_{-4}$ & $^{a}70^{+20}_{-30}$ & $-$ & $>70$ & $116^{+2}_{-2}$ \\
HD131488 & $1.33$ & $-$ & $2.87^{+0.05}_{-0.05}$ & $92^{+1}_{-2}$ & $46^{+7}_{-8}$ & $-$ & $^{b}84^{+2}_{-2}$ & $96.4^{+0.7}_{-0.7}$ \\
HD131835 & $0.890$ & $-$ & $5.38^{+0.10}_{-0.09}$ & $83.7^{+0.9}_{-1.1}$ & $87^{+4}_{-4}$ & $-$ & $73.7^{+0.4}_{-0.4}$ & $59.6^{+0.4}_{-0.4}$ \\
HD138813 & $1.33$ & $-$ & $1.23^{+0.07}_{-0.08}$ & $120^{+10}_{-10}$ & $130^{+20}_{-20}$ & $-$ & $48^{+5}_{-6}$ & $47^{+9}_{-9}$ \\
HD139664 & $1.26$ & $160^{+10}_{-10}$ & $1.8^{+0.2}_{-0.2}$ & $75^{+6}_{-4}$ & $60^{+10}_{-10}$ & $0.14^{+0.02}_{-0.02}$ & $>80$ & $76^{+1}_{-1}$ \\
HD142315 & $1.25$ & $-$ & $0.5^{+0.2}_{-0.1}$ & $^{i}140^{+120}_{-50}$ & $<500$ & $-$ & $^{b}70^{+10}_{-20}$ & $80^{+20}_{-10}$ \\
HD142446 & $1.25$ & $-$ & $0.8^{+0.2}_{-0.2}$ & $110^{+30}_{-20}$ & $^{a}80^{+50}_{-40}$ & $-$ & $^{e}40^{+20}_{-20}$ & $^{f}100^{+40}_{-40}$ \\
HD145560 & $1.25$ & $-$ & $1.7^{+0.1}_{-0.1}$ & $76^{+4}_{-4}$ & $^{a}50^{+20}_{-20}$ & $-$ & $47^{+7}_{-8}$ & $28^{+8}_{-7}$ \\
HD146181 & $1.25$ & $-$ & $0.83^{+0.07}_{-0.06}$ & $74^{+8}_{-8}$ & $<90$ & $-$ & $>50$ & $54^{+6}_{-7}$ \\
HD146897 & $1.25$ & $-$ & $1.26^{+0.08}_{-0.08}$ & $82^{+6}_{-5}$ & $<90$ & $-$ & $>70$ & $114^{+3}_{-3}$ \\
HD147137 & $1.25$ & $-$ & $0.5^{+0.2}_{-0.2}$ & $^{i}120^{+80}_{-30}$ & $^{i}100^{+60}_{-50}$ & $-$ & $-^{h}$ & $^{f}180^{+50}_{-60}$ \\
HD164249 & $1.35$ & $-$ & $1.0^{+0.1}_{-0.1}$ & $62^{+3}_{-2}$ & $^{a}20^{+10}_{-10}$ & $-$ & $<50$ & $^{f}110^{+30}_{-60}$ \\
GSC07396-00759 & $0.880$ & $-$ & $1.9^{+0.1}_{-0.1}$ & $78^{+2}_{-2}$ & $43^{+9}_{-8}$ & $<0.17$ & $>79$ & $148^{+1}_{-1}$ \\
HD172167 & $1.34$ & $2500^{+10}_{-10}$ & $8^{+1}_{-1}$ & $118^{+4}_{-3}$ & $67^{+9}_{-7}$ & $-$ & $<40$ & $-^{h}$ \\
HD181327$^{\star}$ & $0.880$ & $^{a}40^{+30}_{-20}$ & $18.8^{+0.3}_{-0.3}$ & $81.3^{+0.3}_{-0.3}$ & $16.0^{+0.5}_{-0.6}$ & $<0.09$ & $29.9^{+0.5}_{-0.5}$ & $98^{+1}_{-1}$  \\
HD197481 & $1.35$ & $^{j}320^{+20}_{-20}$ & $5.98^{+0.08}_{-0.08}$ & $34.9^{+0.2}_{-0.2}$ & $12.4^{+0.5}_{-0.5}$ & $0.017^{+0.003}_{-0.003}$ & $88.4^{+0.1}_{-0.1}$ & $128.54^{+0.07}_{-0.07}$ \\
HD202628 & $1.25$ & $29^{+5}_{-5}$ & $1.14^{+0.06}_{-0.06}$ & $155.4^{+0.9}_{-0.9}$ & $16^{+3}_{-3}$ & $^{a}0.02^{+0.01}_{-0.01}$ & $57.9^{+0.4}_{-0.4}$ & $129.9^{+0.5}_{-0.5}$ \\
HD206893 & $0.878$ & $40^{+9}_{-9}$ & $2.9^{+0.2}_{-0.2}$ & $108^{+4}_{-3}$ & $100^{+10}_{-10}$ & $-$ & $^{a}24^{+9}_{-11}$ & $60^{+20}_{-20}$ \\
HD207129 & $1.26$ & $60^{+9}_{-9}$ & $3.0^{+0.3}_{-0.3}$ & $150^{+2}_{-2}$ & $39^{+7}_{-6}$ & $^{a}0.06^{+0.03}_{-0.03}$ & $60.7^{+1.0}_{-0.9}$ & $118^{+1}_{-1}$ \\
TYC93404371 & $1.33$ & $-$ & $3.9^{+0.4}_{-0.4}$ & $130^{+20}_{-20}$ & $^{a}100^{+40}_{-60}$ & $-$ & $^{a}40^{+10}_{-20}$ & $130^{+20}_{-20}$ \\
HD216956$^{\star}$ & $1.34$ & $696^{+10}_{-11}$ & $24.9^{+0.2}_{-0.2}$ & $140.11^{+0.10}_{-0.09}$ & $16.5^{+0.2}_{-0.2}$ & $0.0190^{+0.0009}_{-0.0009}$ & $66.87^{+0.05}_{-0.03}$ & $156.61^{+0.03}_{-0.02}$ \\
HD216956C & $0.880$ & $-$ & $0.8^{+0.1}_{-0.1}$ & $26.4^{+0.5}_{-0.6}$ & $<8.6$ & $-^{h}$ & $44^{+3}_{-3}$ & $116^{+4}_{-5}$ \\
HD218396 & $1.34$ & $^{a}20^{+20}_{-10}$ & $3.9^{+0.5}_{-0.5}$ & $290^{+10}_{-10}$ & $250^{+30}_{-30}$ & $-^{h}$ & $39^{+4}_{-4}$ & $51^{+6}_{-7}$ \\
\hline
\end{longtable}
\tablefoot{
\tablefoottext{a}{Marginally resolved/detected, i.e. having a posterior probability distribution with a non-zero peak but consistent with zero at the $3\sigma$ level.}
\tablefoottext{b}{Inclination consistent with 90$^{\circ}$ (perfectly edge-on) to within 3$\sigma$.}
\tablefoottext{c}{Inclination consistent with 0$^{\circ}$ (perfectly face-on) to within 3$\sigma$.}
\tablefoottext{d}{For HD39060, the star was fitted to long baselines only, and its flux density fixed in the fit.}
\tablefoottext{e}{Inclination unconstrained within prior boundaries, but showing a peak in the probability distribution at the reported value.}
\tablefoottext{f}{PA unconstrained within prior boundaries, but showing a peak in the probability distribution at the reported value.}
\tablefoottext{g}{PA unconstrained within prior boundaries, exhibiting a multi-modal posterior probability distribution.}
\tablefoottext{h}{Quantity unconstrained within prior boundaries.}
\tablefoottext{i}{Probability distribution exhibiting a long tail out to large values.}
\tablefoottext{j}{The star exhibits significant variability. Different flux densities were found for different datasets.}
\tablefoottext{$\star$}{Significant residual emission associated with the planetary system after subtraction of best-fit model.}}

\begin{landscape}
\begin{longtable}{ccccccccccccc}
\caption{Stellar and dust properties from REASONS spectral fitting}
\label{tab:sdbprops}\\
\hline\hline
%\hspace{-20mm}
%\tablenum{3}
Target & d & T$_{\ast}$ & R$_{\ast}$ &  L$_{\ast}$ & L$_{\rm dust}$/L$_{\ast}$ & T$_{\rm cold}$ & $\lambda_{0}$ & $\beta$ & Warm & Age & Age & Age \\
 & pc & K & R$_{\odot}$ &  L$_{\odot}$ &  & K & $\mu$m &  & dust & Myr & Notes & Ref. \\
\hline
\endhead
HD105 & 38.83$^{+0.03}_{-0.03}$ & 6010$^{+100}_{100}$ & 1.031$^{+0.007}_{-0.007}$ & 1.252$^{+0.008}_{-0.008}$ & 2.8$^{+0.1}_{-0.1} \times 10^{-4}$ & 48$^{+4}_{-4}$ & 200$^{+300}_{-300}$ & 0.7$^{+0.5}_{-0.5}$ & N & 45$^{+4}_{-4}$ & 1 & 1 \\
GJ14 & 14.688$^{+0.005}_{-0.005}$ & 4100$^{+100}_{100}$ & 0.655$^{+0.004}_{-0.005}$ & 0.1094$^{+0.0007}_{-0.0007}$ & 9.1$^{+0.8}_{-0.8} \times 10^{-5}$ & 28$^{+4}_{-4}$ & 100$^{+200}_{-200}$ & 0.7$^{+0.2}_{-0.2}$ & N & 500$^{+300}_{-300}$ & 3 & 2 \\
HD9672 & 57.2$^{+0.2}_{-0.2}$ & 8750$^{+100}_{100}$ & 1.71$^{+0.02}_{-0.01}$ & 15.4$^{+0.2}_{-0.2}$ & 7.2$^{+0.2}_{-0.2} \times 10^{-4}$ & 57$^{+1}_{-1}$ & 90$^{+10}_{-10}$ & 0.89$^{+0.04}_{-0.04}$ & Y & 45$^{+4}_{-4}$ & 2 & 1 \\
HD10638 & 67.8$^{+0.2}_{-0.2}$ & 7310$^{+100}_{100}$ & 1.65$^{+0.01}_{-0.01}$ & 7.06$^{+0.05}_{-0.06}$ & 2.0$^{+0.2}_{-0.2} \times 10^{-4}$ & 74$^{+3}_{-3}$ & 700$^{+500}_{-500}$ & 0.8$^{+0.9}_{-0.9}$ & N & 100$^{+50}_{-50}$ & 3 & 2 \\
HD10647 & 17.35$^{+0.01}_{-0.01}$ & 6150$^{+100}_{100}$ & 1.099$^{+0.008}_{-0.008}$ & 1.552$^{+0.009}_{-0.008}$ & 6.2$^{+0.4}_{-0.3} \times 10^{-5}$ & 100$^{+4}_{-4}$ & 200$^{+300}_{-300}$ & 0.1$^{+0.2}_{-0.2}$ & Y & 1700$^{+600}_{-600}$ & 0 & 4 \\
HD14055 & 35.7$^{+0.6}_{-0.6}$ & 9210$^{+100}_{100}$ & 2.09$^{+0.04}_{-0.04}$ & 28$^{+1}_{-1}$ & 8.9$^{+0.2}_{-0.2} \times 10^{-5}$ & 75.2$^{+1.0}_{-1.0}$ & 150$^{+10}_{-10}$ & 0.81$^{+0.08}_{-0.08}$ & N & 300$^{+200}_{-200}$ & 3 & 5 \\
HD15115 & 48.77$^{+0.07}_{-0.07}$ & 6740$^{+100}_{100}$ & 1.41$^{+0.01}_{-0.01}$ & 3.67$^{+0.02}_{-0.02}$ & 8$^{+40}_{-1} \times 10^{-5}$ & 130$^{+80}_{-80}$ & 80$^{+70}_{-70}$ & 0.8$^{+0.7}_{-0.7}$ & Y & 45$^{+5}_{-5}$ & 1 & 1 \\
HD15257 & 49.0$^{+0.2}_{-0.2}$ & 6900$^{+100}_{100}$ & 2.61$^{+0.02}_{-0.02}$ & 13.9$^{+0.1}_{-0.1}$ & 6.0$^{+0.7}_{-0.5} \times 10^{-5}$ & 81$^{+6}_{-6}$ & 400$^{+400}_{-400}$ & 0.5$^{+0.6}_{-0.6}$ & N & 1000$^{+800}_{-800}$ & 3 & 3 \\
HD15745 & 71.7$^{+0.1}_{-0.1}$ & 6810$^{+100}_{100}$ & 1.47$^{+0.01}_{-0.01}$ & 4.18$^{+0.03}_{-0.03}$ & 1.87$^{+0.04}_{-0.03} \times 10^{-3}$ & 96.1$^{+0.4}_{-0.4}$ & 770$^{+30}_{-30}$ & 2.8$^{+0.2}_{-0.2}$ & N & 23$^{+3}_{-3}$ & 1 & 1 \\
HD16743 & 57.81$^{+0.06}_{-0.06}$ & 6740$^{+100}_{100}$ & 1.65$^{+0.01}_{-0.01}$ & 5.05$^{+0.03}_{-0.03}$ & 7$^{+1}_{-1} \times 10^{-5}$ & 110$^{+10}_{-10}$ & 200$^{+200}_{-200}$ & 1.5$^{+0.8}_{-0.8}$ & Y & 100$^{+400}_{-60}$ & 0 & 24 \\
HD21997 & 69.7$^{+0.1}_{-0.1}$ & 8400$^{+100}_{100}$ & 1.56$^{+0.01}_{-0.02}$ & 10.86$^{+0.09}_{-0.09}$ & 5.8$^{+0.1}_{-0.1} \times 10^{-4}$ & 64.3$^{+0.5}_{-0.5}$ & 160$^{+20}_{-20}$ & 1.1$^{+0.1}_{-0.1}$ & N & 42$^{+6}_{-4}$ & 1 & 1 \\
HD22049 & 3.220$^{+0.001}_{-0.001}$ & 5152$^{+100}_{100}$ & 0.736$^{+0.003}_{-0.003}$ & 0.344$^{+0.002}_{-0.002}$ & 1.09$^{+0.07}_{-0.06} \times 10^{-4}$ & 74$^{+3}_{-3}$ & 450$^{+50}_{-50}$ & 0.56$^{+0.08}_{-0.08}$ & N & 800$^{+400}_{-400}$ & 0 & 4 \\
HD32297 & 129.7$^{+0.5}_{-0.5}$ & 7430$^{+100}_{100}$ & 1.598$^{+0.009}_{-0.009}$ & 7.02$^{+0.09}_{-0.09}$ & 6.1$^{+0.1}_{-0.1} \times 10^{-3}$ & 80$^{+1}_{-1}$ & 80$^{+10}_{-10}$ & 0.37$^{+0.01}_{-0.01}$ & Y & 30$^{+10}_{-10}$ & 3 & 6 \\
HD35841 & 103.1$^{+0.1}_{-0.1}$ & 6340$^{+100}_{100}$ & 1.26$^{+0.02}_{-0.02}$ & 2.33$^{+0.02}_{-0.03}$ & 1.35$^{+0.07}_{-0.07} \times 10^{-3}$ & 72.0$^{+0.9}_{-0.9}$ & 500$^{+400}_{-400}$ & 1.4$^{+0.9}_{-0.9}$ & N & 42$^{+6}_{-4}$ & 1 & 1 \\
HD36546 & 100.2$^{+0.4}_{-0.4}$ & 8840$^{+100}_{100}$ & 1.59$^{+0.02}_{-0.02}$ & 13.8$^{+0.2}_{-0.2}$ & 4.7$^{+0.1}_{-0.1} \times 10^{-3}$ & 148$^{+2}_{-2}$ & 600$^{+400}_{-400}$ & 1.0$^{+0.8}_{-0.8}$ & N & 6$^{+4}_{-4}$ & 0 & 20 \\
HD38206 & 70.7$^{+0.2}_{-0.2}$ & 10080$^{+100}_{100}$ & 1.69$^{+0.02}_{-0.02}$ & 26.5$^{+0.5}_{-0.5}$ & 1.16$^{+0.04}_{-0.04} \times 10^{-4}$ & 62$^{+1}_{-1}$ & 120$^{+30}_{-30}$ & 0.9$^{+0.1}_{-0.1}$ & Y & 42$^{+6}_{-4}$ & 1 & 21 \\
HD38858 & 15.210$^{+0.007}_{-0.007}$ & 5830$^{+100}_{100}$ & 0.897$^{+0.005}_{-0.006}$ & 0.835$^{+0.005}_{-0.005}$ & 7.7$^{+0.3}_{-0.3} \times 10^{-5}$ & 51.6$^{+0.8}_{-0.8}$ & 200$^{+100}_{-100}$ & 0.4$^{+0.2}_{-0.2}$ & N & 5000$^{+5000}_{-5000}$ & 0 & 7 \\
HD39060 & 19.63$^{+0.06}_{-0.06}$ & 8180$^{+100}_{100}$ & 1.47$^{+0.02}_{-0.02}$ & 8.68$^{+0.08}_{-0.08}$ & 2.5$^{+0.1}_{-0.1} \times 10^{-3}$ & 96$^{+5}_{-5}$ & 60$^{+20}_{-20}$ & 0.74$^{+0.03}_{-0.03}$ & Y & 23$^{+3}_{-3}$ & 1 & 1 \\
HD48682 & 16.61$^{+0.02}_{-0.02}$ & 6030$^{+100}_{100}$ & 1.241$^{+0.009}_{-0.009}$ & 1.84$^{+0.01}_{-0.01}$ & 6.2$^{+0.2}_{-0.2} \times 10^{-5}$ & 55.3$^{+0.6}_{-0.6}$ & 180$^{+20}_{-20}$ & 0.9$^{+0.1}_{-0.1}$ & N & 4000$^{+2000}_{-2000}$ & 0 & 8 \\
HD50571 & 33.93$^{+0.02}_{-0.02}$ & 6560$^{+100}_{100}$ & 1.413$^{+0.010}_{-0.009}$ & 3.34$^{+0.02}_{-0.02}$ & 1.18$^{+0.03}_{-0.04} \times 10^{-4}$ & 41$^{+2}_{-2}$ & 20$^{+20}_{-20}$ & 0.8$^{+0.1}_{-0.1}$ & N & 300$^{+100}_{-100}$ & 0 & 22 \\
HD53143 & 18.340$^{+0.005}_{-0.005}$ & 5430$^{+100}_{100}$ & 0.868$^{+0.005}_{-0.005}$ & 0.587$^{+0.004}_{-0.004}$ & 2.75$^{+0.09}_{-0.10} \times 10^{-4}$ & 73$^{+1}_{-1}$ & 600$^{+400}_{-400}$ & 1.2$^{+0.8}_{-0.8}$ & N & 1000$^{+500}_{-300}$ & 3 & 23 \\
HD54341 & 101.5$^{+0.9}_{-0.9}$ & 9120$^{+100}_{100}$ & 1.94$^{+0.03}_{-0.02}$ & 23.6$^{+0.6}_{-0.7}$ & 2.60$^{+0.11}_{-0.10} \times 10^{-4}$ & 64.9$^{+0.7}_{-0.7}$ & 400$^{+200}_{-200}$ & 1.9$^{+0.7}_{-0.7}$ & N & 200$^{+100}_{-200}$ & 0 & 3 \\
HD61005 & 36.45$^{+0.02}_{-0.02}$ & 5540$^{+100}_{100}$ & 0.858$^{+0.010}_{-0.008}$ & 0.623$^{+0.003}_{-0.003}$ & 2.75$^{+0.05}_{-0.06} \times 10^{-3}$ & 62.4$^{+0.4}_{-0.4}$ & 160$^{+30}_{-30}$ & 0.44$^{+0.04}_{-0.04}$ & N & 45$^{+5}_{-5}$ & 2 & 1 \\
HD76582 & 48.9$^{+0.1}_{-0.1}$ & 7750$^{+100}_{100}$ & 1.75$^{+0.01}_{-0.01}$ & 9.92$^{+0.08}_{-0.09}$ & 2.00$^{+0.07}_{-0.07} \times 10^{-4}$ & 49.5$^{+0.8}_{-0.8}$ & 180$^{+70}_{-70}$ & 1.1$^{+0.3}_{-0.3}$ & N & 1200$^{+900}_{-900}$ & 0 & 9 \\
HD84870 & 88.8$^{+0.2}_{-0.2}$ & 7310$^{+100}_{100}$ & 1.75$^{+0.01}_{-0.01}$ & 7.85$^{+0.08}_{-0.08}$ & 4.3$^{+0.2}_{-0.2} \times 10^{-4}$ & 52$^{+2}_{-2}$ & 150$^{+50}_{-50}$ & 0.7$^{+0.2}_{-0.2}$ & N & 300$^{+200}_{-200}$ & 0 & 10 \\
TWA7 & 34.10$^{+0.03}_{-0.03}$ & 3390$^{+100}_{100}$ & 1.035$^{+0.009}_{-0.009}$ & 0.128$^{+0.001}_{-0.001}$ & 2.10$^{+0.08}_{-0.08} \times 10^{-3}$ & 72$^{+1}_{-1}$ & 600$^{+500}_{-500}$ & 0.6$^{+1.0}_{-1.0}$ & N & 10$^{+3}_{-3}$ & 1 & 1 \\
HD92945 & 21.509$^{+0.009}_{-0.009}$ & 5194$^{+100}_{100}$ & 0.752$^{+0.004}_{-0.004}$ & 0.371$^{+0.002}_{-0.002}$ & 6.6$^{+0.2}_{-0.2} \times 10^{-4}$ & 34$^{+2}_{-2}$ & 30$^{+30}_{-30}$ & 1.0$^{+0.1}_{-0.1}$ & N & 200$^{+100}_{-100}$ & 0 & 4 \\
HD95086 & 86.5$^{+0.1}_{-0.1}$ & 7360$^{+100}_{100}$ & 1.56$^{+0.02}_{-0.02}$ & 6.42$^{+0.08}_{-0.12}$ & 7$^{+9}_{-3} \times 10^{-4}$ & 1000$^{+1000}_{-1000}$ & 200$^{+300}_{-300}$ & 1.2$^{+0.7}_{-0.7}$ & Y & 13.3$^{+1.1}_{-0.6}$ & 0 & 11 \\
HD104860 & 45.19$^{+0.04}_{-0.04}$ & 5960$^{+100}_{100}$ & 1.018$^{+0.006}_{-0.006}$ & 1.181$^{+0.009}_{-0.010}$ & 6.1$^{+0.3}_{-0.2} \times 10^{-4}$ & 41$^{+4}_{-4}$ & 100$^{+200}_{-200}$ & 0.74$^{+0.07}_{-0.07}$ & N & 190$^{+110}_{-100}$ & 0 & 12 \\
HD105211 & 19.69$^{+0.04}_{-0.04}$ & 6910$^{+100}_{100}$ & 1.81$^{+0.01}_{-0.01}$ & 6.75$^{+0.04}_{-0.04}$ & 5.8$^{+0.5}_{-0.4} \times 10^{-5}$ & 46.8$^{+0.9}_{-0.9}$ & 200$^{+200}_{-200}$ & 1.3$^{+0.7}_{-0.7}$ & N & 1000$^{+1000}_{-1000}$ & 3 & 13 \\
HD106906 & 102.4$^{+0.2}_{-0.2}$ & 6530$^{+100}_{100}$ & 1.99$^{+0.02}_{-0.02}$ & 6.46$^{+0.05}_{-0.06}$ & 1.20$^{+0.03}_{-0.03} \times 10^{-3}$ & 102.7$^{+0.9}_{-0.9}$ & 300$^{+100}_{-100}$ & 1.9$^{+0.6}_{-0.6}$ & N & 15$^{+3}_{-3}$ & 1 & 1 \\
HD107146 & 27.47$^{+0.02}_{-0.02}$ & 5890$^{+100}_{100}$ & 0.958$^{+0.008}_{-0.007}$ & 0.999$^{+0.008}_{-0.008}$ & 10$^{+1}_{-9} \times 10^{-4}$ & 50$^{+30}_{-30}$ & 300$^{+300}_{-300}$ & 0.8$^{+0.6}_{-0.6}$ & Y & 150$^{+100}_{-50}$ & 0 & 4 \\
HD109085 & 18.24$^{+0.05}_{-0.05}$ & 6900$^{+100}_{100}$ & 1.57$^{+0.03}_{-0.02}$ & 5.02$^{+0.04}_{-0.04}$ & 1.7$^{+0.4}_{-0.3} \times 10^{-4}$ & 280$^{+40}_{-40}$ & 100$^{+200}_{-200}$ & 1.7$^{+1.0}_{-1.0}$ & Y & 1500$^{+1000}_{-500}$ & 0 & 4 \\
HD109573 & 70.8$^{+0.2}_{-0.2}$ & 9980$^{+100}_{100}$ & 1.673$^{+0.010}_{-0.010}$ & 25.0$^{+0.3}_{-0.3}$ & 4.18$^{+0.07}_{-0.07} \times 10^{-3}$ & 97.4$^{+0.2}_{-0.2}$ & 28.1$^{+0.2}_{-0.2}$ & 0.59$^{+0.01}_{-0.01}$ & N & 10$^{+3}_{-3}$ & 1 & 1 \\
HD110058 & 130.1$^{+0.5}_{-0.5}$ & 7740$^{+100}_{100}$ & 1.60$^{+0.04}_{-0.03}$ & 8.3$^{+0.1}_{-0.1}$ & 9$^{+22}_{-9} \times 10^{-4}$ & 100$^{+300}_{-300}$ & 50$^{+50}_{-50}$ & 0.6$^{+0.7}_{-0.7}$ & Y & 15$^{+3}_{-3}$ & 1 & 1 \\
HD111520 & 108.0$^{+0.2}_{-0.2}$ & 6290$^{+100}_{100}$ & 1.35$^{+0.02}_{-0.02}$ & 2.58$^{+0.02}_{-0.03}$ & 2.30$^{+0.06}_{-0.06} \times 10^{-3}$ & 84.0$^{+0.7}_{-0.7}$ & 600$^{+400}_{-400}$ & 1.0$^{+0.8}_{-0.8}$ & N & 15$^{+3}_{-3}$ & 1 & 1 \\
HD112810 & 133.7$^{+0.3}_{-0.3}$ & 6420$^{+100}_{100}$ & 1.45$^{+0.01}_{-0.01}$ & 3.21$^{+0.02}_{-0.03}$ & 8.9$^{+0.3}_{-0.3} \times 10^{-4}$ & 70.2$^{+0.8}_{-0.8}$ & 500$^{+300}_{-300}$ & 1.1$^{+0.7}_{-0.7}$ & N & 15$^{+3}_{-3}$ & 1 & 1 \\
HD113556 & 100.5$^{+0.2}_{-0.2}$ & 6840$^{+100}_{100}$ & 1.49$^{+0.01}_{-0.01}$ & 4.39$^{+0.03}_{-0.03}$ & 5.1$^{+0.7}_{-0.7} \times 10^{-4}$ & 73$^{+3}_{-3}$ & 150$^{+90}_{-90}$ & 2.2$^{+0.6}_{-0.6}$ & N & 15$^{+3}_{-3}$ & 1 & 1 \\
HD113766 & 120$^{+20}_{-20}$ & 5590$^{+100}_{100}$ & 2.7$^{+0.4}_{-0.3}$ & 7$^{+2}_{-1}$ & 2.2$^{+0.7}_{-0.3} \times 10^{-2}$ & 600$^{+400}_{-400}$ & 300$^{+500}_{-500}$ & 1.3$^{+0.7}_{-0.7}$ & Y & 15$^{+3}_{-3}$ & 1 & 1 \\
HD114082 & 95.1$^{+0.2}_{-0.2}$ & 6620$^{+100}_{100}$ & 1.47$^{+0.01}_{-0.01}$ & 3.74$^{+0.02}_{-0.02}$ & 3.55$^{+0.07}_{-0.08} \times 10^{-3}$ & 112.7$^{+0.7}_{-0.7}$ & 300$^{+200}_{-200}$ & 1.5$^{+0.7}_{-0.7}$ & N & 15$^{+3}_{-3}$ & 1 & 1 \\
HD115600 & 109.0$^{+0.2}_{-0.2}$ & 6900$^{+100}_{100}$ & 1.53$^{+0.01}_{-0.01}$ & 4.75$^{+0.04}_{-0.03}$ & 1.95$^{+0.05}_{-0.04} \times 10^{-3}$ & 109$^{+1}_{-1}$ & 200$^{+100}_{-100}$ & 1.1$^{+0.6}_{-0.6}$ & N & 15$^{+3}_{-3}$ & 1 & 1 \\
HD117214 & 107.4$^{+0.3}_{-0.3}$ & 6390$^{+100}_{100}$ & 1.92$^{+0.02}_{-0.02}$ & 5.54$^{+0.05}_{-0.05}$ & 2.66$^{+0.07}_{-0.06} \times 10^{-3}$ & 112.0$^{+0.9}_{-0.9}$ & 300$^{+200}_{-200}$ & 1.3$^{+0.7}_{-0.7}$ & N & 15$^{+3}_{-3}$ & 1 & 1 \\
HD121191 & 132.3$^{+0.4}_{-0.4}$ & 7710$^{+100}_{100}$ & 1.50$^{+0.01}_{-0.01}$ & 7.21$^{+0.07}_{-0.07}$ & 3.2$^{+0.6}_{-0.5} \times 10^{-3}$ & 550$^{+70}_{-70}$ & 100$^{+200}_{-200}$ & 1$^{+1}_{-1}$ & Y & 16$^{+2}_{-2}$ & 1 & 1 \\
HD121617 & 117.9$^{+0.4}_{-0.4}$ & 8800$^{+100}_{100}$ & 1.60$^{+0.03}_{-0.03}$ & 14.0$^{+0.4}_{-0.3}$ & 5.1$^{+0.3}_{-0.3} \times 10^{-3}$ & 103$^{+2}_{-2}$ & 80$^{+10}_{-10}$ & 0.74$^{+0.05}_{-0.05}$ & N & 16$^{+2}_{-2}$ & 1 & 1 \\
HD127821 & 31.69$^{+0.02}_{-0.02}$ & 6460$^{+100}_{100}$ & 1.368$^{+0.009}_{-0.008}$ & 2.94$^{+0.02}_{-0.02}$ & 1.91$^{+0.07}_{-0.06} \times 10^{-4}$ & 39$^{+3}_{-3}$ & 30$^{+30}_{-30}$ & 1.0$^{+0.1}_{-0.1}$ & N & 1300$^{+500}_{-500}$ & 0 & 4 \\
HD129590 & 136.3$^{+0.4}_{-0.4}$ & 5810$^{+100}_{100}$ & 1.70$^{+0.03}_{-0.02}$ & 2.98$^{+0.07}_{-0.04}$ & 5.5$^{+0.2}_{-0.1} \times 10^{-3}$ & 92.7$^{+0.5}_{-0.5}$ & 500$^{+400}_{-400}$ & 1.0$^{+0.8}_{-0.8}$ & N & 16$^{+2}_{-2}$ & 1 & 1 \\
HD131488 & 152.2$^{+0.8}_{-0.8}$ & 8600$^{+100}_{100}$ & 1.55$^{+0.03}_{-0.03}$ & 11.8$^{+0.4}_{-0.3}$ & 2.9$^{+0.5}_{-0.6} \times 10^{-3}$ & 400$^{+500}_{-500}$ & 100$^{+100}_{-100}$ & 0.2$^{+0.9}_{-0.9}$ & Y & 16$^{+2}_{-2}$ & 1 & 1 \\
HD131835 & 129.7$^{+0.5}_{-0.5}$ & 7580$^{+100}_{100}$ & 1.733$^{+0.010}_{-0.009}$ & 8.92$^{+0.09}_{-0.09}$ & 2.67$^{+0.07}_{-0.07} \times 10^{-3}$ & 71.0$^{+0.9}_{-0.9}$ & 140$^{+20}_{-20}$ & 0.69$^{+0.05}_{-0.05}$ & Y & 16$^{+2}_{-2}$ & 1 & 1 \\
HD138813 & 136.6$^{+0.6}_{-0.6}$ & 8760$^{+100}_{100}$ & 1.87$^{+0.02}_{-0.02}$ & 18.6$^{+0.3}_{-0.2}$ & 7.7$^{+0.2}_{-0.2} \times 10^{-4}$ & 140$^{+1}_{-1}$ & 200$^{+300}_{-300}$ & 1$^{+1}_{-1}$ & Y & 10$^{+3}_{-3}$ & 1 & 1 \\
HD139664 & 17.40$^{+0.04}_{-0.04}$ & 6720$^{+100}_{100}$ & 1.35$^{+0.01}_{-0.01}$ & 3.37$^{+0.02}_{-0.02}$ & 1.25$^{+0.10}_{-0.09} \times 10^{-4}$ & 73$^{+6}_{-6}$ & 200$^{+200}_{-200}$ & 1.0$^{+0.4}_{-0.4}$ & N & 200$^{+200}_{-100}$ & 0 & 4 \\
HD142315 & 144.7$^{+0.6}_{-0.6}$ & 8700$^{+100}_{100}$ & 2.49$^{+0.02}_{-0.02}$ & 31.9$^{+0.3}_{-0.3}$ & 6.6$^{+0.1}_{-0.1} \times 10^{-4}$ & 104.4$^{+1.0}_{-1.0}$ & 300$^{+100}_{-100}$ & 2.1$^{+0.5}_{-0.5}$ & N & 10$^{+3}_{-3}$ & 1 & 1 \\
HD142446 & 135.6$^{+0.4}_{-0.4}$ & 6540$^{+100}_{100}$ & 1.51$^{+0.02}_{-0.02}$ & 3.76$^{+0.04}_{-0.04}$ & 1.8$^{+0.2}_{-0.2} \times 10^{-4}$ & 160$^{+10}_{-10}$ & 200$^{+300}_{-300}$ & 1.5$^{+1.0}_{-1.0}$ & Y & 16$^{+2}_{-2}$ & 1 & 1 \\
HD145560 & 121.2$^{+0.3}_{-0.3}$ & 6380$^{+100}_{100}$ & 1.48$^{+0.02}_{-0.03}$ & 3.25$^{+0.04}_{-0.03}$ & 2.0$^{+0.3}_{-0.3} \times 10^{-3}$ & 50$^{+5}_{-5}$ & 300$^{+300}_{-300}$ & 1.3$^{+0.7}_{-0.7}$ & Y & 16$^{+2}_{-2}$ & 1 & 1 \\
HD146181 & 127.5$^{+0.3}_{-0.3}$ & 6360$^{+100}_{100}$ & 1.38$^{+0.02}_{-0.03}$ & 2.82$^{+0.03}_{-0.03}$ & 1.9$^{+0.2}_{-0.2} \times 10^{-3}$ & 85$^{+2}_{-2}$ & 500$^{+500}_{-500}$ & 0.7$^{+0.7}_{-0.7}$ & N & 16$^{+2}_{-2}$ & 1 & 1 \\
HD146897 & 132.2$^{+0.4}_{-0.4}$ & 6100$^{+100}_{100}$ & 1.64$^{+0.02}_{-0.02}$ & 3.37$^{+0.04}_{-0.04}$ & 8.0$^{+0.4}_{-0.3} \times 10^{-3}$ & 91$^{+1}_{-1}$ & 400$^{+300}_{-300}$ & 1.3$^{+0.8}_{-0.8}$ & N & 10$^{+3}_{-3}$ & 1 & 1 \\
HD147137 & 143.5$^{+0.5}_{-0.5}$ & 6230$^{+100}_{100}$ & 1.78$^{+0.01}_{-0.01}$ & 4.31$^{+0.04}_{-0.04}$ & 4.4$^{+0.1}_{-0.2} \times 10^{-4}$ & 163$^{+3}_{-3}$ & 200$^{+200}_{-200}$ & 1.6$^{+1.0}_{-1.0}$ & Y & 10$^{+3}_{-3}$ & 1 & 1 \\
HD158352 & 63.8$^{+0.3}_{-0.3}$ & 7450$^{+100}_{100}$ & 2.77$^{+0.03}_{-0.03}$ & 21.4$^{+0.3}_{-0.3}$ & 7.1$^{+0.5}_{-0.5} \times 10^{-5}$ & 52$^{+2}_{-2}$ & 130$^{+100}_{-100}$ & 0.6$^{+0.2}_{-0.2}$ & Y & 900$^{+700}_{-700}$ & 3 & 5 \\
HD161868 & 29.7$^{+0.2}_{-0.2}$ & 9070$^{+100}_{100}$ & 2.01$^{+0.02}_{-0.02}$ & 24.5$^{+0.5}_{-0.5}$ & 6.3$^{+0.5}_{-0.6} \times 10^{-5}$ & 57$^{+8}_{-8}$ & 100$^{+90}_{-90}$ & 1.1$^{+0.1}_{-0.1}$ & Y & 300$^{+200}_{-200}$ & 0 & 4 \\
HD164249 & 49.30$^{+0.06}_{-0.06}$ & 6340$^{+100}_{100}$ & 1.48$^{+0.02}_{-0.02}$ & 3.20$^{+0.03}_{-0.03}$ & 8.4$^{+0.2}_{-0.2} \times 10^{-4}$ & 71.3$^{+0.5}_{-0.5}$ & 130$^{+20}_{-20}$ & 1.0$^{+0.1}_{-0.1}$ & N & 23$^{+3}_{-3}$ & 1 & 1 \\
GSC07396-00759 & 71.8$^{+0.1}_{-0.1}$ & 3620$^{+100}_{100}$ & 0.94$^{+0.01}_{-0.01}$ & 0.136$^{+0.001}_{-0.001}$ & 8$^{+97}_{-6} \times 10^{-5}$ & 10$^{+10}_{-10}$ & 200$^{+400}_{-400}$ & 1.2$^{+0.9}_{-0.9}$ & N & 23$^{+3}_{-3}$ & 1 & 1 \\
HD170773 & 36.93$^{+0.04}_{-0.04}$ & 6620$^{+100}_{100}$ & 1.44$^{+0.01}_{-0.01}$ & 3.56$^{+0.02}_{-0.02}$ & 4.9$^{+0.1}_{-0.1} \times 10^{-4}$ & 39.3$^{+0.7}_{-0.7}$ & 70$^{+10}_{-10}$ & 0.87$^{+0.04}_{-0.04}$ & N & 200$^{+100}_{-100}$ & 3 & 3 \\
HD172167 & 7.68$^{+0.02}_{-0.02}$ & 10710$^{+100}_{100}$ & 2.407$^{+0.008}_{-0.007}$ & 69$^{+1}_{-1}$ & 4.27$^{+0.04}_{-0.04} \times 10^{-6}$ & 134.4$^{+0.5}_{-0.5}$ & 900$^{+40}_{-40}$ & 0.7$^{+0.1}_{-0.1}$ & N & 460$^{+10}_{-10}$ & 0 & 14 \\
HD181327 & 47.78$^{+0.07}_{-0.07}$ & 6450$^{+100}_{100}$ & 1.35$^{+0.01}_{-0.01}$ & 2.83$^{+0.02}_{-0.02}$ & 2.71$^{+0.05}_{-0.05} \times 10^{-3}$ & 80.0$^{+0.3}_{-0.3}$ & 340$^{+60}_{-60}$ & 0.56$^{+0.07}_{-0.07}$ & N & 23$^{+3}_{-3}$ & 1 & 1 \\
HD182681 & 70.7$^{+0.4}_{-0.4}$ & 9540$^{+100}_{100}$ & 1.87$^{+0.02}_{-0.02}$ & 26.1$^{+0.5}_{-0.5}$ & 2.8$^{+0.1}_{-0.1} \times 10^{-4}$ & 80$^{+7}_{-7}$ & 100$^{+200}_{-200}$ & 0.7$^{+0.1}_{-0.1}$ & N & 110$^{+90}_{-90}$ & 0 & 15 \\
HD191089 & 50.11$^{+0.05}_{-0.05}$ & 6460$^{+100}_{100}$ & 1.32$^{+0.01}_{-0.01}$ & 2.73$^{+0.02}_{-0.02}$ & 1.57$^{+0.03}_{-0.03} \times 10^{-3}$ & 94.3$^{+0.8}_{-0.8}$ & 220$^{+70}_{-70}$ & 0.7$^{+0.2}_{-0.2}$ & N & 23$^{+3}_{-3}$ & 1 & 1 \\
HD197481 & 9.714$^{+0.002}_{-0.002}$ & 3597$^{+100}_{100}$ & 0.808$^{+0.005}_{-0.005}$ & 0.0984$^{+0.0009}_{-0.0009}$ & 3.9$^{+0.2}_{-0.1} \times 10^{-4}$ & 49$^{+3}_{-3}$ & 200$^{+300}_{-300}$ & 0.22$^{+0.08}_{-0.08}$ & N & 23$^{+3}_{-3}$ & 1 & 1 \\
HD202628 & 23.79$^{+0.01}_{-0.01}$ & 5820$^{+100}_{100}$ & 0.974$^{+0.006}_{-0.006}$ & 0.980$^{+0.005}_{-0.005}$ & 1$^{+0}_{-1} \times 10^{-4}$ & 40$^{+50}_{-50}$ & 100$^{+100}_{-100}$ & 0.9$^{+0.6}_{-0.6}$ & Y & 1100$^{+400}_{-400}$ & 0 & 16 \\
HD205674 & 55.74$^{+0.09}_{-0.09}$ & 6640$^{+100}_{100}$ & 1.38$^{+0.01}_{-0.01}$ & 3.37$^{+0.02}_{-0.02}$ & 3.5$^{+0.1}_{-0.1} \times 10^{-4}$ & 54.8$^{+1.0}_{-1.0}$ & 170$^{+50}_{-50}$ & 0.9$^{+0.2}_{-0.2}$ & N & 800$^{+600}_{-600}$ & 3 & 17 \\
HD206893 & 40.77$^{+0.06}_{-0.06}$ & 6500$^{+100}_{100}$ & 1.33$^{+0.01}_{-0.01}$ & 2.84$^{+0.02}_{-0.02}$ & 2.70$^{+0.09}_{-0.09} \times 10^{-4}$ & 47$^{+4}_{-4}$ & 100$^{+100}_{-100}$ & 0.9$^{+0.2}_{-0.2}$ & N & 160$^{+20}_{-20}$ & 0 & 25 \\
HD207129 & 15.56$^{+0.01}_{-0.01}$ & 5880$^{+100}_{100}$ & 1.062$^{+0.008}_{-0.008}$ & 1.219$^{+0.007}_{-0.007}$ & 9.3$^{+0.3}_{-0.3} \times 10^{-5}$ & 44.7$^{+0.8}_{-0.8}$ & 75$^{+8}_{-8}$ & 0.82$^{+0.07}_{-0.07}$ & N & 3000$^{+2000}_{-1000}$ & 0 & 4 \\
TYC93404371 & 36.72$^{+0.02}_{-0.02}$ & 4000$^{+100}_{100}$ & 0.911$^{+0.005}_{-0.005}$ & 0.191$^{+0.001}_{-0.001}$ & 1.04$^{+0.03}_{-0.03} \times 10^{-3}$ & 45.5$^{+1.0}_{-1.0}$ & 40$^{+40}_{-40}$ & 0.02$^{+0.01}_{-0.01}$ & N & 23$^{+3}_{-3}$ & 1 & 1 \\
HD216956 & 7.70$^{+0.03}_{-0.03}$ & 8470$^{+100}_{100}$ & 1.89$^{+0.01}_{-0.01}$ & 16.5$^{+0.2}_{-0.2}$ & 1.41$^{+0.02}_{-0.02} \times 10^{-5}$ & 134.8$^{+0.7}_{-0.7}$ & 500$^{+800}_{-800}$ & 0.01$^{+0.04}_{-0.04}$ & N & 440$^{+40}_{-40}$ & 0 & 18 \\
HD216956C & 7.676$^{+0.002}_{-0.002}$ & 3059$^{+100}_{100}$ & 0.249$^{+0.001}_{-0.001}$ & 0.00489$^{+0.00004}_{-0.00004}$ & 1.7$^{+0.2}_{-0.2} \times 10^{-4}$ & 24$^{+5}_{-5}$ & 100$^{+200}_{-200}$ & 1.3$^{+0.5}_{-0.5}$ & N & 440$^{+40}_{-40}$ & 0 & 18 \\
HD218396 & 40.88$^{+0.08}_{-0.08}$ & 7290$^{+100}_{100}$ & 1.47$^{+0.01}_{-0.01}$ & 5.49$^{+0.05}_{-0.05}$ & 3.7$^{+0.3}_{-0.3} \times 10^{-5}$ & 176$^{+6}_{-6}$ & 31$^{+2}_{-2}$ & 2.2$^{+0.7}_{-0.7}$ & Y & 30$^{+10}_{-10}$ & 0 & 19 \\
\hline
\end{longtable}
\tablefoot{Distances are from \textit{Gaia} DR3 \citep{Gaia2016b,Gaia2023j}, or from \textit{Hipparcos} if Gaia measurement not available \citep{vanLeeuwen2007}. Effective temperature uncertainties derived here from multi-band photometry have been assigned a floor of $\pm100$ K to account for systematics; the value of 100 K was chosen as a reasonable estimate from testing photometric against spectroscopic temperature measurements \citep{Pearce2022}. Age notes corresponding to: 1. Membership from BANYAN Sigma \citep[][]{Gagne2018}, or moving group age also from \citet[][]{Gagne2018} - 2. Membership from BANYAN Sigma \citep[][]{Gagne2018}, but moving group age from \citet[][]{Zuckerman2019} - 3. Age uncertainty (symmetric) conservatively estimated as 25\%$\times$Age$\times$log$_{10}$(Age/Myr) \citep[][]{Pawellek2015}.
\tablebib{Age references corresponding to: 1. This work - 2. \citet[][]{Chen2014} - 3. \citet[][]{Rhee2007a} - 4. \citet[][and refs. therein]{Pearce2022} - 5. \citet[][]{ZorecRoyer2012} - 6. \citet[][]{Kalas2005} - 7. \citet[][]{Kennedy2015} - 8. \citet[][]{Eiroa2013} - 9. \citet[][]{Marshall2016} - 10. \citet[][]{Vican2016}   - 11. \citet[][]{Booth2021} - 12. \citet[][]{Brandt2014} - 13. \citet[][]{DodsonRobinson2016} - 14. \citet[][]{Yoon2010} - 15. \citet[][]{Gray2006} - 16. \citet[][]{Faramaz2019} - 17. \citet[][]{Casagrande2011} - 18. \citet[][]{Mamajek2012} - 19. \citet[][]{Baines2012} - 20. \citet[][]{Currie2017} - 21. \citet[][]{Torres2008} - 22. \citet[][]{Moor2011b} - 23. \citet[][]{Decin2000} - 24. \citet[][]{Desidera2015} - 25. \citet[][]{Hinkley2023}  }}
\end{landscape}

\section{Undetected, Unresolved, and/or Contaminated Targets}
\label{sec:confused}

\begin{figure*}
\hspace{-3mm}
\includegraphics[scale=0.37]{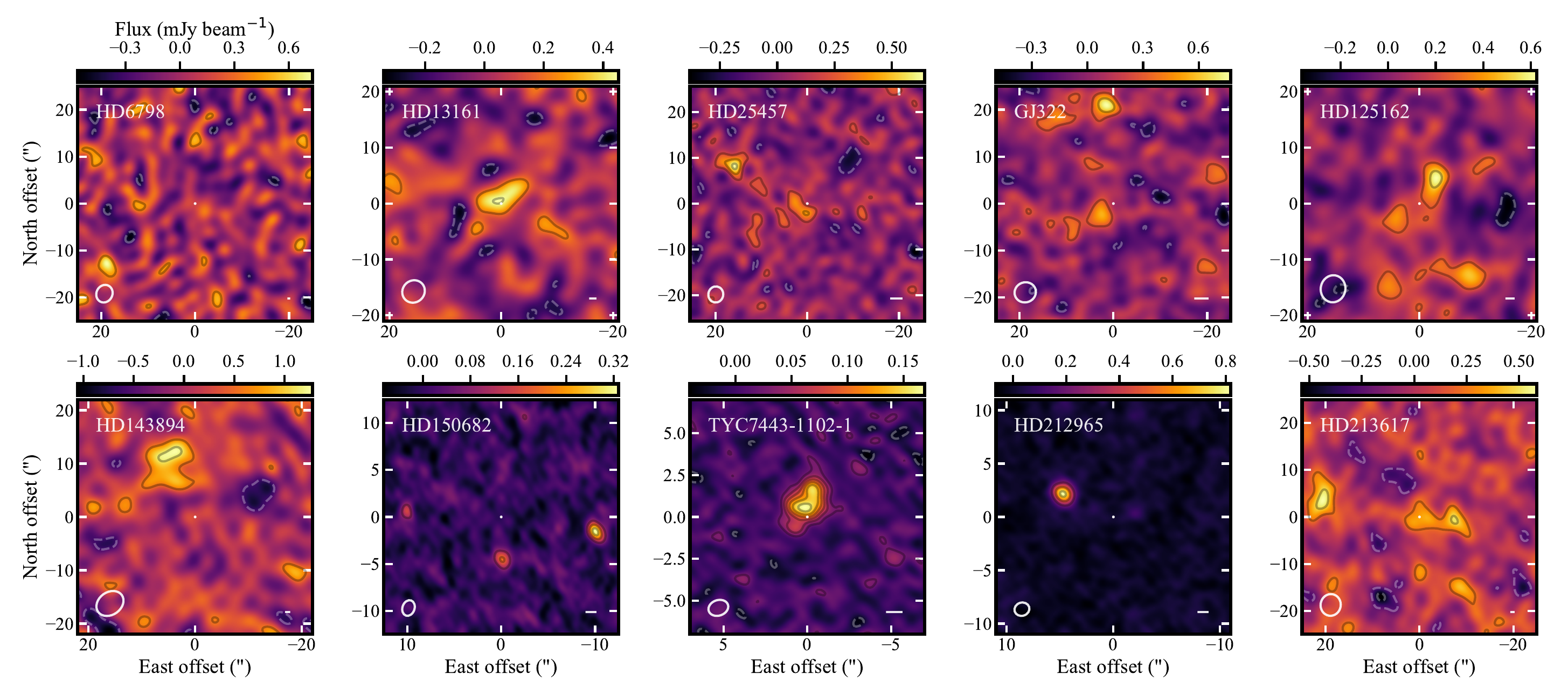}
%\vspace{0mm}
%\end{subfigure} \\
%\vspace{-2mm}
\caption{Millimetre continuum images for the 10 systems in the REASONS observing programme where a belt was not detected or significantly resolved, ordered by source RA. North is up, East is left. Bars indicate a physical scale of 50 au, and ellipses represent the synthesized beam of the observations. Images were obtained with the CLEAN algorithm as described in Sect. \ref{sec:obs}, with weighting parameters, resulting RMS noise levels and beams listed in the tables available on \textsc{Zenodo} (see Sect. \ref{sec:dataavail}). All images are in a linear colour scale, stretching from 0 (black) to the maximum intensity of the image. Contours are set at [-4,-2,2,4,..] times the RMS noise level of each image.}
\label{fig:galleryconfused}
\end{figure*}

Figure \ref{fig:galleryconfused} shows continuum images for the 10/25 belts in the REASONS observing programme that were unexpectedly not detected and/or spatially resolved. We here comment on possible reasons for this in each system.

\textit{HD6798}:
We marginally detect a point source (SMA 011224.5+794012.3) at the 3$\sigma$ level SE of the stellar location, with a primary beam-corrected peak surface brightness of $1.3\pm0.4$ mJy/beam at 1.31 mm. An image-plane Gaussian fit leads to a centroid position that is offset by $18.8\arcsec\pm0.3\arcsec$ and $-13.8\arcsec\pm0.5\arcsec$ from the map center along the RA and Dec directions, respectively. 

Previous JCMT 850 $\mu$m observations detected a point source with a location consistent with the stellar location, and a flux density of $7.2\pm1.0$ mJy \citep{Holland2017}. The SMA nondetection sets a 3$\sigma$ upper limit on the flux density for an unresolved point source of $<$0.6 mJy, which would indicate a steep mm spectral slope of $\gtrsim5.7$ when compared to the JCMT flux alone. The offset point source SMA 011224.5+794012.3 marginally detected here is not detected in the JCMT map nor in archival \textit{Herschel} PACS maps at 100 and 160 $\mu$m. 

HD6798 was observed by \textit{Herschel} at far-IR wavelengths; its emission is consistent with the stellar location, significantly brighter than the expected stellar contribution and unresolved in the $\sim$7$\arcsec$ resolution PACS 100 $\mu$m maps. This constrains the diameter of the belt to $\lesssim5\arcsec$, and rules out the possibility that the belt was interferometrically resolved out by the SMA observations. Then, either there really is a break in the spectral slope between 850 $\mu$m and 1.3 mm, which is unlikely, or the source seen by the JCMT at the field center corresponds to SMA 011224.5+794012.3, which would imply a very large systematic pointing offset. Either way, the \textit{Herschel} observations imply that a belt is most likely present around HD6798, but we cannot reconcile its mm flux of $<$0.6 mJy measured by the SMA at 1.31 mm with the previous 850 $\mu$m JCMT detection for a typical collisionally evolving planetesimal belt. \\

\begin{figure*}
\hspace{-20mm}
\includegraphics[scale=0.28]{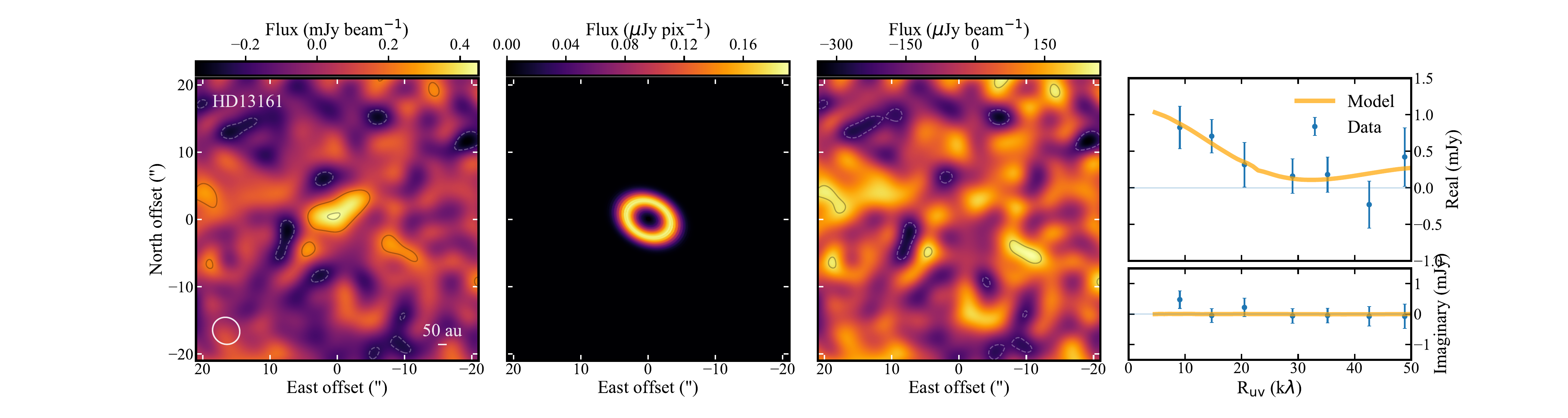}
%\vspace{0mm}
%\end{subfigure} \\
%\vspace{-8mm}
\caption{HD13161 ($\beta$ Tri) SMA modelling results. Left to right panels: SMA image (contours at intervals of 2x the RMS noise level of 0.11 mJy beam$^{-1}$), full resolution best-fit model image of the belt, residual image after subtraction of best-fit visibilities, and real and imaginary parts of the azimuthally averaged complex visibility profiles, for both the data and the best-fit model. The best-fit model belt's inclination and position angle, largely unconstrained by the SMA data, were fixed to the \textit{Herschel} results of \citet{Booth2013}. The belt width, also largely unconstrained by the model, was fixed to 100 au.}
\label{fig:HD13161combo}
\end{figure*}

\textit{HD13161 ($\beta$ Tri)}:
We detect emission at a location consistent with that of the star, which appears marginally resolved from visual inspection of the real part of the visibility function (Fig. \ref{fig:HD13161combo}, rightmost panel, de-projected assuming inclination and position angle from \textit{Herschel} resolved data). The very low surface brightness prevents us from drawing a definitive conclusion on the origin of the emission in the image plane. We expect the star itself - when extrapolating from IR data assuming Rayleigh-Jeans emission - to have a flux density of $\sim$0.2 mJy, and significantly contribute to the emission (given the SMA map's noise level of 0.1 mJy).

We therefore fit the visibilities with a model comprising a belt and a point source, both centred at the phase center of the SMA observations (corresponding to the stellar location). The fit results indicate marginal detection of a belt with total flux density of $0.8\pm0.4$ mJy, and of the star with flux density $0.28\pm0.12$ mJy. Peaks in the posterior probability distributions are suggestive of a broad belt having a radius of $\sim$140 au and width $\sim$100 au, with mostly unconstrained inclination and position angle. The location of peak probability in belt radius is consistent with previous \textit{Herschel} resolved imaging \citep{Booth2013}. In Fig. \ref{fig:HD13161combo} we plot the residuals obtained by fixing the inclination ($\sim41.3^{\circ}$) and PA ($\sim64.3^{\circ}$) from the \textit{Herschel} results, showing that these values also produce a reasonable fit to the data given the large uncertainties. In conclusion, we confirm the presence of a belt around HD13161, which is however only marginally detected and resolved by the SMA observations. \\

\textit{HD25457}:
The field centred on HD25457 appears significantly confused. The image obtained with natural weighting of the visibilities (Fig. \ref{fig:galleryconfused}) shows a strong (6.7$\sigma$) detection of a source, SMA 040238.99-001604.8, $15.9\arcsec\pm0.3\arcsec$ and $8.0\arcsec\pm0.3\arcsec$ in the RA and Dec directions, respectively, NE of the star. The measured primary beam corrected peak flux is $0.9\pm0.2$ mJy beam$^{-1}$. Another source elongated over $\sim$two beams in the NE-SW direction (SMA 040237.10-001613.9), is detected closer, but again offset ($2\arcsec$, or $\sim$0.6 times the size of a beam) SE of the center of the map. This source has a flux (spatially integrated within a 5$\arcsec$ radius of the image center) of $0.9\pm0.2$ mJy. In previous JCMT observations of this field, the emission attributed to a belt around HD25457 was also found to be $\sim2\arcsec$ offset in the same direction \citep{Holland2017} as that of our latter source (SMA 040237.10-001613.9). Compared to the JCMT flux of $6.2\pm1.4$ mJy at 850 $\mu$m, our measured flux of $0.9\pm0.2$ mJy at 1.27 mm would suggest a relatively steep spectral slope of $\alpha=4.8\pm0.7$.

\begin{figure}
\centering
\includegraphics[scale=0.45]{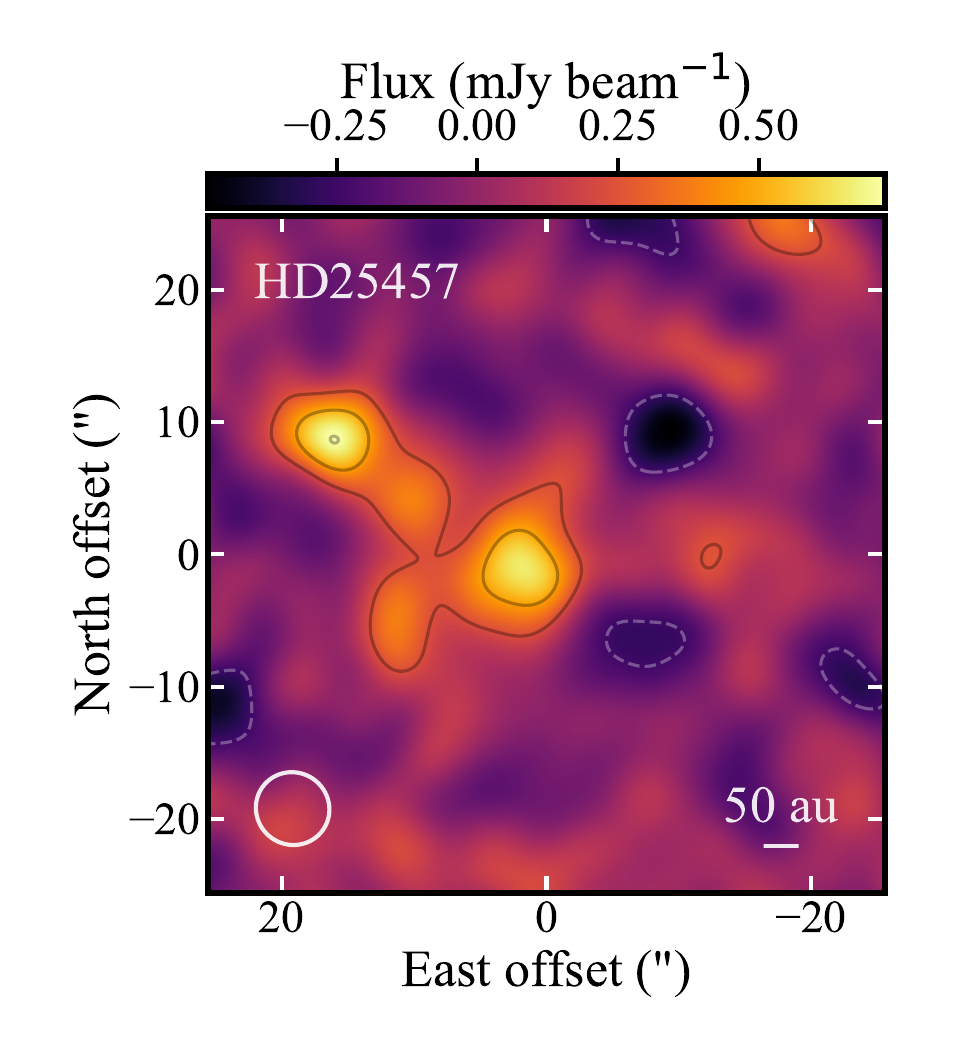}
%\vspace{0mm}
%\end{subfigure} \\
%\vspace{-5mm}
\caption{SMA continuum image of the field around the star HD25457, after application of a 5$\arcsec$ u-v taper. Contours are set at intervals of 2x the RMS noise level of 0.12 mJy beam$^{-1}$.}
\label{fig:HD25457_5astap}
\end{figure}

Inspection of the u-v radial profile after subtraction of the strongest, most offset source (SMA 040238.99-001604.8) indicates a larger amount of emission on larger scales ($\lesssim15$ k$\lambda$) than captured in the image obtained with natural weights. Figure \ref{fig:HD25457_5astap} shows an image obtained after applying a $5\arcsec$ u-v taper, with an RMS noise level of 120 $\mu$Jy and a resolution of $\sim5.5\arcsec$. This suggests that the two sources are part of an extended complex of emission to the E of the stellar location, which comprises two further $\sim3\sigma$ peaks in the tapered image. The presence of such extended emission was also noted in the lower resolution, JCMT 850 $\mu$m map, while it could not be resolved at IR wavelengths with IRAS and MIPS at 60 and 70 $\mu$m. We therefore cannot robustly conclude on the nature of the emission in the surroundings of HD25457; higher resolution and SNR observations are necessary. However, the complex, multi-component nature of the emission, and the moderate offset of the central peak from the predicted stellar location argue against a belt nature for the observed mm emission. \\

\textit{GJ322}:
A compact source (SMA J085200.95+660815.8) is detected at the 5.5$\sigma$ level N of the stellar position (Fig. \ref{fig:galleryconfused}), with a primary beam-corrected peak surface brightness of $1.2\pm0.2$ mJy/beam. A 2D Gaussian image-plane fit yields a centroid that is offset by $1.9\arcsec\pm0.4\arcsec$ and $20.6\arcsec\pm0.3\arcsec$ along the RA and Dec directions, respectively, from the center of the map. We also find a 4$\sigma$ peak nearer the stellar location, with primary-beam corrected peak surface brightness of $0.56\pm0.14$ mJy/beam. However, this source is also significantly offset from the stellar location, by $2.9\arcsec\pm0.5\arcsec$ and $-2.5\arcsec\pm0.6\arcsec$ along the RA and Dec direction. 

\begin{figure}
\hspace{-0mm}
\centering
  \includegraphics[scale=0.35]{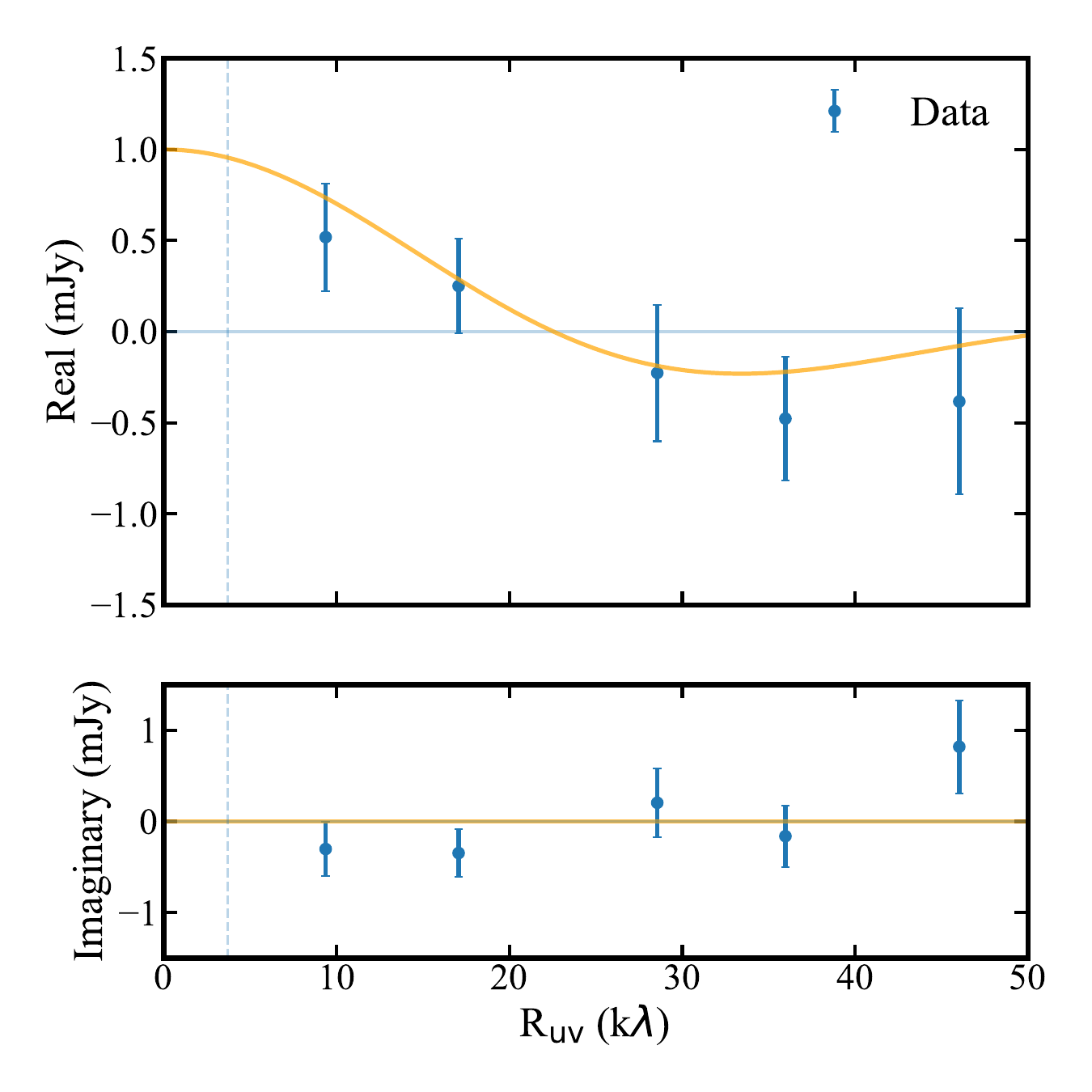}
%\vspace{0mm}
%\end{subfigure} \\
%\vspace{-5mm}
\caption{u-v radial profile of measured SMA complex-valued visibility function (blue error bars), at the phase center of the observations (corresponding to the location of star GJ322). In orange is a model of a face-on belt with a Gaussian radial surface density distribution, with $\sim$3.5$\arcsec$ radius, $\sim$1.0$\arcsec$ width, and $\sim$1 mJy total flux density, showing that the offset peak seen in Fig. \ref{fig:galleryconfused} does not strongly rule out the presence of a belt around GJ322.}
\label{fig:GJ322deprojvis}
\end{figure}

The previous 850 $\mu$m JCMT detection was located at a position consistent with that of the star, with a flux density of $7.3\pm1.4$ mJy, while the stronger source to the N of the star was not detected. We identify a few possibilities to explain this discrepancy: 1) the JCMT detection corresponds to the point source peaking close to the center of the SMA map, which would however imply an unlikely steep spectral slope of $6.4\pm0.5$ for the source. 2) the JCMT detection corresponds to the N source (SMA J085200.95+660815.8), and appears in the center of the JCMT map due to a systematic pointing offset, which is however also unlikely \citep[see Fig. 32 of][]{Holland2017}. 3) Both sources in the SMA map are quasars or other sources with significant time variability to explain the SMA-JCMT discrepancy. 4) The central peak in the SMA map is part of a belt whose flux was significantly resolved out interferometrically. Figure \ref{fig:GJ322deprojvis} shows the u-v radial profile of the real and imaginary part of the SMA visibilities, after u-v plane subtraction of the N point source SMA J085200.95+660815.8. As proof of concept, this shows that, for example, we cannot strongly rule out a face-on belt model with $\sim$3.5$\arcsec$ radius, $\sim$1.0$\arcsec$ width, and $\sim$1 mJy total flux density. 

While deeper observations are needed to conclude whether a belt is truly associated with GJ322, we here take the SMA data point at the shortest u-v distance in Fig. \ref{fig:GJ322deprojvis} to derive a 3$\sigma$ upper limit of $<1.5$ mJy on the total flux density of the belt. This might still require a spectral slope too steep to be reasonable for a planetesimal belt, but the current uncertainties on the JCMT and SMA fluxes do not allow us to reach this conclusion. \\

\begin{figure*}
\hspace{-20mm}
\includegraphics[scale=0.28]{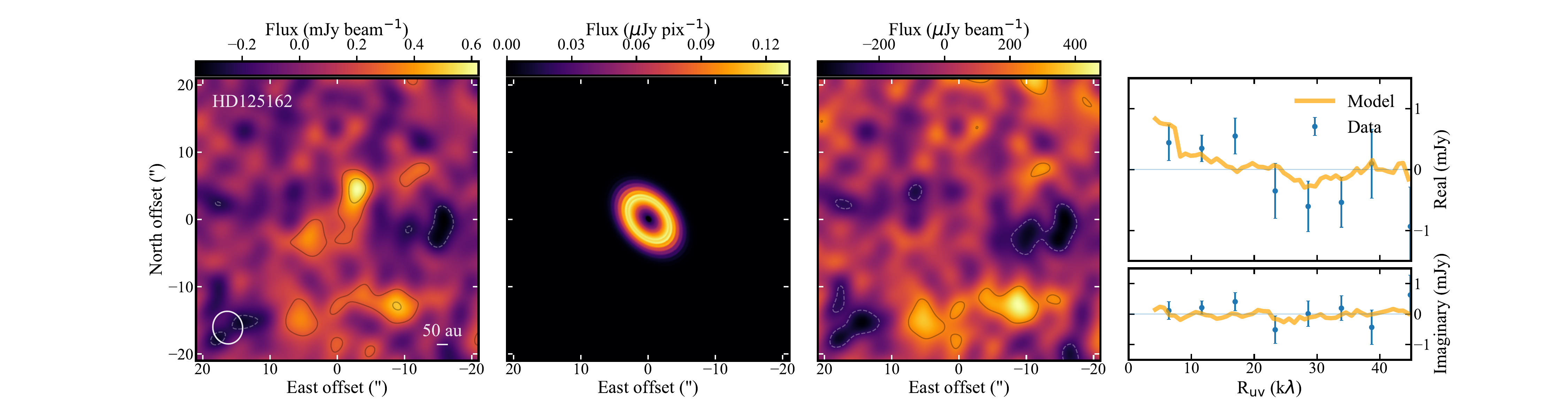}
%\vspace{0mm}
%\end{subfigure} \\
%\vspace{-8mm}
\caption{HD125162 ($\lambda$ Boo) SMA modelling results. Left to right panels: SMA image (contours at intervals of 2x the RMS noise level of 0.12 mJy beam$^{-1}$), full resolution best-fit model image of the belt, residual image after subtraction of best-fit visibilities, and real and imaginary parts of the azimuthally averaged complex visibility profiles, for both the data and the best-fit model. The best-fit model belt's inclination and position angle, largely unconstrained by the SMA data, were fixed to the \textit{Herschel} results of \citet{Booth2013}. The belt width, also unconstrained by the model, was fixed to 80 au.}
\label{fig:HD125162combo}
\end{figure*}

\textit{HD125162 ($\lambda$ Boo)}:
A point-like emission peak is detected at the 5.5$\sigma$ level NW of the stellar location (SMA J141622.4+460525.2), with additional low level 1-2$\sigma$ emission seen over 2-3 beams across the map center, in the NW-SE direction (Fig. \ref{fig:galleryconfused}). We also detect a 4$\sigma$ peak at a larger offset, $\sim$15$\arcsec$ to the SW of the star. Inspection of the u-v radial profile of the real part of the SMA visibilities (see e.g. Fig. \ref{fig:HD125162combo}, rightmost panel), indicates marginal detection of extended emission (at the shortest u-v spacings), likely corresponding to the low level emission seen at the stellar location, and extended in the NW-SE direction.

We therefore attempt to model the visibilities from the HD125162 field as a belt + offset point source combination. The brightest, NW point source (SMA J141622.4+460525.2) is significantly recovered with a flux density of $0.45^{0.14}_{-0.15}$ mJy and an offset of $-2.7\arcsec\pm0.4\arcsec$ and $4.4\arcsec\pm0.7\arcsec$ in the RA and Dec directions, respectively. The HD125162 belt is marginally detected, with posterior probability distributions with a peak in belt radius at $\sim$120 au, which would be broadly consistent with previous \textit{Herschel} resolved images \citep{Booth2013}, flux density of $\sim0.7^{+0.4}_{-0.3}$
mJy at 1.27 mm, and largely unconstrained inclination and position angle. Given the belt is barely detected and resolved, we do not include HD125162 in our resolved sample; deeper observations are needed to confirm the presence of a resolved belt and disentangle it from nearby contaminating sources. \\

\textit{HD143894 (44 Ser)}: A marginally resolved source (SMA J160218.04+224826.5) is significantly detected NE of the stellar position (Fig. \ref{fig:galleryconfused}). An image-plane 2D Gaussian fit to the emission in the image plane yields a spatially integrated, primary beam corrected flux density of $3.9\pm1.3$ mJy at 1.27 mm, and best-fit RA and Dec offsets, respectively, of $4.7\arcsec\pm0.8\arcsec$ and $10.0\arcsec\pm0.7\arcsec$.
Although noticeably smaller, a hint of such offset was already present in the source detected in the 850 $\mu$m JCMT data, where the total flux measured within a 40$\arcsec$ aperture was $10.1\pm1.2$ mJy \citep{Holland2017}. Combined with the SMA flux density reported here, the spectral slope is $2.4\pm0.3$, consistent with the expectation of thermal emission from dust grains. Given the large SMA offset, this source of mm emission is not associated with the star. We set a $3\sigma$ upper limit to unresolved ($\lesssim5\arcsec$) emission at the location of the star of 0.9 mJy.

This new evidence indicates that emission across the IR to mm spectrum is likely dominated by the unrelated, offset source SMA J160218.04+224826.5, with no significant evidence for a cold belt associated with the star HD143894. \\

\textit{HD150682}:
Three compact sources are significantly detected in the field centred on HD150682 observed by ALMA at 1.27 mm (Fig. \ref{fig:galleryconfused}), all significantly offset from the stellar location. 2D Gaussian fits in the image plane yield the following characteristics for the three sources: ALMA J164135.96+265458.3, offset by $-9.93\arcsec\pm0.04\arcsec$ in RA, $-1.67\arcsec\pm0.06\arcsec$ in Dec from the star, with primary beam-corrected flux density of $0.52\pm0.06$ mJy; ALMA J164136.70+265455.4, offset by $-0.01\arcsec\pm0.07\arcsec$ in RA, $-4.59\arcsec\pm0.09\arcsec$ in Dec from the star, with primary beam-corrected flux density of $0.28\pm0.05$ mJy; and ALMA J164137.46+265500.3, offset by $10.09\arcsec\pm0.04\arcsec$ in RA, $0.29\arcsec\pm0.16\arcsec$ in Dec from the star, with primary beam-corrected flux density of $0.23\pm0.06$ mJy.
The only evidence for excess associated with the star is from a \textit{Spitzer} detection at 70 $\mu$m with a spatial resolution of 18$\arcsec$. We therefore conclude that the MIPS emission is likely attributable to the three offset sources detected by ALMA, and that there is therefore no significant evidence of excess emission around the star HD150682. \\

\textit{TYC7443-1102-1}:
The ALMA 1.27 mm data shows strong emission significantly offset with a peak N of the stellar location, and complex structure elongated in the SE-NW direction (Fig. \ref{fig:galleryconfused}). This is consistent with the structure seen at higher resolution by ALMA at 0.87 mm, which resolved the same extended structure into two separate sources \citep{Tanner2020}. Their analysis confirms that the emission detected by \textit{Herschel} is not associated with the star, which therefore does not show significant evidence for an IR and/or mm excess. \\

\textit{HD212695}:
A bright source, ALMA J222614.72-024719.8 is detected to the ENE of the location of HD212695 in the ALMA 1.27 mm data (Fig. \ref{fig:galleryconfused}). A 2D Gaussian fit in the image plane indicates that the source is marginally resolved, with deconvolved FWHM of $\sim1.2\arcsec$, inclination of $\sim50^{\circ}$ and PA of $\sim16^{\circ}$. The best-fit, primary beam-corrected spatially integrated flux density is $1.38\pm0.06$ mJy, and the source is significantly offset by the stellar location, by $4.76\arcsec\pm0.02\arcsec$ in RA, $2.04\arcsec\pm0.02\arcsec$ in Dec.
As for HD150682, the only evidence for excess associated with the star is from a \textit{Spitzer} MIPS detection at 70 $\mu$m with a spatial resolution 18$\arcsec$, which could not resolve the above offset. Therefore, it is likely that both the IR and mm emission are associated with unrelated source ALMA J222614.72-024719.8, and we conclude that there is no significant evidence for excess emission associated with the star HD212695. \\

\begin{figure}
\hspace{-0mm}
\centering
  \includegraphics[scale=0.35]{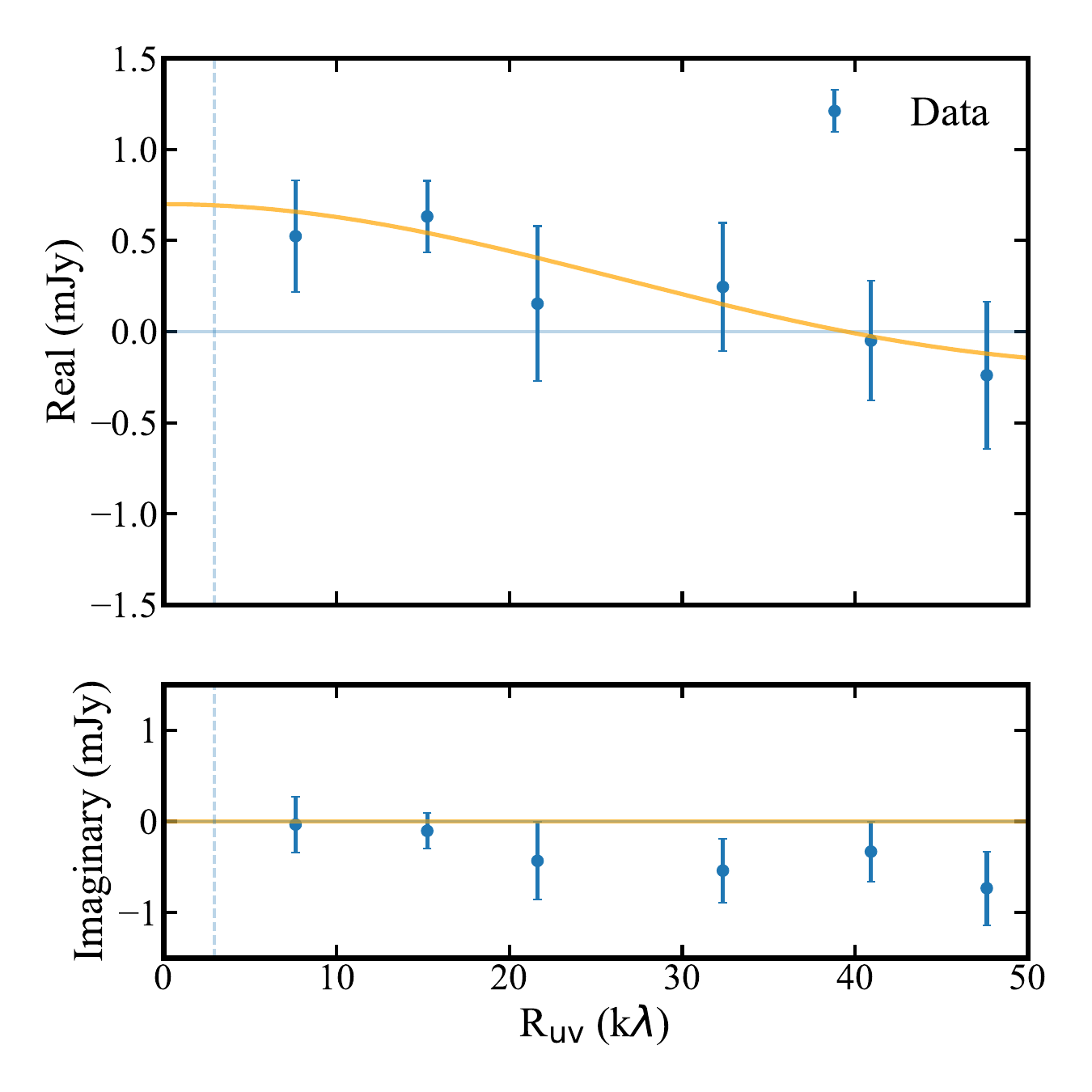}
%\vspace{0mm}
%\end{subfigure} \\
%\vspace{-5mm}
\caption{u-v radial profile of measured SMA complex-valued visibility function (blue error bars), at the phase center of the observations (corresponding to the location of star HD213617), after subtraction of the offset sources SMA 223237.17+201352.3 and SMA 223235.16+201347.6. In orange is a model of a face-on belt with a Gaussian radial surface density distribution, with 2.0$\arcsec$ radius, 0.5$\arcsec$ width, and 0.7 mJy total flux density, showing that the faint, marginal emission seen in Fig. \ref{fig:galleryconfused} at the phase center could plausibly originate from a belt around HD213617.}
\label{fig:HD213617deprojvis}
\end{figure}

\textit{HD213617}:
At least two sources are detected in the field around the star HD213617 (Fig. \ref{fig:galleryconfused}). The strongest source (SMA 223237.17+201352.3), detected at the 4.5$\sigma$ level with a primary beam corrected peak flux of 0.8 mJy beam$^{-1}$, appears $20.8\arcsec\pm0.5\arcsec$ and $3.6\arcsec\pm0.9\arcsec$ away from the star in the RA and Dec directions, respectively. The source is marginally resolved; a 2D Gaussian fit in the image plane yields a FWHM of $\sim$6.1$\arcsec$ $\times$ 3.4$\arcsec$ and position angle of -2$^{\circ}$, and a spatially integrated, primary beam corrected flux density of 1.8$\pm$0.6 mJy.
A point source (SMA 223235.16+201347.6) is also detected at the 4$\sigma$ level $7.5\arcsec$ to the W of the star, with a peak flux of 0.5 mJy beam$^{-1}$. 

Marginal residual emission can be seen at the phase center of the observations after subtraction of these two sources from the visibilities (leading to the de-projected u-v profile, constructed assuming face-on azimuthally symmetric emission, in Fig. \ref{fig:HD213617deprojvis}). The emission is clearest on the shortest u-v distances ($\lesssim20$ k$\lambda$), with a potential decrease at longer baselines, hinting at the presence of resolved emission at the phase center. This is in agreement with the image center, which shows emission at the 2-3$\sigma$ level. We conclude that our data, while indicating that the field is contaminated by two other sources, does not allow us to conclusively rule out the presence of emission originating from a belt around the star. This is demonstrated by the expected de-projected visibility curve of, for example, a 2.0$\arcsec$ radius, 0.5$\arcsec$ width, and 0.7 mJy flux density belt in Fig. \ref{fig:HD213617deprojvis}, which would describe the visibility curve well. We therefore use the shortest set of baselines to set a conservative upper limit of $<1.5$ mJy on the total flux density of any belt that may be present around HD213617.

We note that previous JCMT observations detected $4.6\pm1.3$ mJy of emission at 850 $\mu$m, at a location consistent with that of the star \citep{Holland2017}. Our new SMA dataset indicates that this emission is likely contaminated by nearby source SMA 223235.16+201347.6, which would have been unresolved when observed at the JCMT's spatial resolution. Detection of offset source SMA 223237.17+201352.3, on the other hand, was not reported, though potentially present as a low S/N peak in the JCMT image (Fig. A14 of that manuscript). \\

\end{appendix}
\end{document}